\documentclass{jpp}
\usepackage{graphicx}
\usepackage{epstopdf, epsfig}
\usepackage{hyperref}
\usepackage{amsmath,amsfonts,amssymb}
\usepackage{subfigure}
\usepackage{ragged2e}

\usepackage{comment}

\usepackage{mathptmx}
\usepackage{etoolbox}
\usepackage{ragged2e}

\newcommand{\bs}[1]{\boldsymbol{#1}}

\newcommand{\mc}[1]{\mathcal{#1}}

\newcommand{\half}[0]{\frac{1}{2}}

\newcommand{\paren}[1]{\left(#1\right)}

\newcommand{\wt}{\widetilde}

\newcommand{\avg}[1]{\langle #1 \rangle}

\newcommand{\mr}[1]{\mathrm{#1}}

\newcommand{\vth}{v_\mr{th}}

\newcommand{\lth}{\lambda_\mr{th}}

\newcommand{\fm}{f_\mr{M}}

\renewcommand{\avg}[1]{\left\langle #1 \right\rangle}

\renewcommand{\bs}[1]{\mathbf{#1}}

\newcommand{\reacF}[1]{#1}

\newcommand{\reactop}{\Sigma}

\shorttitle{Fusion-power amplification by compressive hydrodynamic fluctuations}
\shortauthor{H. Fetsch and N. J. Fisch}

\title{Fusion-power amplification by compressive hydrodynamic fluctuations}

\author{Henry Fetsch\aff{1}\corresp{\email{hfetsch@princeton.edu}} and Nathaniel J. Fisch\aff{1}}

\affiliation{\aff{1}Department of Astrophysical Sciences, Princeton University, Princeton, NJ 08540, USA}

\begin{document}

\maketitle

\begin{abstract}
\noindent Compressive fluctuations in hot plasma, including acoustic waves and compressible turbulence, increase the rate of fusion reactions. This power amplification comprises hydrodynamic, ``two-temperature,'' and kinetic components, the first resulting from the clumping of hot ions in the peaks of the fluctuations, the second from the unequal heating of ions and electrons as fluctuations dissipate, and the third from the long mean free paths of fast ions near the Gamow peak, which allow these ions to stream across gradients in fluctuating hydrodynamic fields before colliding. In many cases, the increase in fusion power produced by waves exceeds that produced if the wave energy were instead used for heating. 
Response functions describing the modification to fusion power by compressive fluctuations are obtained in magnetized and unmagnetized fusion plasmas. 
Comparison to the related shear flow reactivity enhancement effect, a kinetic mechanism that increases fusion power in some divergence-free flows, illustrates a fundamental distinction between compressible and solenoidal turbulence in fusion plasmas.
\end{abstract}

\section{Introduction}

In plasmas undergoing thermonuclear fusion, the reaction rate increases superlinearly with temperature and density. As a result, compressive fluctuations tend to increase the volume-averaged rate of energy production. Quantifying this increase is the subject of this work. 
Solenoidal flows, too, even without variations in density or temperature, can enhance fusion reactivity through a recently identified kinetic mechanism \citep{Fetsch_Fisch_2025a,Fetsch_Fisch_2025b, Fetsch_Fisch_2026_dpp}. This ``shear flow reactivity enhancement'' (SFRE) effect provides a boost to fusion power that may be sufficient, in some inertial confinement fusion (ICF) systems, to allow plasma laden with fine-scale solenoidal turbulence to ignite under conditions where thermalized plasma would fail to ignite \citep{Fetsch_Fisch_2026}.  
In light of these results, it is natural to ask whether compressive effects permit a further relaxation of the ignition threshold under some conditions. 

%Surprisingly, the direct effect of fluctuations on fusion power has received little attention. 
Surprisingly, the direct relationship between compressive fluctuations and fusion power has received little attention. 
With a focus on magnetic confinement fusion (MCF) plasmas, \cite{Tidman_1972} demonstrated the instability of magnetosonic waves when the increased fusion power produced in high-pressure regions causes perturbations to grow more rapidly than thermal conduction smooths them out. The same phenomenon, as well as the instability of unmagnetized or field-aligned acoustic waves, was derived independently by \cite{Jackson_1986}, who termed it the ``Sound Amplification by Stimulated Emission of (Acoustic) Radiation'' (SASER) effect. 
Under typical MCF conditions, however, it has been argued by \cite{Bishop_Fitzpatrick_Hastie_Jackson_1989} that nuclear thermoacoustic instabilities tend to be insignificant in fusion devices for two reasons: first, the long slowing-down distance of alpha particles means that self-heating is typically spread over many periods of an acoustic wave and so does not provide localized positive feedback in regions of enhanced fusion rate that would amplify perturbations over time; second, thermal conduction is generally sufficient to suppress the instability. 
By contrast, in stellar cores, particularly those of dense stars, acoustic fluctuations can become unstable because the densities and length scales in these environments -- many times greater than those in MCF -- allow fusion products to deposit their energy locally and increase the heating rate in high-pressure regions \citep{Ledoux_1941,Gabriel_1964,Dilke_Gough_1972}.  
ICF plasmas fall between these regimes, although their short confinement time may limit the impact of growing perturbations \citep{Atzeni_Meyer-Ter-Vehn_2004}.
More generally, thermal instabilities in reacting media have received attention outside of the context of controlled fusion and astrophysics. 
In fact, study of these instabilities long predates the discovery of fusion, beginning with the observation of sound waves in flammable gases driven unstable by an increase in combustion rate in high-pressure regions (cf. ``singing flames'') \citep{Rijke_1859,Richardson_1922,Clavin_1994,Silva_2023}.

\begin{figure}
    \centering
    \includegraphics[width=0.6\columnwidth]{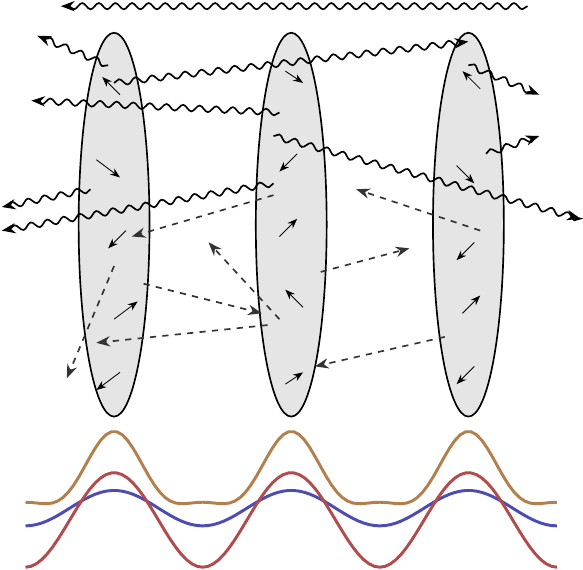}
    \caption{Acoustic wave in a fusion plasma. In compressed regions, elevated density and temperature increase the fusion rate. Thermal ions (solid arrows) have short mean free paths. Fast ions (dotted arrows) have mean free paths comparable to the perturbation wavelength. Alpha particles (wavy lines) have mean free paths much longer than the perturbation wavelength. The perturbed ion temperature is shown in red, the perturbed electron temperature in blue, and the perturbed fusion rate in orange. }
    \label{fig_acoustic_wave} 
\end{figure}

Unaddressed in these treatments of local thermoacoustic stability is the global effect of compressive fluctuations on fusion power. 
It will be shown here that small-amplitude perturbations to density and temperature tend to increase the volume-averaged power produced in fusion plasmas. 
This amplification consists of three parts: a purely hydrodynamic component originating from the nonlinear response of fusion power to variations in density and temperature; a ``two-temperature'' component originating from the unequal heating of ions and electrons as the perturbation dissipates; and a kinetic component originating from non-Maxwellian perturbations to the ion distribution function generated when fast ions, which have low collision frequencies and long mean free paths, traverse spatial gradients of background hydrodynamic quantities or respond sluggishly to temporal variations of the background. 
These effects are represented qualitatively in an acoustic wave in Figure~\ref{fig_acoustic_wave}. At the peaks of the wave, the higher density and temperature produce a higher fusion rate. In this example, however, the fluctuating electron temperature is larger than the fluctuating ion temperature, meaning that the fusion-power increase is smaller than it would otherwise be. Thermal ions (solid lines) have short mean free paths, allowing the wave to be well described by hydrodynamics. Alpha particles (wavy lines) cross multiple periods of the wave before slowing down, preventing the growth of thermoacoustic instabilities. However, fast ions (dotted lines) travel a distance comparable to the perturbation wavelength between collisions, leading to deviations from equilibrium in the tail of the ion distribution function.

The remainder of this work develops a quantitative version of the above picture. Consideration is restricted to fluctuations that are hydrodynamic (or magnetohydrodynamic), meaning that the plasma harboring the fluctuations remains close to local thermodynamic equilibrium at each point in space. 
Particular attention is paid to acoustic waves, but results are presented in generality to apply whenever the two-point correlations between the fluctuations are known. 
The paper is organized as follows. 
Section~\ref{sec_fusion} reviews the physics of fusion reactivity and introduces the response functions used in this work to quantify modifications to fusion power. A distinction is made between the impulse response, describing the transient behavior after a fusion plasma is suddenly perturbed, and the time-independent response function appearing at late times as the decay of the fluctuations reaches a quasi-steady state. 
Section~\ref{sec_hydro} obtains a general formula for the fusion-power amplification in purely hydrodynamic systems and then specializes this formula to acoustic fluctuations in magnetized and unmagnetized fusion plasmas. Section~\ref{sec_two_temp} describes the effect on fusion power of perturbations that primarily dissipate their energy into ions or electrons, producing a transient ``hot-ion'' or ``hot-electron'' mode. Section~\ref{sec_kinetic} derives the kinetic effect of fluctuations on fusion rate, providing, in essence, the compressive analogue to the SFRE. Section~\ref{sec_comparison} discusses the physical interpretation of the results of the preceding section and, by consideration of limiting cases, provides physical explanations for these results. Finally, Section~\ref{sec_discussion} considers the fusion-power response to compressive fluctuations in relation to known expressions for the SFRE. It is shown, that, in general, compressive fluctuations dominate on large spatial scales, whereas solenoidal fluctuations dominate on small spatial scales. These results illustrate a previously unappreciated sensitivity of fusion power to the energy partition between solenoidal and dilatational modes in compressible turbulence.

\section{Fusion power}
\label{sec_fusion}

In a plasma that is everywhere in local thermodynamic equilibrium, the power per unit volume $P_f$ produced by fusion reactions at each point in space is a function of the local ion density $n$ and ion temperature $T$ and can be written as
\begin{equation}
    \label{eq_Pf_start}
    P_f =  \epsilon_f n^2 \overline{\sigma v} ,
\end{equation}
where $\epsilon_f$ is the average energy released per reaction and $\overline{\sigma v}$ is the reactivity. For notational convenience, $\overline{\sigma v}$ is defined in this work to include combinatorial factors accounting for relative abundances of the reactants. Only two-body reactions are considered in this work. The reactivity is a function of temperature and can, in some cases including plasmas whose Coulomb coupling strength is not vanishingly small, depend on density as well.

\subsection{Response functions}
\label{sec_fusion_response}

If an initially uniform plasma is subjected to a spatially and temporally varying perturbation, $P_f$ becomes a function of space and time. Heuristically, $P_f$ increases in regions of high density because clumping of reactants leads to more chances for interactions and in regions of high temperature because reactivity usually increases with temperature; likewise, of course, $P_f$ tends to decrease in regions of low density and low temperature. 
In general, if the wavelength of the perturbation is shorter than the stopping distance of the fusion products, or if the temporal period of the perturbation is much shorter than the fusion self-heating time, the variations in fusion power will not grow over time and lead to runaway local heating. 
Nevertheless, density and temperature fluctuations can modify the global power balance. 
For functions that are periodic in space and time, let $\avg{\cdot}$ denote a Reynolds average taken over space and time (unless otherwise specified) such that
\begin{equation}
    \label{eq_Pf_avg}
    \avg{P_f} = \frac{1}{V\tau} \int_0^\tau dt \int_V d^3 x P_f(t,\bs x) ,
\end{equation}
where $V$ is the system volume and $\tau$ is a time interval equal to an integer number of oscillation periods. For functions that are decaying in time, \eqref{eq_Pf_avg} applies if $\tau$ can be chosen to be much shorter than the dissipation time so that the $\avg{\cdot}$ operation averages over any oscillatory behavior but preserves the decay of the ``envelope.'' For functions whose oscillation period is not widely separated from the decay time, a more general averaging operation is described in Appendix~\ref{sec_app_averaging}.

Formally, the question addressed in this work is the following: what is the linear response of $\avg{P_f}$ to perturbations in density, temperature, and flow velocity with a given spectrum? 
In other words, suppose that a uniform, unperturbed plasma with fusion power density $P_{f,0}$ is subjected to perturbations carrying energy $E$ per ion. Provided that the energy of the perturbation is much less than the thermal energy of the plasma, how does $\avg{P_f}$ deviate from $P_{f,0}$? 
The deviation can be expressed in the form
\begin{equation}
    \label{eq_Pf_pert_setup}
    \avg{P_f}(t) = \wt P_{f,0}(t) \bigg[ 1 + \Big(\reacF H + \Delta \reacF R + \reacF K + \reacF G\Big) \wt E(t) \bigg] ,
\end{equation}
where $\reacF H$, $\Delta \reacF R$, $\reacF K$, and $\reacF G$ are the linear response functions relating the amplitude of the perturbation to its effect on fusion power. 
If the fusion power is being evaluated at time $t$, $\wt E(t)$ is the energy of the perturbation at that time; in general, $\wt E$ decreases over time from its initial value $E$. The unperturbed fusion power is shown as a function of time $\wt P_{f,0}(t)$ because the temperature of the background plasma increases as the perturbation dissipates and its energy thermalizes. 

In \eqref{eq_Pf_pert_setup}, $\reacF H$ and $\reacF K$ describe, respectively, the hydrodynamic and kinetic components of the fusion-power response. The forms of $\reacF H$ and $\reacF K$ depend on the spectrum of the perturbations, meaning their frequencies and wavelengths as well as the energy partition among degrees of freedom including flows and temperature and density fluctuations. The response function $\Delta \reacF R$ describes the change in fusion power due to the heating of the background as the perturbation decays, relative to the change in fusion power that would be produced if the same amount of heating occurred but the heat were instantly equipartitioned so that ions and electrons had the same temperature. We hence refer to $\Delta \reacF R$ as the ``two-temperature component'' of the fusion-power response.

The function $\reacF G$ describes the change in fusion power due to the SFRE \citep{Fetsch_Fisch_2025a}. By definition, $\reacF G$ is zero for purely dilatational flows, which involve no shear (i.e. the vorticity is zero). A quantitative theory of $\reacF G$ was developed by \cite{Fetsch_Fisch_2025b,Fetsch_Fisch_2026_dpp}. 
Although the SFRE is a kinetic effect, semantic convenience leads us in this work to use the ``kinetic component'' of the fusion-power response to refer to all kinetic effects modifying fusion power \textit{except} those resulting from solenoidal flows, which are referred to as the ``SFRE component.'' The kinetic response function $K$ therefore records contributions from all variations in density, temperature, and pressure. In an unmagnetized plasma, $K$ and $G$ can be readily distinguished by a Helmholtz decomposition; $G$ records contributions from the solenoidal component of the flow field, which satisfies ${\nabla \times \bs u = 0}$, while $K$ records contributions from the dilational component, which satisfies ${\nabla \cdot \bs u = 0}$ and from any associated density and temperature fluctuations.

In many systems described by \eqref{eq_Pf_pert_setup}, the response functions can be treated as constant in time; even when the perturbation is evolving, all time dependence of the fusion power is captured by the decaying amplitude of the perturbation. In this case, the fusion-power perturbation at time $t$ is instantaneously proportional to $\wt E(t)$.
%, where $\wt E(t)$ is the time-varying energy of the perturbation evaluated at time $t$. 
In these systems, if electrons and ions have the same temperature, then ${\Delta \reacF R= 0}$ by construction.
%; since $T_0$ is defined to be the temperature in the absence of the perturbation at the instant where $P_f$ is being evaluated, any heating that occurred at earlier times is absorbed into the definition of $T_0$. 
If, however, the interspecies equilibration rate is comparable to the decay time, a separation between electron and ion temperatures may persist throughout the decay of the perturbation, and $\Delta \reacF R$ can be nonzero. Examples include waves that damp primarily on electrons, or flows that viscously dissipate and primarily heat ions. The explosive behavior of viscous dissipation in some cases can suddenly convert kinetic energy to ion thermal energy and generate large temperature separations \cite{Davidovits_Fisch_2016a,Davidovits_Fisch_2016b,Campos_Morgan_2019}. At moderate Coulomb coupling strengths, compression and expansion can drive a separation between electron and ion temperatures \cite{Fetsch_Foster_Fisch_2023}.

In reality, many rapidly decaying perturbations do not admit time-independent fusion-power response functions. When the physics of the problem produces components in the response that vary in time independently of the perturbation amplitude, it is useful to recast \eqref{eq_Pf_pert_setup} in the alternative form
\begin{equation}
    \label{eq_Pf_pert_time_dep}
    \avg{P_f}(t) = P_{f,0} \bigg[ 1 + \Big(\wt {\reacF H}(t) + \wt {\reacF R}(t) + \wt {\reacF K}(t) + \wt {\reacF G}(t)\Big) E \bigg] ,
\end{equation}
where $\wt {\reacF H}(t)$, $\wt {\reacF K}(t)$, $\wt {\reacF R}(t)$, and $\wt {\reacF G}(t)$ are time-dependent functions describing the evolution of the fusion power immediately after a perturbation is driven instantaneously with energy $E$ at $t=0$ and then allowed to decay so that $\wt E(t)$ is given by
\begin{equation}
    \label{eq_E_t_decay}
    \wt E(t) = \begin{cases}
        0 & t < 0 \\
        E e^{-t/t_\mr{decay}} & t \geq 0 ,
    \end{cases}
\end{equation}
where $t_\mr{decay}$ is the characteristic decay time of the perturbation. In other words, $\wt {\reacF H}(t)$, $\wt {\reacF K}(t)$, $\wt {\reacF R}(t)$, and $\wt {\reacF G}(t)$ represent the impulse response of the fusion power following an event that suddenly creates some perturbation in the plasma. As we will see in \S\ref{sec_hydro}, \S\ref{sec_two_temp}, and \S\ref{sec_kinetic}, the impulse response consists of many transient components that decay at various rates, typically governed by collisional time scales. One of these components decays like $\wt E(t)$. 
If all of the other temporal components of the response die out faster than $\wt E(t)$, the time-independent response functions are restored in the limit
\begin{equation}
    \label{eq_H_setup_time_indep_lim}
    H = E \lim_{t \to \infty} \frac{\wt {\reacF H}(t) }{\wt E(t)} ,
\end{equation}
and likewise for $\reacF K$ and $\reacF G$. 
The relation between $\Delta \reacF R$ and $\wt {\reacF R}(t)$ is more complicated because these quantities represent different things. The time-independent $\Delta \reacF R$ describes the difference between $\avg{P_f}$ and $P_{f,0}$ due to a quasi steady-state temperature difference ${\Delta T = T - T_e}$ from unequal partition of the dissipated energy between electrons and ions. ($T_e$ is the electron temperature.) By contrast $\wt {\reacF R}(t)$ describes the \textit{total increase} in fusion power due to background heating from decay of the perturbation. 

If fusion power is evaluated at time $t=t_0$, the background temperature $T_0$ at which $P_{f,0}$ is evaluated can be defined to be the average of the electron and ion temperatures: 
\begin{equation}
    \label{eq_T0_temps_avg_setup}
    T_0 = \frac{T(t_0) - S_2(t_0) + Z T_e(t_0) - Z S_{e,2}(t_0)}{1+Z} ,
\end{equation}
where the reversible temperature shifts $S_2$ and $S_{e,2}$ associated with the perturbation have been subtracted (cf. Appendix~\ref{sec_app_acoustic}). 
Thus, even if the perturbation has already been decaying for some time and heating the background in the process, the heating from times $t < t_0$ is absorbed into the definition of $T_0$, except insofar as a residual effect is to leave ${\Delta T \neq 0}$. 
Regardless of the initial partition of the dissipated energy between electrons and ions, $\wt {\reacF R}(t)$ must be positive at sufficiently late times, when the perturbation has decayed and electrons and ions have had time to equilibrate. To reconcile $\Delta \reacF R$ and $\wt {\reacF R}(t)$, it is helpful to define ${\wt {\reacF R}^\mr{(eq)}(t) = (E - \wt E(t))/c_V}$, where $c_V$ is the plasma's heat capacity per ion at constant volume. Then, letting 
\begin{equation}
    \label{eq_Delta_R_t_def}
    \Delta \wt{\reacF R}(t) = \wt {\reacF R}(t) - \wt {\reacF R}^\mr{(eq)}(t) ,
\end{equation}
we can define ${\Delta \reacF R = E \lim_{t \to \infty} \Delta \wt{\reacF R}(t)/\wt E(t)}$, mirroring the form of \eqref{eq_H_setup_time_indep_lim}.

When the limit in \eqref{eq_H_setup_time_indep_lim} does not exist -- typically for rapidly decaying perturbations that leave in their wake a non-equilibrium plasma that relaxes on a longer time scale -- the time-independent response functions are not well defined, and we are forced to resort to the time-dependent description in \eqref{eq_Pf_pert_time_dep}. Both classes of response functions are considered in this work. Sometimes, for the sake of comparison, it will be useful to write a characteristic time-independent expression for response functions even in cases where the limit in \eqref{eq_H_setup_time_indep_lim} does not exist. In these cases, we adopt the heuristic
\begin{equation}
\label{eq_H_tdecay_heuristic} 
\reacF H \sim \frac{E \wt{\reacF H}(t = t_\mr{decay})}{\wt E(t_\mr{decay})}
\end{equation}
and its equivalents for $\reacF K$ and $\reacF R$, noting that this expression transitions smoothly into \eqref{eq_H_setup_time_indep_lim} as ${t_\mr{decay} \to \infty}$.

The multiplicity of temporal components in the fusion-power response is a distinctive feature of plasmas, arising from the additional degrees of freedom in the partition of energy between species with widely separated masses as well as in the distribution function of each species. Even when a perturbation is chosen to be a normal mode of the system -- that is, to satisfy a dispersion relation obtained by linearizing the fluid equations -- the slow equilibration between electrons and ions (owing to their large mass ratio) can lead to the second-order temperature perturbations failing to keep pace with the wave and producing a long-lived component of the fusion-power response. Similarly, the large difference in collisionality in different regions of velocity space means that kinetic modes, with decay rates different from those of the primary fluid perturbation, can be excited and produce long-lived components of the response. It is to the consequence of these non-equilibrium kinetic fluctuations that we now turn our focus.

\subsection{Kinetic effects on reactivity}
\label{sec_fusion_kinetic_effects}

The relative importance of hydrodynamic and kinetic effects depends on the length and time scales of the perturbations relative to distances and times over which ions travel between collisions. In the limit where perturbations are long and slow relative to these scales, the ions remain everywhere close to local thermodynamic equilibrium, kinetic corrections to the distribution function vanish, and $\reacF K \to 0$. On the other hand, for perturbations with sufficiently short wavelengths or high frequencies, some ions can transmit information between regions of the perturbation where the background fluid quantities are different, or some ions may retain information about earlier states of the system. The resulting deviations from local Maxwellian distributions perturb the fusion reactivity, leading to nonzero $\reacF K$.

Consider a perturbation with characteristic length scale $L$ and characteristic rate of variation $\omega$. Taking for simplicity a single ion species with mass $m$, let ${\vth = \sqrt{T/m}}$ be the ion thermal velocity and let $\nu_0$ be the collision frequency of ions at the thermal velocity. The thermal mean free path is ${\lth = \vth/\nu_0}$. The importance of kinetic effects is typically estimated by evaluating the Knudsen number $\mr{Kn}$ defined as
\begin{equation}
    \label{eq_kn_def}
    \mr{Kn} = \frac{\lth}{L} .
\end{equation}
When $\mr{Kn} \ll 1$, the distribution function is close to Maxwellian everywhere, and a fluid description of the plasma is appropriate. Gradients in background fluid quantities produce corrections to the distribution away from Maxwellian. The Chapman-Enskog procedure yields an asymptotic series in powers of $\mr{Kn}$ describing these corrections; their effect on scalar quantities, such as viscous dissipation and heat conduction, appears at order $\mc O(\mr{Kn}^2)$. 
%It would be natural to expect that any effects on fusion reactivity also appear at order $\mc O(\mr{Kn}^2)$ and are therefore negligible except in the vicinity of very strong gradients. 
It would then be natural to assume that, when $\mr{Kn} \ll 1$, the reactivity of the plasma is well approximated by evaluating the reactivity of a Maxwellian distribution at each point.

In reality, this assumption can break down because the handful of fast ions that are responsible for most fusion reactions have much longer mean free paths than their thermal counterparts; the Knudsen number thus underestimates the effect of gradients on the fast-ion population \citep{Molvig_Hoffman_Albright_Nelson_Webster_2012,McDevitt_Tang_Guo_2017,Davidovits_Fisch_2014,Yin_Albright_Vold_Nystrom_Bird_Bowers_2019,Fetsch_Fisch_2025a}. 
Fusion rates are determined by the fast-ion population because the fusion cross section is a rapidly increasing function of the relative velocity of colliding ions, scaling as 
\begin{equation}
    \label{eq_sigma_model}
    \sigma(|\bs v - \bs v'|) \sim \frac{A S(|\bs v - \bs v'|)}{v^2} e^{-b \sqrt{2}\vth/{|\bs v - \bs v'|}}
\end{equation}
where $|\bs v - \bs v'|$ is the relative velocity of the colliding ions, $A$ is a constant, and the unitless astrophysical S-factor $S(|\bs v - \bs v'|)$ is a slowly varying function. The ``Gamow parameter'' ${b = \sqrt{2E_G/T}}$ is a constant describing the height of the Coulomb barrier (which is quantified by the Gamow energy $E_G$) relative to the thermal energy. Under most conditions of interest in laboratory and astrophysical fusion plasmas, $b \gg 1$. The Gamow peak is the region of velocity space where the product of the Maxwellian distribution and the fusion cross section is maximized. Most fusion reactions occur in a narrow region around this peak. Let ${v_* = b^{1/3} \vth/\sqrt{2}}$ be the Gamow velocity such that ${|\bs v - \bs v'| = 2v_*}$ at the Gamow peak. 
The sensitivity to the ion tail is significant because, in plasma, collision frequency depends strongly on velocity. For fast ions scattering off of other ions, the collision frequency scales as $\nu(v) \propto v^{-3}$ when $v \gg \vth$. It follows that the ``Gamow mean free path'' $\lambda_* = v_*/\nu(v_*)$ is much longer than $\lth$. The effect of gradients on the ion population responsible for fusion reactions is therefore not well captured by the Knudsen number; a more appropriate estimate can be obtained in terms of the ``Gamow-Knudsen number'' $\mr{Gk}$ defined as
\begin{equation}
    \label{eq_gk_def}
    \mr{Gk} = \frac{\lambda_*}{L} ,
\end{equation}
which, for $b \gg 1$, scales as ${\mr{Gk} \sim \mc O(b^{4/3} \mr{Kn})}$. When $\mr{Gk}$ is close to unity, the distribution can be expected to deviate significantly from Maxwellian in the vicinity of the Gamow peak, producing kinetic corrections to the fusion reactivity. Notably, this is true even when ${\mr{Kn} \ll 1}$ and the bulk of the ion distribution is very close to Maxwellian. 
Perturbations with ${\mr{Kn} \sim \mc O(1)}$ tend to dissipate quickly via collisional and collisionless mechanisms including viscosity, thermal conduction, and Landau damping and may not survive long enough to produce a significant change in reactivity. However, perturbations satisfying ${\mr{Kn} \ll 1 \sim \mr{Gk}}$ survive longer times because thermal particles experience weaker gradients but can produce significant effects on reactivity. 
Practically, the separation of scales between $\mr{Kn}$ and $\mr{Gk}$ allows us in this work to consider the kinetic response to hydrodynamic perturbations -- for example, acoustic waves satisfying a dispersion relation obtained from fluid equations -- because kinetic effects are negligible for the evolution of plasma density, momentum, and temperature, even when they are important in determining the reactivity.

Even when gradients are weak, kinetic effects can become important if the plasma is subjected to rapid temporal variations. Let $\nu_* = \nu(v_*)$ be the collision frequency of ions at the Gamow velocity. Then kinetic effects on the reactivity can be neglected under the condition that $\omega \ll \nu_*$. Since $\nu_* \ll \nu_0$, this means that some perturbations that are slow enough to be considered collisional in general ($\omega \ll \nu_0$) can still be fast enough to affect the fusion reactivity. 

Notwithstanding the cases considered here in which kinetic effects are important, a large class of perturbations exist in which length and time scales are long enough to permit a purely hydrodynamic treatment of the plasma, including the ions near the Gamow peak. This regime is the subject of the following section.

\section{Hydrodynamic response}
\label{sec_hydro}

We begin by considering the hydrodynamic component of the effect of compressive fluctuations on fusion power. This is the component that survives in the limit of high collisionality ($\mr{Gk} \ll 1$ and $\omega \ll \nu_*$), where the plasma is everywhere in local thermodynamic equilibrium and the distribution function is Maxwellian at the local density and temperature. 
In this limit, the fusion reactivity is a function $\overline{\sigma v}(T, n)$ of temperature and, at least in principle, of density. 
According to \eqref{eq_Pf_start} and \eqref{eq_Pf_avg}, the averaged fusion power density $P_f$ has the form
\begin{equation}
    \label{eq_Pf_start}
    \avg{P_f} =  \epsilon_f \avg{ \paren{n_0 + \delta n}^2 \overline{\sigma v}(T_0 + \delta T, n_0 + \delta n) } ,
\end{equation}
where $n_0$ and $T_0$ are the uniform unperturbed density and temperature, and $\delta n$ and $\delta T$ are the fluctuations.
In most laboratory fusion systems, $\overline{\sigma v}$ is independent of density and increases superlinearly with $T$. By Jensen's inequality, fluctuations satisfying ${\avg{\delta n} = \avg{\delta T} = 0}$ in these systems must increase the total fusion power.

\subsection{General perturbations}

To make this notion more precise, consider a statistically homogeneous plasma with small spatially and temporally varying fluctuations in ion density, ion temperature, and flow velocity about uniform mean values $n_0$, $T_0$, and $\bs u_0 = 0$ respectively. 
Let $\mc M$ (so named in reference to the Mach number) be a small parameter characterizing the amplitude of the fluctuations, and let subscripts 1 and 2 denote perturbed hydrodynamic quantities to first and second order in $\mc M$, \textit{viz.}
\begin{equation}
    \label{eq_M_ordering_1}
    \frac{n_1}{n_0}, \frac{T_1}{T_0}, \frac{u_1}{\vth} \sim \mc O(\mc M) 
\end{equation}
and
\begin{equation}
    \label{eq_M_ordering_2}
    \frac{n_2}{n_0}, \frac{T_2}{T_0}, \frac{u_2}{\vth} \sim \mc O(\mc M^2) .
\end{equation}

%Let ${T_1/T_0 \sim n_1/n_0 \sim u_1/\vth}$, where $\vth$ is the ion thermal velocity, and let ${T_2 \sim T_1^2/T_0}$ be the second-order temperature and pressure perturbations. The temperatures of all ion species, and of electrons, are assumed to be equal, and the ideal gas law $p = nT$ is assumed to apply. 
Even in two-body reactions (three-body reactions and so on are not considered in this work), the reactivity can depend on density through interactions between the reacting ions and the plasma background. Electron screening, for example, produces a percent-level enhancement of some reactions in the Sun \citep{Salpeter_1954,Bahcall_Chen_Kamionkowski_1998}. 
In the classical and weakly coupled limit, however, $\overline{\sigma v}$ is a function of temperature alone. In the following analysis, we take the weakly coupled limit. 
To second order in the fluctuation amplitude, the reactivity can be expanded about $T_0$ as
\begin{equation}
    \label{eq_reac_expansion}
    \overline{\sigma v} \sim \paren{1 + \overline{\sigma v}^\prime (T_1 + T_2) + \half \overline{\sigma v}^{\prime\prime} T_1^2} \overline{\sigma v}_0,
\end{equation}
where $\overline{\sigma v}_0 = \overline{\sigma v}(T_0,n_0)$, and $\overline{\sigma v}^\prime$ and $\overline{\sigma v}^{\prime\prime}$ are the first and second logarithmic derivatives of $\overline{\sigma v}$ with respect to temperature.
\begin{comment}It follows that the fusion power density averaged over the system volume is
\begin{equation}
    \label{eq_Pf_pert_general}
\begin{split}
    \avg{P_f} \approx \epsilon_f n_0^2 \overline{\sigma v}_0 \left [ 1 + \frac{\avg{n_1^2}}{n_0^2} +\half s_{tt} \avg{T_1^2} + s_t \frac{2\avg{n_1 T_1} + \avg{T_2}}{n_0 T_0}\right ] ,
\end{split}
\end{equation}
where $P_{f,0} = \epsilon_f n_0^2 \overline{\sigma v}_0$ is the fusion power density of the unperturbed plasma.
\end{comment}
In general, at least over limited temperature ranges, reactivity can be approximated by a power-law ${\overline{\sigma v} \approx \overline{\sigma v}_0 (T/T_0)^\alpha}$, for a constant $\alpha > 0$. 
%In this case, \eqref{eq_Pf_pert_general} becomes
It follows that the fusion power density averaged over the system volume is 
\begin{equation}
    \label{eq_Pf_pert_alpha}
    \avg{P_f} \sim \epsilon_f n_0^2 \overline{\sigma v}_0 \left [ 1 + \frac{\avg{n_1^2}}{n_0^2} + \frac{\alpha(\alpha - 1)}{2} \frac{\avg{T_1^2}}{T_0^2} + \alpha \frac{2\avg{n_1 T_1} + n_0\avg{T_2}}{n_0 T_0}\right ] .
\end{equation}

Before applying \eqref{eq_Pf_pert_alpha} to specific perturbations, we note that inspection of \eqref{eq_Pf_pert_alpha} is enough to anticipate several qualitative features. Unsurprisingly, isothermal density fluctuations enhance $P_f$ in all cases because ${\avg{n^2} \geq \avg{n}^2}$. 
Likewise, because ${\avg{T^2} \geq \avg{T}^2}$, isochoric temperature fluctuations increase the volume-averaged fusion power, provided that ${\alpha > 1}$ (which is the case under most laboratory and astrophysical conditions of interest). 
%However, the power lost to bremsstrahlung decreases. This is a result of the weak temperature dependence of the radiated power and means that the balance of self-heating and losses becomes more favorable in the presence of fluctuations than might otherwise be supposed. 
%When density and temperature perturbations are in phase with each other -- or when one quantity is not perturbed -- the net effect on power balance is favorable. 

To compute the hydrodynamic fusion-power response function $H$, we need an expression for the energy of the perturbation. 
For simplicity, let the ion and electron temperatures be equal everywhere. 
Assuming that the plasma can be described as an ideal gas (i.e. classical, non-relativistic, and weakly coupled) with all species in local thermodynamic equilibrium, the perturbed energy per ion is, to second order in the perturbation amplitude \citep{Chu_1965},
\begin{equation}
    \label{eq_E_pert}
    E \sim \half m \frac{\avg{u_1^2}}{T_0} + \frac{(1 + Z)}{2}\frac{\avg{n_1^2}}{n_0^2} + \frac{1+Z}{2(\gamma - 1)} \frac{\avg{T_1^2}}{T_0^2} 
\end{equation}
where $\gamma$ is the ratio of specific heats, $Z$ is the average ionization state, and $m$ is the average ion mass. Electron mass is neglected in computing the kinetic energy. The generalization of \eqref{eq_E_pert} to the case of different ion and electron temperatures is straightforward and is given in Appendix~\ref{sec_app_acoustic_energy}.
%Noting that the energy can equivalently be written as ${E = \avg{\half m u_1^2/T_0 + (1 + Z) (n_1T_1 + n_0T_2)/(\gamma - 1)}}$, it follows from \eqref{eq_E_pert} that the second-order temperature perturbation is
%\begin{equation}
%    \label{eq_T2_pert}
%    \frac{\avg{T_2}}{T_0} = \frac{\gamma - 1}{2} \frac{\avg{n_1^2}}{n_0^2} + \half \frac{\avg{T_1^2}}{T_0^2} - \frac{\avg{n_1T_1}}{n_0T_0}.
%\end{equation}

According to \eqref{eq_Pf_pert_alpha} and \eqref{eq_E_pert}, the leading-order contributions to both the change in fusion power and the energy of the perturbation are ${\mc O(\mc M^2)}$, i.e. of second order in the fluctuation amplitude. It is for this reason that $\reacF H$, $\Delta \reacF R$, $\reacF K$, $\reacF G$ are universal functions dependent on the spectrum of the perturbations but independent of their amplitude (that this is true of $\reacF K$ will be shown in \S\ref{sec_kinetic}). Moreover, because the volume-averaged and time-averaged product of any two perturbations with different wavenumbers of frequencies is zero, the response functions can be computed independently for each mode in the spectrum and then summed to obtain the total response.

\subsection{Isobaric perturbations}
\label{sec_hydro_isobaric}

Different behavior appears when density and temperature perturbations are out of phase with each other to a sufficient degree that $\avg{n_1 T_1} < 0$. In isobaric systems, the density and temperature must be anticorrelated so that $nT$ is constant. We consider here perturbations for which $n T = n_0 T_0$ everywhere, meaning that
%${\avg{n_1^2}/n_0^2 = \avg{T_1^2}/T_0^2 = \avg{T_2}/T_0 = -\avg{n_1T_1}}/n_0T_0$
\begin{equation}
    \label{eq_correlations_isobaric}
    \frac{\avg{n_1^2}}{n_0^2} = \frac{\avg{T_1^2}}{T_0^2} = \frac{\avg{T_2}}{T_0} = -\frac{\avg{n_1 T_1}}{n_0 T_0},
\end{equation}
and, therefore, that
\begin{equation}
    \label{eq_Pf_pert_isobaric}
    \avg{P_f} \sim \epsilon_f n_0^2 \overline{\sigma v}_0 \left [ 1 + \paren{1 - \frac{3}{2}\alpha + \half \alpha^2} \frac{\avg{n_1^2}}{n_0^2} \right ] .
\end{equation}

Isobaric configurations of the type described by \eqref{eq_correlations_isobaric} can appear in heated or compressed systems whose initial conditions include density or temperature variations. Thermally bistable systems can also develop isobaric structures, albeit typically with larger perturbation amplitudes than those considered here \citep{Field_1965,Waters_Proga_2023,Waters_Stricklan_2025}. In general, regardless of the mechanism generating spatial inhomogeneities, the plasma rearranges toward a state of uniform pressure on a time scale on the order of $L/\vth$; when $\mr{Kn} \ll 1$, the subsequent diffusive relaxation toward global thermal equilibrium occurs on a time scale much longer than $L/\vth$ (with rare exceptions, e.g. where thermal conduction leads to rapid rearrangement by triggering thermal instabilities \citep{Fetsch_Fisch_2024}). 
When the evolution of the perturbation is slow, the kinetic-energy term in \eqref{eq_E_pert} can be neglected, so the energy of the perturbation is ${E \sim (1+Z)\half \gamma/(\gamma - 1) \avg{n_1^2}/n_0^2}$. Taking ${\gamma = 5/3}$, the hydrodynamic response function is
\begin{equation}
    \label{eq_H_isobaric}
    \reacF H^\mr{(isobar)} \sim \frac{4}{5(1 + Z)} \paren{1 - \frac{3}{2}\alpha + \half \alpha^2} .
\end{equation}
It follows from \eqref{eq_correlations_isobaric} that, in the range ${1 < \alpha < 2}$, isobaric perturbations decrease the volume-averaged fusion power. 
For $\alpha < 1$, fusion power increases because the reactivity reduction in the cool, dense regions is offset by the advantage of ``clumping'' of fuel ions. For $\alpha > 2$, fusion power increases because the increased reactivity in high-temperature regions outweighs the fact that the density in those regions is low. 
This dependence on $\alpha$ is a well-known result; it is for this reason that fusion devices typically include a hot core in pressure equilibrium with a cooler, denser periphery, and that MCF designs aim to maximize $\overline{\sigma v}/T^2$ \citep{Greenwald_2002}.
%For most conditions of interest for MCF and ICF, however, $\alpha$ falls within this range, and so isobaric perturbations are unfavorable. This is unsurprising -- it is well established that cold chunks of high-density material typically are deleterious in the core of fusion devices. 
It is evident from \eqref{eq_Pf_pert_alpha}, however, that larger increases in fusion power are possible when density and temperature perturbations are in phase. Since their pressure is nonuniform, such perturbations must be dynamical; acoustic waves are a natural example of such perturbations and are considered in the following section.

\subsection{Acoustic waves in limiting cases}
\label{sec_hydro_acoustic}

We proceed to compute the hydrodynamic fusion-power response to acoustic waves in certain interesting limits that allow for transparent analytical results. In particular, we consider here waves in which the dissipation rate is negligible, all particle species remain in local thermodynamic equilibrium with each other, and all perturbed fluid quantities are in phase with each other. Dispersion relations and expressions for second-order fluctuation quantities in acoustic waves are derived in Appendix \ref{sec_app_acoustic}.

\begin{figure}
    \centering
    %\vspace{0.5cm}
    \includegraphics[width=0.6\columnwidth]{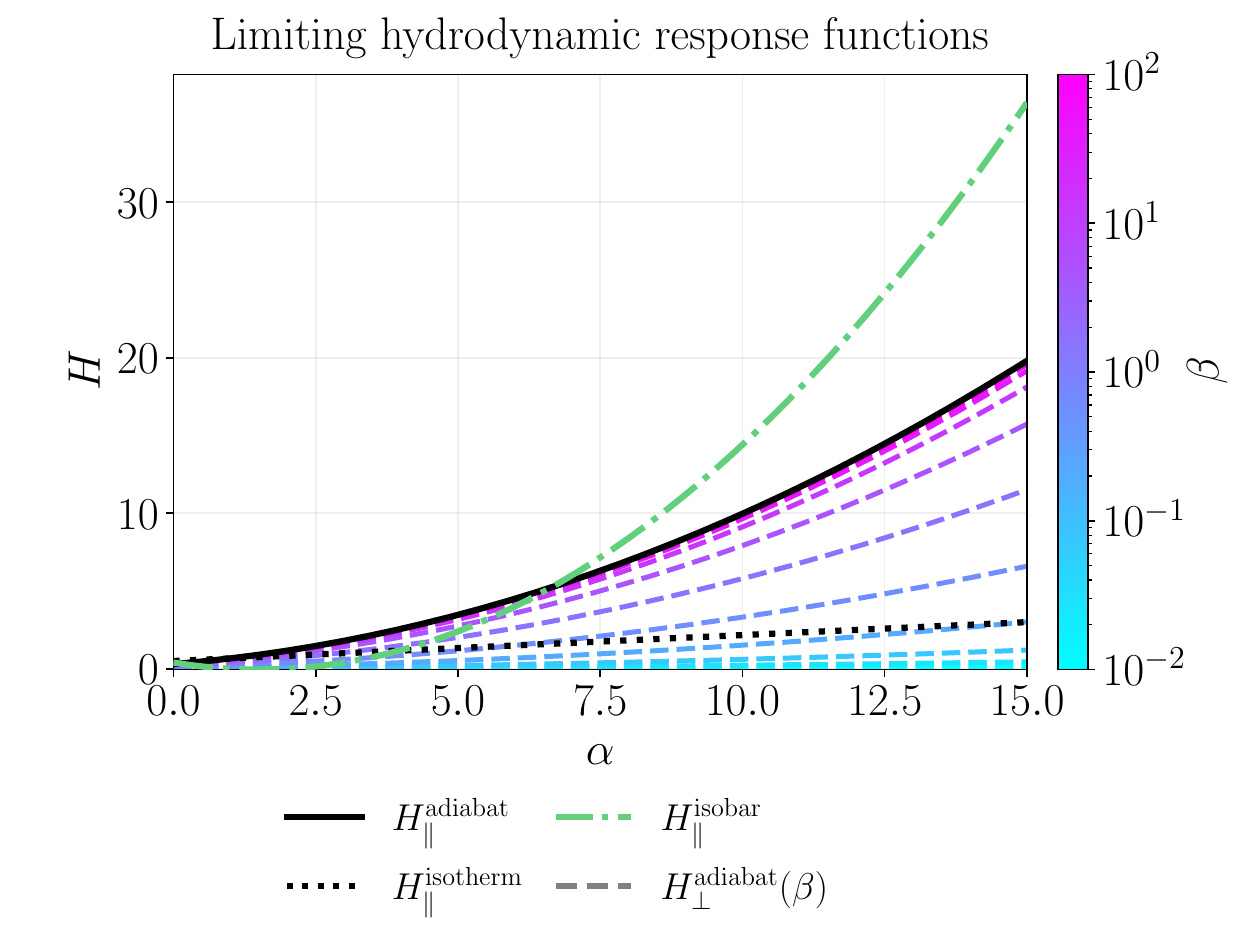}
    \caption{\justifying
    Hydrodynamic fusion-power response functions for acoustic waves in limiting cases. The ``parallel'' subscript denotes waves oriented parallel to the magnetic field or simply in an unmagnetized plasma. The ``perpendicular'' subscript denotes waves perpendicular to the field. The ratio $\beta$ of thermal to magnetic pressure is varied from $0.01$ to $100$.
    }
    \label{fig_hydro_limiting}
\end{figure}

For a linear acoustic wave satisfying these conditions, the density and temperature perturbations are related by ${T_1 / T_0 = (g- 1) n_1 / n_0}$ and ${\avg{T_2}/T_0 = \half\big[(g-1)(g-3) + \gamma - 1\big]\avg{n^2}/n_0^2}$, where $g$ is the effective polytropic index of the wave. Using \eqref{eq_Pf_pert_alpha}, the perturbed fusion power is 
\begin{equation}
    \label{eq_Pf_alpha_g}
    \avg{P_f} \sim \epsilon_f n_0^2 \overline{\sigma v}_0 \left [1 + \paren{1 + \frac{2g + \gamma - 3}{2}\alpha + \frac{(g-1)^2}{2}\alpha^2}\frac{\avg{n_1^2}}{n_0^2}\right ] ,
\end{equation}
and from \eqref{eq_E_pert}, normalized energy of the wave is simply
\begin{equation}
    \label{eq_E_acoustic_parallel}
    E_\parallel \sim (Z+1) g \frac{\avg{n_1^2}}{n_0^2} .
\end{equation}
In the following special cases, we adopt $\gamma = 5/3$, appropriate for a fully ionized, weakly coupled, classical plasma.
Consider an acoustic wave propagating in unmagnetized plasma, or in magnetized plasma parallel to the magnetic field. For an adiabatic wave,  $g=\gamma$, and the response function is
\begin{equation}
    \label{eq_H_par_adiabat}
    \reacF H^\text{(adiabat)}_\parallel \sim \frac{3}{5(Z+1)}\paren{1 + \alpha + \frac{2}{9}\alpha^2}  .
\end{equation}
In the isothermal limit, where the wavelength is small enough to allow both species to preserve a spatially uniform temperature, $g=1$, the response function becomes simply
\begin{equation}
    \label{eq_H_par_isotherm}
    \reacF H^\text{(isotherm)}_\parallel \sim \frac{1 + \frac{1}{3}\alpha}{(Z+1)} . 
\end{equation}
Finally, in magnetized plasma, the temperature perturbation of an adiabatic acoustic wave propagating perpendicular to the magnetic field is still described by an effective adiabatic index $g=5/3$, but the normalized wave energy is ${E_\perp \sim (1 + Z)(g + 2/\beta)\avg{n_1^2}/n_0^2}$, where $\beta$ is the ratio of thermal to magnetic pressure. The response function in then
\begin{equation}
    \label{eq_H_perp_adiabat}
    \reacF H^\text{(adiabat)}_\perp \sim \frac{\beta}{2 + \frac{5}{3}\beta}\frac{1}{(Z+1)}\paren{1 + \alpha + \frac{2}{9}\alpha^2} .
\end{equation}
Because heat conduction is suppressed perpendicular to the magnetic field, we neglect the isothermal limit for perpendicular acoustic waves. The response functions derived here in limiting cases are shown in Fig.~\ref{fig_hydro_limiting}.
As expected, $\reacF H^\text{(adiabat)}_\perp$ reduces to $\reacF H^\text{(adiabat)}_\parallel$ in the limit high-$\beta$ limit. Notably, at low $\beta$, the response to perpendicular magnetosonic waves is substantially reduced because the magnetic field arrests the compression that is responsible for the increase in fusion power at the peaks of acoustic waves.

\subsection{General acoustic waves}
\label{sec_hydro_acoustic_general}

While the limiting cases considered above are useful for estimating the impact of acoustic waves on fusion power, the assumptions of \S\ref{sec_hydro_acoustic} are often violated.  
In dissipative waves, the density and temperature perturbations develop a phase shift, which, per \eqref{eq_Pf_pert_alpha}, tends to reduce the clumping of high-density, high-temperature ions that increase fusion power. Moreover, in rapidly dissipating waves, the temperature perturbations of electrons and ions can become decoupled due to preferential heating of one species combined with slow equilibration between species. In such regimes, decay is sometimes so rapid that no time-independent response function exists -- the limit in \eqref{eq_H_setup_time_indep_lim} diverges -- and it is instead necessary to compute the time-dependent impulse response.
However, provided that the hydrodynamic limit still holds, \eqref{eq_Pf_pert_alpha} is exact and can be used to compute the hydrodynamic component of the response along with the fluctuating quantities in acoustic waves derived in Appendix \ref{sec_app_acoustic}. 

The background-heating component is given by 
\begin{equation}
    \label{eq_hydro_R_t}
    \wt {\reacF R}(t) = \alpha \frac{\avg{Q_2(t)}}{T_0 E} ,
\end{equation}
where $Q_2$ is the second-order temperature increase in the background plasma given by \eqref{eq_app_Q2_Delta_soln} and \eqref{eq_app_Q2_tot_soln}. It is additionally interesting to compare $\wt {\reacF R}(t)$ to the quantity $\wt {\reacF R}^\mr{(eq)}(t)$, defined as change in the reactivity produced if the energy deposited in the background were to be equipartitioned among electrons and ions. This quantity is given by
\begin{equation}
    \label{eq_hydro_R_eq_t}
    \wt {\reacF R}^\mr{(eq)}(t) = \frac{\alpha}{1+Z}  \frac{\avg{\overline{Q}_2(t)}}{T_0 E} ,
\end{equation}
where $\overline{Q}_2$ is the second-order ``total'' temperature increase -- the weighted sum of the electron and ion temperatures -- given by \eqref{eq_app_Q2_tot_soln}.

\begin{figure}
    \centering
    \vspace{0.5cm}
    \includegraphics[width=0.6\columnwidth]{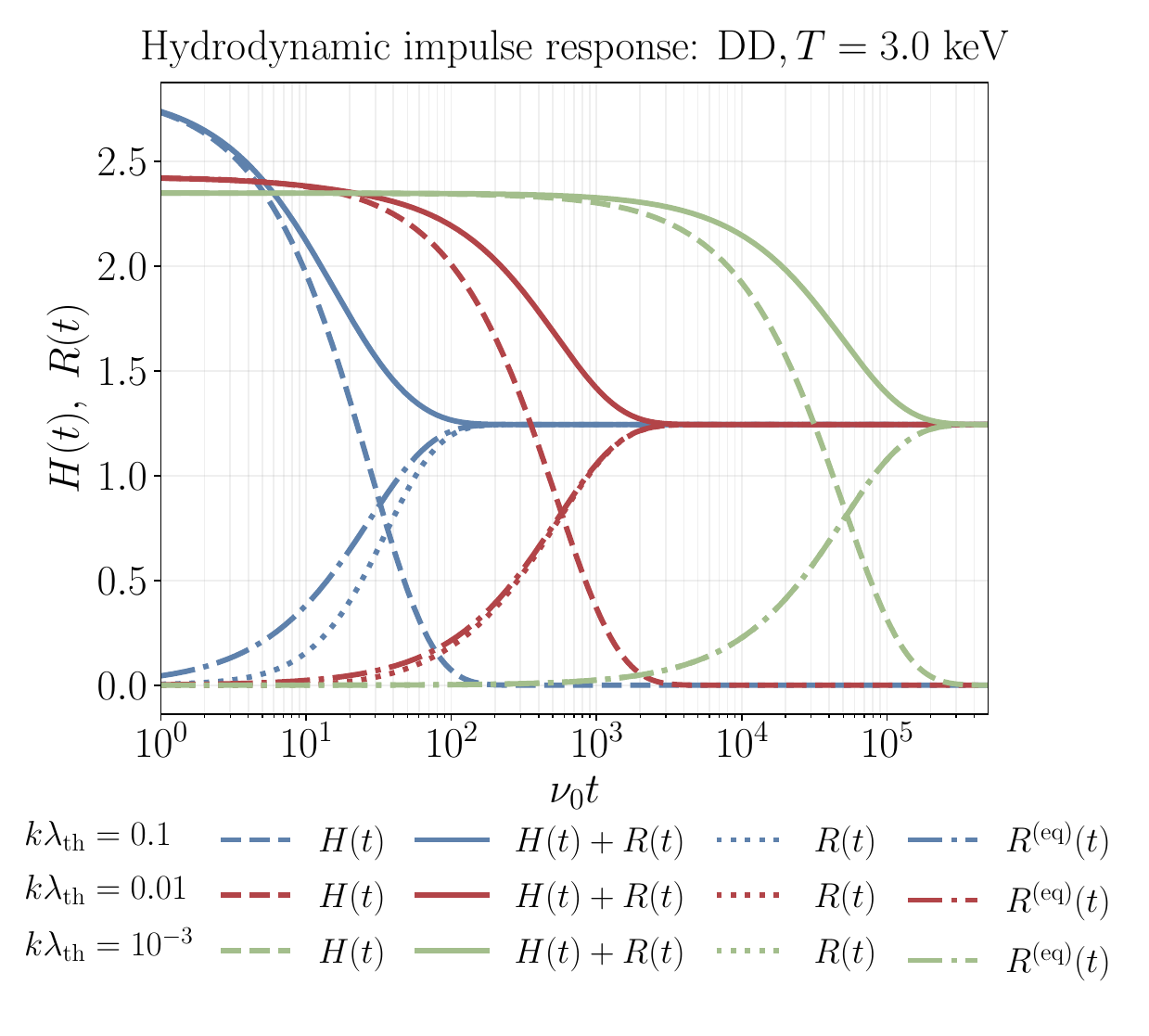}
    \caption{\justifying
    Time evolution of the hydrodynamic fusion power in an unmagnetized deuterium plasma at $T = 3~\mr{keV}$. Acoustic waves are generated suddenly at $t=0$ with wavenumbers of $k\lth = 0.1$ (blue), $k\lth = 0.01$ (red), and $k\lth = 0.001$ (green). 
    }
    \label{fig_hydro_t_DD}
\end{figure}

\begin{figure}
    \centering
    \vspace{0.5cm}
    \includegraphics[width=0.6\columnwidth]{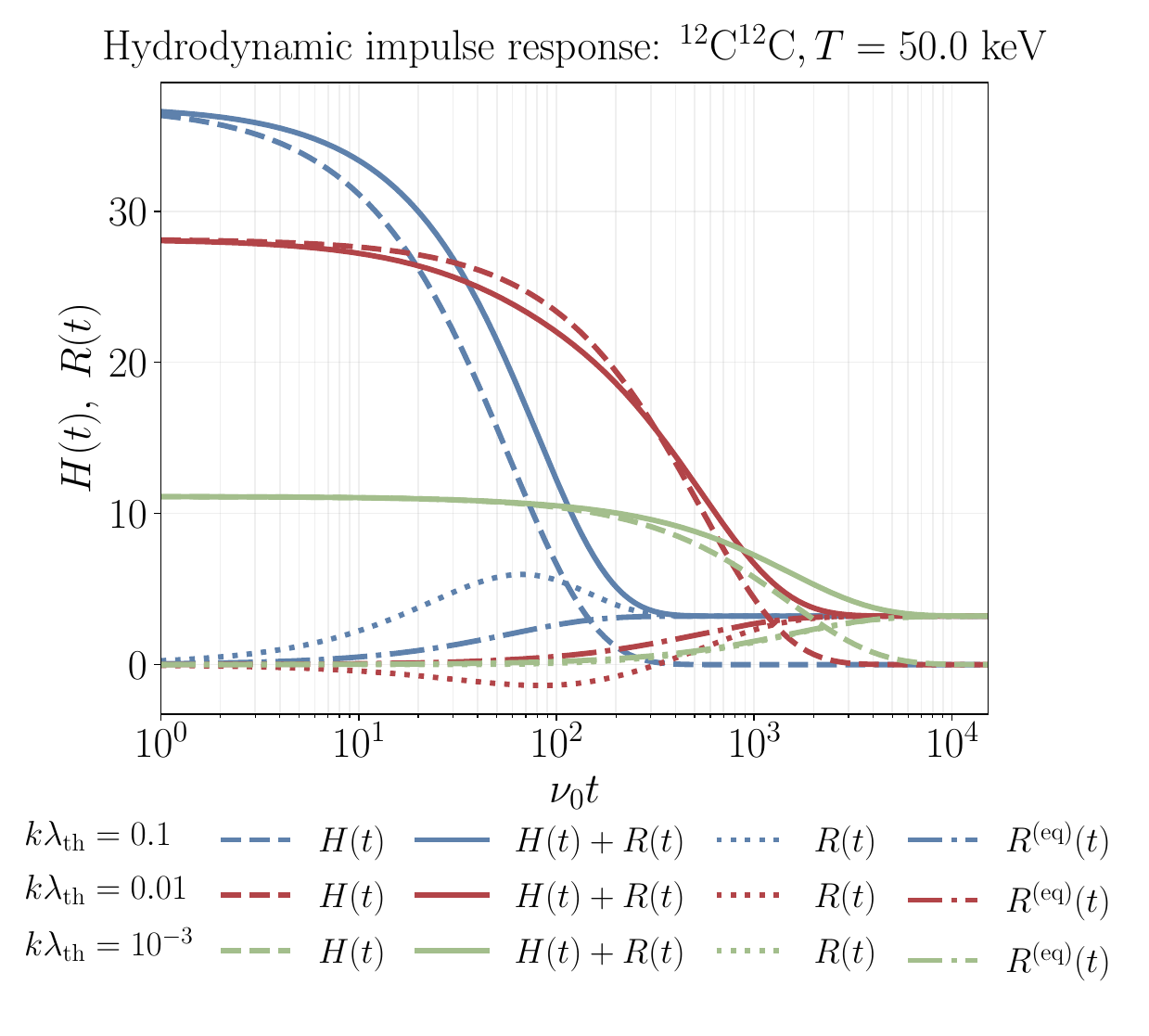}
    \caption{\justifying
    Time evolution of the hydrodynamic fusion power in an unmagnetized \textsuperscript{12}C plasma at $T = 50~\mr{keV}$. Acoustic waves are generated suddenly at $t=0$ with wavenumbers of $k\lth = 0.1$ (blue), $k\lth = 0.01$ (red), and $k\lth = 0.001$ (green). 
    }
    \label{fig_hydro_t_CC}
\end{figure}

\begin{figure}
    \centering
    \vspace{0.5cm}
    \includegraphics[width=0.6\columnwidth]{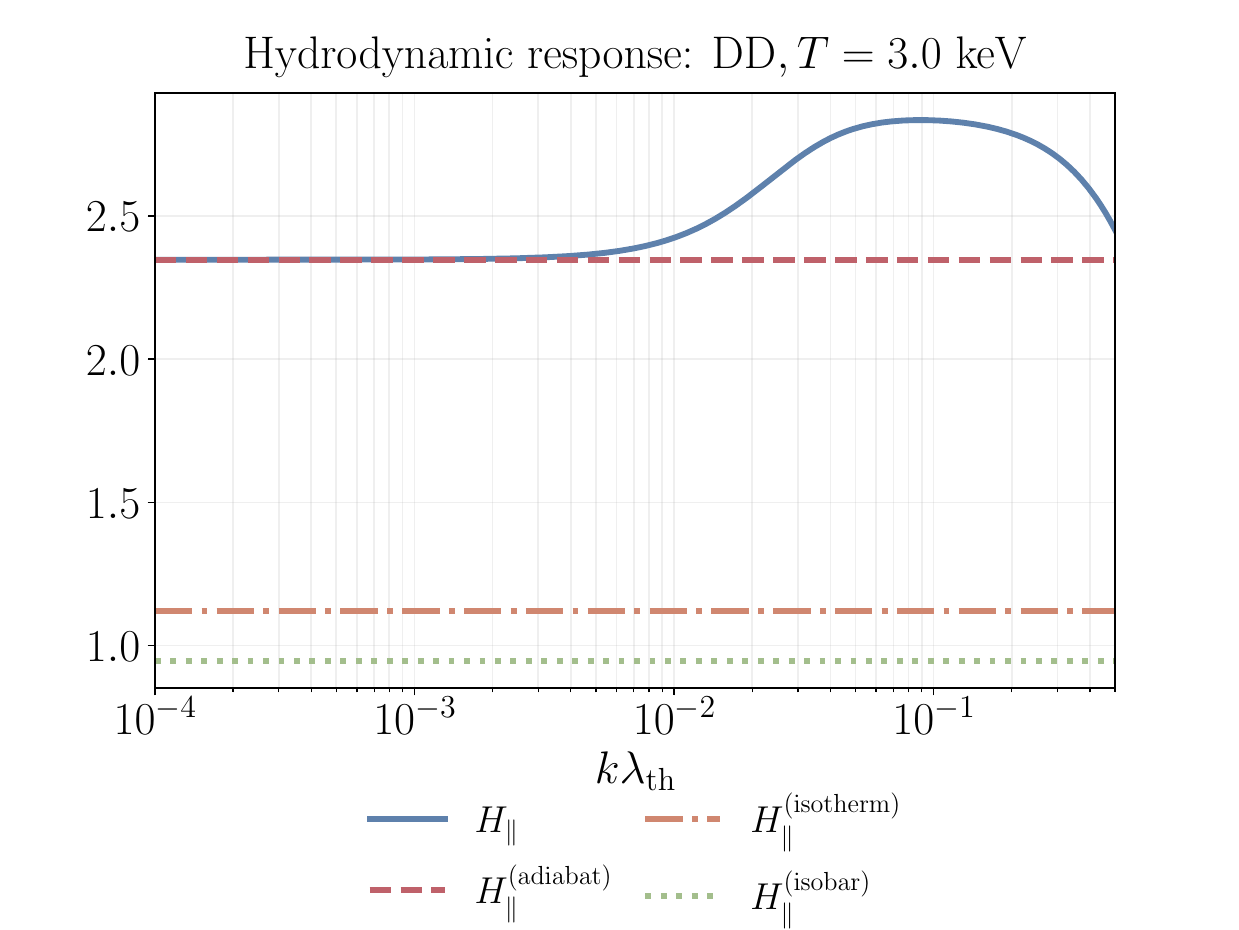}
    \caption{\justifying
    Comparison of the time-independent hydrodynamic response functions for adiabatic and isothermal acoustic waves, as well as isobaric perturbations, in an unmagnetized deuterium plasma at $T = 3~\mr{keV}$. Additionally shown is the time-independent response function computed from the fluctuating quantities in acoustic waves derived in Appendix \ref{sec_app_acoustic}.
    }
    \label{fig_hydro_steady}
\end{figure}

The hydrodynamic impulse response is shown in Fig.~\ref{fig_hydro_t_DD} for acoustic waves with wavenumbers ${k\lth \in (10^{-1}, 10^{-2}, 10^{-3})}$ in an unmagnetized deuterium plasma at $T = 3~\mr{keV}$. The former regime is highly dissipative, with significant separation between the ion and electron temperatures. This is particularly notable in the separation between the $\wt R(t)$ and $\wt R^\mr{(eq)}(t)$ curves; evidently, depending on the length scale of the wave, dissipation can drive a ``hot-ion mode'' or a ``hot-electron mode.'' While the shortest-wavelength perturbations produce a slightly larger hydrodynamic response because a larger fraction of their energy is partitioned into the ions (the electrons are closer to isothermal at short wavelengths and have a lower effective polytropic index), the value of $H$ depends only weakly on the wavenumber. This can be seen more clearly in Fig.~\ref{fig_hydro_steady}, where the time-independent hydrodynamic response functions derived in \S\ref{sec_hydro_acoustic} in limit cases are compared to the response function computed from the fluctuating quantities in acoustic waves derived in Appendix \ref{sec_app_acoustic}. As described in the appendix, the time-independent response function $H$ can be extended, in an approximate way, beyond its strict regime of applicability (where \eqref{eq_H_setup_time_indep_lim} converges) by evaluating $H(t)$ at $t = 1/2\mu$, where $\mu$ is the decay rate of the wave. This evaluation scheme is used in Fig.~\ref{fig_hydro_steady} to extend the domain to short wavelengths, where the interspecies collision rate is slower than the wave dissipation rate.

\section{Two-temperature response}
\label{sec_two_temp}

We saw in \S\ref{sec_hydro_acoustic_general} that the impulse response to the driving of acoustic waves is affected by the decoupling of ion and electron temperatures. Depending on the mechanics of wave dissipation, the ion temperature may be transiently amplified or suppressed relative to what it would be if the dissipated wave energy were to be placed instantly into equipartition. Naturally, this variation in ion temperature affects the fusion reactivity. 
In this section, we determine the time-independent response function $\Delta R$ associated with unequal heating of electrons and ions in plasmas haboring decaying perturbations. While these effects can appear in magnetized plasma, we focus in this section on unmagnetized plasma, or on perturbations parallel to the magnetic field.

It is important to note that the increases in the background electron and ion temperatures via dissipation, and hence also $\Delta R$, are distinct from any reversible temperature shift associated with the perturbation itself. Given first-order fluctuating quantities satisfying some dispersion relation, for example the acoustic relations described in Appendix~\ref{sec_app_acoustic}, solving the fluid equations with those first-order quantities as givens yields a set of second-order fluctuating quantities. The average temperature shift, denoted by $\avg{S_2}$ in Appendix~\ref{sec_app_acoustic} is a quantity that decays at the same rate as the wave energy and, because it is a consequence of reversible (non-entropy generating) processes, survives even in the limit of zero dissipation. To compute the hydrodynamic response in \S\ref{sec_hydro}, the change in the average temperature was taken to be precisely this reversible shift, i.e. $\avg{T_2} = \avg{S_2}$. This relation holds exactly if the fusion power is evaluated at $t=0^+$ and initial conditions are imposed such that $T_0 = T(t=0^-) = T_e(t=0^-)$, where $t=0^-$ and $t=0^+$ indicate the moments just before and just after the perturbation is generated, respectively. 
At times $t > 0$, both temperatures will shift upwards as the wave energy is dissipated. The shifts in the electron and ion temperatures are described by $\avg{Q_{e,2}(t)}$ and $\avg{Q_{i,2}(t)}$, respectively, in Appendix~\ref{sec_app_acoustic}. Using \eqref{eq_hydro_R_t} and \eqref{eq_hydro_R_eq_t} combined according to \eqref{eq_Delta_R_t_def}, the impulse response in the fusion power due to two-temperature effects is given by
\begin{equation}
    \label{eq_Delta_R_t}
    \Delta \wt {\reacF R}(t) = \alpha \frac{\avg{Q_{2}(t)}}{T_0 E} - \alpha \frac{\avg{\overline{Q}_2(t)}}{(1+Z)T_0 E} .
\end{equation}
Noting that ${\overline{Q}_2(t) = (\gamma-1)(E - \wt E(t))/(1+Z)}$, and that $\avg{Q_2(t)}$ is given by the terms in \eqref{eq_app_T2_t} that are not associated with $\avg{S_2}$, we have
\begin{equation}
    \label{eq_Delta_R_t_L}
    \Delta \wt {\reacF R}(t) = \frac{\alpha}{1+Z} \bigg[ e^{-2\mu t} - e^{-\nu_\mr{eff} t} \bigg] \frac{\mc L - Z \mc L_e}{(\nu_\mr{eff} - 2\mu)E} |\wt n|^2 ,
\end{equation}
where, as in Appendix~\ref{sec_app_acoustic}, $\nu_\mr{eff} = \nu_\mr{ie} (\gamma-1)(1 + 1/Z)$ and $\nu_\mr{ie}$ is the ion-electron energy exchange rate. The functions $\mc L$ and $\mc L_e$ are defined in \eqref{eq_app_L_def} and \eqref{eq_app_Le_def} respectively. Then, defining ${\widehat \nu_\mr{eff} = \nu_\mr{eff}/\mu}$ and ${\widehat \nu_\mr{ie} = \nu_\mr{ie}/\mu}$ and writing $\Delta \wt {\reacF R}(t)$ in terms of correlations between first-order fluctuating quantities, we have
\begin{equation}
\label{eq_Delta_R_t_correl}
\begin{split}
    \Delta \wt {\reacF R}(t) = \frac{\alpha}{1+Z}  \frac{e^{-2\mu t} - e^{-\nu_\mr{eff} t}}{(\widehat \nu_\mr{eff} - 2)E} &\Bigg[\frac{1}{\gamma - 1}\frac{\avg{T_1^2}}{T_0^2} - \frac{Z}{\gamma-1} \frac{\avg{T_{e,1}^2}}{T_0^2} - \frac{\avg{n_1 T_1}}{n_0 T_0} + Z\frac{\avg{n_1 T_{e,1}}}{n_0 T_0} 
    \\
     & - \frac{\omega}{\mu} \frac{\avg{n_1(z) T_1(z - \frac{L}{4})} - Z\avg{n_1(z) T_{e,1}(z - \frac{L}{4})}}{n_0 T_0} 
    \\
     & + \frac{4\eta k^2}{3\mu}\frac{\avg{u_1^2}}{\vth^2} + 2 \widehat \nu_\mr{ie} \frac{\avg{\paren{ 2 n_1 + \nu_{\mr{ie},T_e} \frac{T_{e,1}}{T_0}n_0}\paren{T_{e,1} - T_1}}}{n_0 T_0}\Bigg] .
\end{split}
\end{equation}
\begin{comment}
\begin{equation}
\label{eq_Delta_R_t_correl}
\begin{split}
    \Delta \wt {\reacF R}(t) = \frac{\alpha}{1+Z}  \frac{e^{-2\mu t} - e^{-\nu_\mr{eff} t}}{(\widehat \nu_\mr{eff} - 2)E} &\Bigg[\frac{1}{\gamma - 1}\frac{\avg{T_1^2}}{T_0^2} - \frac{\avg{n_1 T_1}}{n_0 T_0} - \frac{Z\gamma}{\gamma-1} \frac{\avg{n_1^2}}{n_0^2} 
    \\
     & + \frac{\gamma+1}{\gamma-1} \frac{\avg{n_1 p_{e,1}}}{n_0^2 T_0} - \frac{1}{Z(\gamma-1)} \frac{\avg{p_{e,1}^2}}{n_0^2 T_0^2}
    \\
     & - \frac{\omega}{\mu} \left( \frac{\avg{n_1(z) T_1(z - \frac{L}{4})}}{n_0 T_0} - \frac{\avg{n_1(z) p_{e,1}(z - \frac{L}{4})}}{n_0^2 T_0} \right)
    \\
     & + \frac{4\eta k^2}{3\mu}\frac{\avg{u_1^2}}{\vth^2} 
    \\
     & + 2 \widehat \nu_\mr{ie} \avg{\paren{ (2 - \nu_{\mr{ie},T_e}) \frac{n_1}{n_0} + \frac{\nu_{\mr{ie},T_e}}{Z} \frac{p_{e,1}}{n_0 T_0}}\paren{\frac{p_{e,1}}{Zn_0 T_0} - \frac{n_1}{n_0} - \frac{T_1}{T_0}}}\Bigg] .
\end{split}
\end{equation}
\end{comment}
From \eqref{eq_Delta_R_t_correl}, the time-independent two-temperature response function $\Delta R$ can be immediately obtained by replacing the prefactor with $\alpha/[(1+Z)(\widehat \nu_\mr{eff} - 2) E]$. In cases where it is necessary to resort to the heuristic described in \eqref{eq_H_tdecay_heuristic}, where $\Delta \reacF R \sim \Delta \wt {\reacF R}(t_\mr{decay}) E/\wt E(t_\mr{decay})$, we can take $t_\mr{decay} = 1/2\mu$ and obtain an estimate of $\Delta \reacF R$ from \eqref{eq_Delta_R_t_correl} by replacing the prefactor with ${\alpha (1 - e^{-2\widehat \nu_\mr{eff}})/[(1+Z)(\widehat \nu_\mr{eff} - 2) E]}$. 

\begin{figure}
    \centering
    %\vspace{0.5cm}
    \includegraphics[width=0.6\columnwidth]{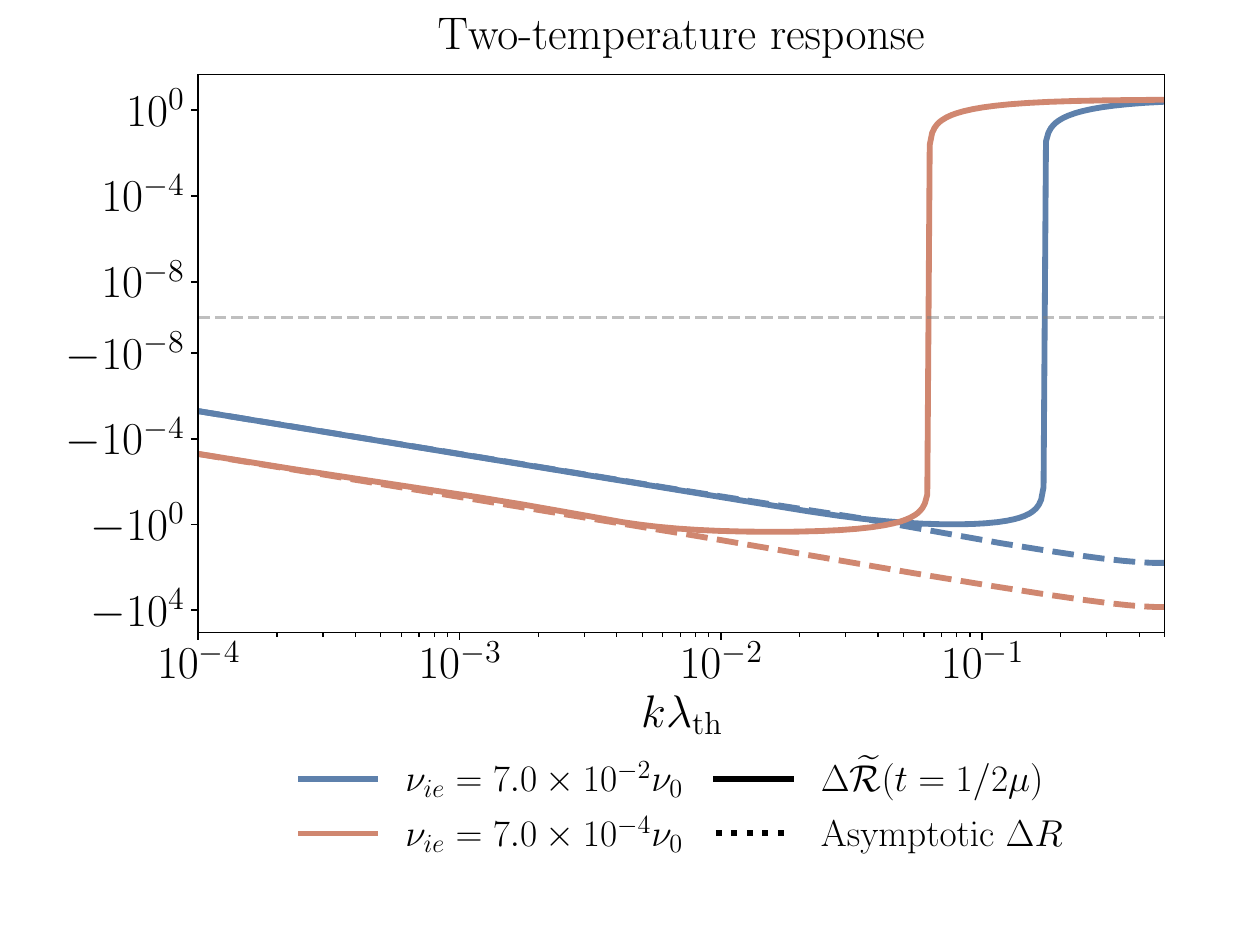}
    \caption{\justifying
    Two-temperature response function for acoustic waves as a function of wavenumber. The transport coefficients correspond to a deuterium plasma, but $\nu_\mr{ie}$ is artificially varied to illustrate the dependence of $\Delta R$ on the ion-electron energy exchange rate. The response function is evaluated at $t = 1/2\mu$ to extend the domain of applicability to rapidly decaying waves, as described in \S\ref{sec_fusion_response}. 
    }
    \label{fig_DeltaR_t}
\end{figure}

Finally, noting that $\nu_\mr{ie} \gg \mu$ for perturbations that are not very rapidly decaying, it is useful to expand $\Delta \reacF R$ in the limit where $\widehat \nu_\mr{eff} \gg 1$ to obtain a more transparent asymptotic expression for the two-temperature response. Taking the time-independent limit of \eqref{eq_Delta_R_t_correl} and then expanding to leading order in $1/\widehat \nu_\mr{eff}$ yields
\begin{equation}
\label{eq_Delta_R_t_asymptotic}
\begin{split}
    \Delta \reacF R \sim \frac{\alpha Z}{(1+Z)^2(\gamma-1)\widehat \nu_\mr{ie} E}\Bigg[ & (1-Z) \left( \frac{1}{\gamma-1}\frac{\avg{T_1^2}}{T_0^2} - \frac{\avg{n_1 T_1}}{n_0 T_0} \right) - \frac{\omega}{\mu}(1-Z)\frac{\avg{n_1(z) T_1(z - \frac{L}{4})}}{n_0 T_0} 
    \\
    & + \frac{4\eta k^2}{3 \mu}\frac{\avg{u_1^2}}{\vth^2} + 2 \frac{Z}{Z+1} \frac{(\kappa_i - \kappa_e)k^2}{\mu} \avg{ \paren{ 2 \frac{n_1}{n_0} + \nu_{\mr{ie},T_e} \frac{T_1}{T_0} } \frac{T_1}{T_0} } \Bigg] ,
\end{split}
\end{equation}
where correlations involving the electron-temperature perturbation have been eliminated using \eqref{eq_app_lin_three} and \eqref{eq_app_lin_four}. 
The asymptotic result in \eqref{eq_Delta_R_t_asymptotic} is not specific to acoustic wave -- it would also apply, for instance, to an isobaric configuration with clouds of cold, dense plasma balanced by hot, diffuse plasma. Some insight can be gained, however, by considering the behavior of $\Delta \reacF R$ in acoustic waves. In the limit of nearly adiabatic waves, the first two terms in the brackets of \eqref{eq_Delta_R_t_asymptotic} vanish, leaving only the viscous and thermal conductive terms to drive temperature separation. Assuming that $\eta$, $\kappa_i$, and $\kappa_e$ obey diffusive scalings, the much greater mobility of electrons relative to ions means that the $\kappa_e$ term dominates the two-temperature response. Assuming that $\nu_{\mr{ie},T_e} = -3/2$, the average in the final term of \eqref{eq_Delta_R_t_asymptotic} is positive, meaning that ${\Delta \reacF R < 0}$. In the long-wavelength limit where waves approach adiabaticity, the small amount of dissipation is dominated by electron thermal conduction, leading to a slightly higher electron temperature and thus a negative two-temperature response function. In the isothermal limit, the first two terms in the brackets of \eqref{eq_Delta_R_t_asymptotic} again vanish. Now, however, the thermal conduction term also vanishes because the temperature perturbation approaches zero. The viscous term remains, as it must in any dynamical perturbation, and it follows that $\Delta \reacF R > 0$. Because viscous heating is primarily deposited into ions, the plasma is driven into a transient hot-ion mode during the decay of the wave, generating a positive two-temperature response function. 
%${T_1/T_0 = (\gamma-1)n_1/n_0}$, $\gamma = 5/3$
These scalings are illustrated in Fig.~\ref{fig_DeltaR_t}, where the time-independent two-temperature response function for acoustic waves is displayed as a function of wavenumber at two values of $\nu_\mr{ie}$. The heuristic \eqref{eq_H_tdecay_heuristic} is used to generate a smooth estimate of $\Delta R$. At small $k$, it can be seen that $\Delta R \propto k^{-2}$ and that \eqref{eq_Delta_R_t_correl} and \eqref{eq_Delta_R_t_asymptotic} are in good agreement. As $k$ increases, and so $\mu$ also increases, \eqref{eq_Delta_R_t_asymptotic} fails to capture the behavior of $\Delta R$, which eventually becomes positive, indicating a hot-ion mode with increased fusion reactivity. Note that the apparently sudden jump in the the value of $\Delta R$ around $k\lth \approx 0.07$ is an artifact of the logarithmic scales of the axes in Fig.~\ref{fig_DeltaR_t}. 
If the domain of the plot is extended to $k\lth \sim 1$ to illustrate the nominal behavior of \eqref{eq_Delta_R_t_asymptotic} at very short wavelengths, the asymptotic form of $\Delta R$ also becomes positive. However, the fluid equations break down in this $\mr{Kn} \sim 1$ regime, as does the assumption that $\widehat \nu_\mr{eff} \gg 1$.

\section{Kinetic response}
\label{sec_kinetic}

We proceed to evaluate kinetic corrections to the fusion rate generated by compressive fluctuations. 
For the sake of tractability, we restrict consideration to the ``nearly hydrodynamic'' regime, where $\mr{Kn} \ll 1$ and $\omega \ll \nu_0$ such that the bulk of the plasma behaves as a fluid. The relation between the perturbation scales and fast-ion collisional scales, however, is left arbitrary ($\mr{Gk}$ and $\omega/\nu_*$ can be of any order). 

\subsection{Reactivity enhancement}

Physically, the kinetic component of the fusion-power response can be understood in several ways. In non-isothermal acoustic wave, for example, a fraction of the ions from the hot peaks will stream across temperature gradients into the cooler troughs; hence, while the hydrodynamic response involves a concentration of fusion power at the wave peaks, kinetic effects lead to smearing of the fusion power across the wave. 
Fluid flows, which typically carry half of the wave energy, have no direct effect on reactivity in the purely hydrodynamic limit -- adding a drift velocity to all ions in one patch of fluid does not change the relative velocities of colliding ions in that patch. However, as fast ions with long mean free paths stream through the background with inhomogeneous flows, they gain and lose velocity relative to the local fluid. On average, this increases the population of tail ions and therefore increasing the fusion reactivity; the SFRE describes this effect in solenoidal flows, and the following calculations describe the analogous effect in dilatational flows. 
Finally, even in the absence of gradients, periodic heating and cooling in some perturbations may be too rapid for the tail of the distribution to keep pace; in such a wave, the ``effective temperature'' of ions near the Gamow peak can differ from the thermal temperature, leading to a change in reactivity.

To compute corrections to the fusion reactivity, suppose that the distribution function $f$ can be expanded in powers of the perturbation amplitude $\mc M$ as
\begin{equation}
    \label{eq_f_expansion}
    f = f_0 + f_1 + f_2 + \ldots ,
\end{equation}
where $f_0$ is a Maxwellian distribution at the local temperature and flow velocity, and the kinetic corrections are $f_1 \sim \mc O(\mc M)$, $f_2 \sim \mc O(\mc M^2)$, and so on. Let $\reactop[f_a,f_b]$ be the reactivity operator defined such $\overline{\sigma v} = \varsigma \reactop[f,f]$, where $\varsigma$ is a combinatorial factor equal to one for reactions between different species and equal to $1/2$ for reactions between particles of the same species, and 
\begin{equation}
    \label{eq_reac_op_def}
    \reactop[f_a,f_b] = \iint d^3\bs v d^3\bs v' f_a(\bs v) f_b(\bs v') \sigma(|\bs v - \bs v'|) |\bs v - \bs v'| ,
\end{equation}
where $\sigma$ is the fusion cross section. To second order in $\mc M$, the reactivity can be written as
\begin{equation}
    \label{eq_reac_expansion_f}
    \overline{\sigma v} \sim \reactop[f_0,f_0] + 2 \reactop[f_0,f_1] + \reactop[f_1,f_1] + 2 \reactop[f_0,f_2] .
\end{equation}
The first term in \eqref{eq_reac_expansion_f} is the Maxwellian reactivity. The second term describes collisions between ions sampled from $f_1$ and ions from the background Maxwellian; this term is irrelevant in the systems studied in \cite{Fetsch_Fisch_2026} and \cite{Fetsch_Fisch_2026_dpp} because of its vanishing spatial average, but it is significant in the following derivation because it is multiplied by fluctuating density and thermal reactivity profile. 
The third term in \eqref{eq_reac_expansion_f} describes collisions between pairs of ions from $f_1$, and the fourth term describes collision between ions from $f_2$ and ions from the background Maxwellian; for most perturbations, these are the terms that determine the kinetic correction to the reactivity. 

%Following Ref.~\citep{Fetsch_Fisch_2025a,Fetsch_Fisch_2025b,Fetsch_Fisch_2026_dpp}, we define a fusion-power enhancement factor $\Phi = \avg{P_f}/P_{f,0}$, noting that $\Phi = 1 + HE + KE$. Then, 
From \eqref{eq_reac_expansion_f}, the kinetic response function $\reacF K$ can be written as
\begin{equation}
    \label{eq_K_expansion_f}
   \reacF K = \frac{ \avg{\paren{4\frac{\delta n}{n_0} + 2 \alpha \frac{\delta T}{T_0}} \reactop[f_0,f_1]} + \avg{\reactop[f_1,f_1]}  + 2\avg{\reactop[f_0,f_2]} }{E \reactop[f_0,f_0]} ,
\end{equation}
where, as in \S\ref{sec_hydro}, $E$ is the normalized energy of the perturbation and $\alpha$ is the logarithmic derivative of the reactivity with respect to temperature. It will occasionally be useful to work in terms of the reactivity enhancement factor ${\Phi = E\reacF K}$.
To compute the terms of \eqref{eq_K_expansion_f}, the following calculation derives formulas for $f_1$ and $f_2$ in a plasma harboring small-amplitude compressive fluctuations.

\subsection{Kinetic model}
\label{sec_kinetic_model}

Consider, for simplicity, a plasma consisting of a single ion species of mass $m$, temperature $T$, number density $n$, and charge state $Z$, as well as electrons of the same temperature and of number density $n_e = Z n$. We restrict consideration to perturbations with wavelengths much longer than the Debye length, which allows us to ignore plasma oscillations and to assume that the electrostatic potential energy makes a negligible contribution to the total energy of the perturbation. 
In terms of the peculiar velocity $\bs s = \bs v - \bs u$, the kinetic equation for the ion distribution function $f$ is
\begin{equation}
    \label{eq_kinetic_v}
    \partial_t (nf) + (\bs s + \bs u) \cdot \nabla (nf) + n\paren{\bs a - \frac{d \bs u}{dt} - \bs s \cdot \nabla \bs u} \cdot \frac{\partial f}{\partial \bs s} = nC[f] ,
\end{equation}
where $\bs a$ is the acceleration of particles due to external forces, and $C[f]$ is the collision operator. We will assume that the plasma is sufficiently collisional that $f$ is close to a Maxwellian distribution. Perturbations from Maxwellian are generated by fluctuations in density, temperature, and flow velocity about uniform equilibrium values $n_0$, $T_0$, and ${\bs u_0 = 0}$. We define the thermal velocity $\vth$ to be a constant ${\vth = \sqrt{T_0/m}}$. 
The acceleration of the ions is given by $a = Ze E/m_i + \mc R$, where $e$ is the elementary charge and $\mc R$ is an effective electrothermal force. 
Because the electron mass is small, such that electrons respond quickly to perturbations in the ion density, the electric field is given by $\bs E = -\nabla p_e/(e n_e) - \mc R$, where $p_e$ is the electron pressure and $n_e$ is the electron number density. As required to conserve momentum, $\mc R$ therefore disappears, leaving $a = - Z \nabla p_e / (m_i n_e)$.

It is convenient to write the distribution function in terms of the normalized peculiar velocity $\bs w = \bs s/(\vth \tau)$, where ${\tau = \sqrt{T/T_0}}$, and let ${\widehat {\bs u} = \bs u/\vth}$ be the normalized flow velocity. 
The kinetic equation \eqref{eq_kinetic_v} becomes
\begin{equation}
    \label{eq_kinetic_w}
    \begin{split}
    &\partial_t f + \vth \tau \paren{\bs w + \widehat{\bs u}} \cdot \nabla f + \paren{\frac{d \ln n}{d t} + \vth\tau \bs w \cdot \nabla \ln n} f - 3 \paren{\frac{d \ln \tau}{dt} + \vth \tau \bs w \cdot \nabla \ln \tau} f
    \\& \qquad  - \left[\frac{Z\nabla p_e}{\vth\tau m n_e} + \frac{1}{\tau}\frac{d \widehat{\bs u}}{d t} +\frac{d \ln \tau}{d t} \bs w + \vth\bs w \cdot \nabla \widehat{\bs u} + \vth \tau \bs w \cdot \paren{\nabla \ln \tau} \bs w \right] \cdot \frac{\partial f}{\partial \bs w}
    = C[f] ,
    \end{split}
\end{equation}
where the convective derivative operator is ${d/dt = \partial_t + {\bs u}\nabla}$. 
To obtain a tractable solution to \eqref{eq_kinetic_w}, we approximate the effect of collisions using a velocity-dependent Krook operator, \textit{viz. }
\begin{equation}
    \label{eq_krook}
    C[f] = -\nu (f - \fm) ,
\end{equation}
While \eqref{eq_krook} is a crude approximation -- in particular, it does not conserve moments of the distribution function -- it is an invaluable tool for approximating kinetic dynamics occurring on the tail of the ion distribution \citep{Fetsch_Fisch_2025a,Fetsch_Fisch_2025b}. For simplicity, we approximate the velocity dependence of the collision operator by the model used in \cite{Fetsch_Fisch_2026_dpp}, \textit{viz.}
\begin{equation}
  \label{eq_nu_p}
  \widehat \nu(w) = \frac{1}{1 + w^3} + \frac{1}{Z}\sqrt{\frac{m_e}{m}} ,
\end{equation}
where $m_e$ is the electron mass. The factor of $1/Z$ comes from the $Z^2$ scaling of the ion-electron frequency normalized to the $Z^4$ scaling of the ion-ion collision frequency, with an additional factor of $Z$ accounting for the number of electrons. 
Comparison of a BGK operator of the form \eqref{eq_krook} and a more intricate Fokker-Planck operator in simulations of the SFRE has shown qualitatively similar behavior but with a reduction in the magnitude of the reactivity enhancement when pitch-angle scattering is included \cite{Guo_Wu_Zhang_2026}. 
Thus, while refinements of the model adopted here could improve the quantitative accuracy, it is suitable for assessing how the kinetic component of the fusion-power response scales with the parameters of the systems of interest.

We expand the distribution function as $f = f_0 + f_1 + f_2 + \ldots$, where $f_0 = \fm$, and $f_1$, $f_2$, etc. represent higher-order kinetic perturbations to the distribution function. 
For a perturbation of wavenumber $k$ and real frequency $\omega$, we will write 
\begin{align}
    \label{eq_n_expansion}
    n =& n_0 + n_1 + n_2 + \dots
    \\
    \label{eq_T_expansion}
    T =& T_0 + T_1 + T_2 + \dots ,
    \\
    \label{eq_u_expansion}
    \bs u =& \bs u_1 + \bs u_2 + \dots ,
    \\
    \label{eq_Pe_expansion}
    p_e =& Zn_0 T_0 + p_{e,1} + p_{e,2} + \dots,
\end{align}
where $n_1$, $T_1$, and $\bs u_1$ are the $\mc O(\mc M)$ terms representing the first-order perturbations to each fluid quantity. 
The final term in each of \eqref{eq_n_expansion}, \eqref{eq_T_expansion}, \eqref{eq_u_expansion}, and \eqref{eq_Pe_expansion} represents the second-order perturbation generated by nonlinear interactions. 
For each first-order quantity $Q_1$ ($= n_1$, $T_1$, $\bs u_1$, or $p_{e,1}$) in the perturbation, we consider a single mode of wavevector $\bs k$ and real frequency $\omega$, and define the normalized complex perturbation amplitude $\wt Q$ so that
\begin{equation}
    \label{eq_Q1_fourier}
    Q_1 = Q_0 \Bigg[ \wt Q e^{i\bs k \cdot \bs x - i \omega t - \mu t} + \wt Q^*e^{-i\bs k \cdot \bs x + i \omega t - \mu t}\Bigg] ,
\end{equation}
where $\wt Q^*$ is the complex conjugate of $\wt Q$ and $\mu$ is the damping rate. Each quantity ${S = S_0 + S_1 + S_2 + \dots}$, with perturbation amplitude $\wt S$ defined as in \eqref{eq_Q1_fourier}, satisfies the relation
\begin{equation}
    \label{eq_QT_avg_relation}
    \avg{S_1Q_1} = 2 \mr{Re} \big\{ \wt S^* \wt Q\big\} e^{-2\mu t} ,
\end{equation}
where $\mr{Re}\{\cdot\}$ denotes the real part of the argument.

When computing the averaged fusion power to second order in $\mc M$, the only part of the second-order perturbations ($n_2$, $T_2$, and so on) that will survive the volume-averaging operation is the component with ${\bs k = 0}$ and ${\omega = 0}$.
By conservation of particles, $\avg{n_2} = 0$, and by conservation of momentum, $\partial_t \avg{\bs u_2} = 2\mu\avg{n_1 \bs u_1}/n_0$. 
Let
\begin{equation}
    \label{eq_Gamma_def}
    \Gamma = -\frac{\partial_t \avg{T_2}}{T_0 } \frac{1}{2|\wt n|^2}
\end{equation}
describe the time evolution of the second-order temperature perturbation. 

\subsection{Amplitude expansion}

To first order in the perturbation amplitude, $f_1$ can be computed independently for each Fourier mode of the perturbation. Generically, according \eqref{eq_kinetic_w}, the evolution of the Fourier components of $f_1$ satisfies an equation of the form
\begin{equation}
    \label{eq_f1_evol_y}
    \wt y_1'(t) = \wt A e^{-qt} - \varpi \wt y_1(t) ,
\end{equation}
in terms of constants $\wt A$, $q$, and $\nu$, with the solution 
\begin{equation}
    \label{eq_f1_evol_y_soln}
    \wt y_1(t) = \frac{\wt A}{\varpi - q} \Big[e^{-qt} - Y_0 e^{-\varpi t}  \Big] ,
\end{equation}
where $Y_0$ is a constant of integration. Imposing the initial condition $\wt y_1(0) = 0$ sets $Y_0 = 1$., provided that $\varpi \neq q$. In the problem at hand, $\wt y_1 \to \wt f_1$, $q \to i\omega + \mu$, and $\varpi \to \nu + i \bs k \cdot \bs w \vth$. The collision frequency $\nu$ is constant in time but varies with velocity. The time-independent forcing constant $\wt A$ is the amalgamation of terms in \eqref{eq_kinetic_w} that contribute at first order in $\mc M$. Imposing the initial condition $\wt f_1(0) = 0$ is equivalent to assuming that an initially quiescent, Maxwellian plasma is suddenly subjected to a perturbation at time $t = 0$. Then the solution $\wt f_1(t)$ incorporates the transient response of the distribution function to the perturbation on the time scale $1/\nu$, as well as the typically longer-time scale behavior on the time scale $1/\mu$, with a magnitude that is quasi-stationary with respect to the decaying amplitude of the perturbation. 

It would be convenient to discard the $\exp(-(\nu + i \bs k \cdot \bs w \vth) t)$ term and keep only the the $\exp(-i\omega t - \mu t)$ term, equivalent to considering waves that have been present for sufficient time to damp out the initial transient response. This approach fails, however, when $\mu$ becomes larger than $\nu$, which can occur in rapidly decaying perturbations or for sufficiently weakly collisional particles in the extreme tail of the distribution. In such cases, the transient response to the imposition of the perturbation actually lasts longer than the perturbation itself and can therefore not be neglected. Mathematically, this manifests as the pole in the solution \eqref{eq_f1_evol_y_soln} at $\varpi = q$. (The pole generally lies off of the real axis in the solution for $\wt f_1$, provided that the frequency has a real component, but it remains suggestive of pathological behavior if the $\exp(-\nu t)$ term is ignored when $\nu \sim \mu$.)
We therefore write the amplitude $\wt F_1$ of the first-order perturbation such that
\begin{equation}
    \label{eq_F1_time_dep_def}
    \wt f_1 = \wt F_1 e^{i\bs k \cdot \bs x}\Big[e^{- i \omega t - \mu t} - e^{-\nu t - i\bs k \cdot \bs w \vth t} \Big] .
\end{equation}
where the amplitude $\wt F_1$ of the first-order perturbation satisfies the kinetic equation
\begin{equation}
    \begin{split}
    \label{eq_kinetic_eqn_f1}
    & (-i\omega - \mu) \wt F_1 + i \bs k \cdot \bs w \vth \wt F_1 + \paren{-i\omega -\mu + i \bs k \cdot \bs w \vth}\wt n \fm - \frac{3}{2}(-i\omega - \mu + i\bs k \cdot \bs w \vth) \wt T \fm
    \\& + \left[i \bs k \cdot \bs w \vth Z \wt p_e
    + (-i\omega - \mu) \wt{\bs u} \cdot \bs w
    + \half (-i\omega - \mu) \wt T w^2
    + i\bs k \cdot \bs w \vth \wt{\bs u} \cdot \bs w
    + \half i \bs k \cdot \bs w \vth \wt T w^2
    \right] \fm
    \\& = C[\wt F_1] - \nu e^{-\nu t - i\bs k \cdot \bs w \vth t+ i\omega t + \mu t}\wt F_1.
    \end{split}
\end{equation}
Without loss of generality, we can align the coordinate system such that $\bs k = k \bs{e}_z$, where $\bs{e}_z$ is the unit vector in the $z$ direction. For a longitudinal perturbation, ${\wt{\bs u} = \wt u_z \bs{e}_z}$ and continuity requires that ${\wt u_z = \wt n i(-i\omega - \mu)/k\vth}$. Let $w_z = \bs w \cdot \bs{e}_z$ be the component of the normalized velocity in the $z$ direction, and let $\xi = w_z/w$. Let $\vartheta = \wt T/\wt n$ be the ratio of temperature perturbation amplitude to density perturbation amplitude and let $\varrho = \wt p_e/\wt n$ be the ratio of pressure perturbation amplitude to density perturbation amplitude. 
Then\eqref{eq_kinetic_eqn_f1} becomes
\begin{equation}
    \label{eq_f_soln}
    \begin{split}
    & 0=\nu \wt F_1 -  (i\omega + \mu) \wt F_1 + i k w \xi \vth \wt F_1 + (-i\omega - \mu + i k w \xi \vth) \wt n \fm - \frac{3}{2} (-i\omega - \mu + ikw\xi \vth) \vartheta \wt n \fm
    \\
    & + \Bigg[ i k w \xi \vth Z \varrho + \frac{i(i\omega + \mu)^2}{k\vth} w\xi  + \half (-i\omega - \mu) \vartheta w^2
    - i kw^2\xi^2 \vth \frac{i(i\omega + \mu)}{k\vth} + \half i k w^3 \vth \xi \vartheta  \Bigg] \wt n \fm  ,
    \end{split}
\end{equation}
from which we can solve for $\wt F_1$. 
\begin{comment} to find
\begin{equation}
    \label{eq_f1_soln}
    \begin{split}
    \wt F_1 = \frac{ i k \vth \wt n \fm }{\nu + i k w \xi \vth - i\omega - \mu}\Bigg[&\frac{\omega - i\mu}{k\vth} \paren{1 - \frac{3}{2}\vartheta}  
    - \paren{1 -\frac{3}{2}\vartheta + Z\varrho - \frac{(\omega - i\mu)^2}{k^2\vth^2}}w \xi 
    \\& - \paren{\xi^2 - \half \vartheta}\frac{\omega - i\mu}{k\vth}w^2
    - \half \vartheta w^3 \xi
    \Bigg]
    \end{split}
\end{equation}
\end{comment}
Let 
\begin{equation}
    \label{eq_c_def}
    c = \frac{\omega - i\mu}{k\vth}
\end{equation}
be the normalized complex phase velocity of the perturbation and let 
\begin{equation}
    \label{eq_N_def}
    N = \frac{\nu}{k\vth}
\end{equation}be the inverse Knudsen number. Then \eqref{eq_f_soln} reduces to
\begin{equation}
    \label{eq_f1_soln_c}
    \wt F_1 = \frac{i \wt n \fm}{N + iw\xi - ic} \Bigg[ \paren{1 - \frac{3}{2}\vartheta} c - \Big(1 - \frac{3}{2}\vartheta + Z\varrho - c^2\Big)w\xi - \paren{\xi^2 - \half \vartheta} c w^2  - \half \vartheta w^3 \xi \Bigg] .
\end{equation}
\begin{comment}
\begin{equation}
    \label{eq_f1_soln_c}
    \wt F_1 = \frac{i \wt n \fm}{N + iw\xi - ic} \Bigg[ c - \paren{i\frac{4\eta k c}{3\vth} - \vartheta}w\xi - \paren{\xi^2 - \half \vartheta} c w^2  - \half \vartheta w^3 \xi \Bigg] .
\end{equation}
\end{comment}
At second order in the perturbation amplitude, the volume-averaged reactivity depends only on the $\bs k =0$ component of $f_2$. Keeping only terms that contribute to $F_2(\bs w; \bs k')$, \eqref{eq_kinetic_w} becomes
\begin{equation}
    \label{eq_f2_kinetic_start}
    \begin{split}
    &\partial_t \avg{f_2} + \avg{\paren{\half \frac{T_1}{T_0} \vth \bs w + \vth \widehat{\bs u}_1} \cdot \nabla f_1} 
    \\&+ \avg{\paren{\partial_t \paren{\frac{n_1}{n_0}} + \vth \bs w \cdot \nabla \paren{\frac{n_1}{n_0}} - \frac{3}{2}\partial_t\paren{\frac{T_1}{T_0}} - \frac{3}{2} \vth \bs w \cdot \nabla \paren{\frac{T_1}{T_0}}      } f_1}
    \\
    & + \Bigg\langle \Bigg[-\half \partial_t \paren{\frac{n_1^2}{n_0^2}} + \vth \widehat{\bs u}_1 \cdot \nabla \paren{\frac{n_1}{n_0}} + \vth \half \frac{T_1}{T_0} \bs w \cdot \nabla \paren{\frac{n_1}{n_0}} 
    \\ 
    & \qquad + \frac{3}{4}\partial_t\paren{\frac{T_1}{T_0}}^2 - \frac{3}{2}\partial_t\paren{\frac{T_2}{T_0}} - \frac{3}{2}\vth \widehat{\bs u}_1 \cdot \nabla \paren{\frac{T_1}{T_0}}  - \vth \frac{3}{4} \frac{T_1}{T_0} \bs w \cdot \nabla \paren{\frac{T_1}{T_0}}\Bigg] f_0 \Bigg\rangle
    \\
    & - \Bigg\langle\Bigg[\frac{1}{\vth m n_0}\paren{-\half\frac{T_1}{T_0} - \frac{n_1}{n_0}}\nabla p_{e,1}
    - \half\frac{T_1}{T_0} \vth \partial_t \widehat{\bs u}_1 + \vth \partial_t \widehat{\bs u}_2
    + \half \frac{\partial_t T_2}{T_0} \bs w 
    \\& - \frac{1}{4}\partial_t \paren{\frac{T_1}{T_0}}^2 \bs w + \half \vth \widehat{\bs u}_1 \cdot \nabla \paren{\frac{T_1}{T_0}} \bs w  \Bigg] \cdot \frac{\partial f_0}{\partial \bs w}\Bigg\rangle
    \\ % (note we deleted - \frac{1}{8} \vth \bs w \cdot \nabla \paren{\frac{T_1}{T_0}}^2 \bs w because its spatial average is zero. Same deal for \vth \widehat{\bs u}_1\cdot \nabla \widehat{\bs u}_1 )
    & - \avg{\Bigg[ \frac{1}{\vth m n_0} \nabla p_{e,1} + \partial_t \widehat{\bs u}_1 + \vth \bs w \cdot \nabla \widehat{\bs u}_1 + \half \partial_t \paren{\frac{T_1}{T_0}} \bs w  + \half \vth \bs w \cdot \nabla \paren{\frac{T_1}{T_0}} \bs w\Bigg] \cdot \frac{\partial f_1}{\partial \bs w}}
    \\
    & = -\nu \avg{f_2} - \nu \avg{\frac{n_1}{n_0} f_1} - \nu_T \nu \avg{\frac{T_1}{T_0} f_1} ,
    \end{split}
\end{equation}
where the last two terms on the right-hand side come from requiring that the collision frequency be linear in the density and defining $\nu_T = d \ln \nu/\ln T$, evaluated at $T_0$.
Let $f_2(0,\bs w; \bs k')$ denote the $\bs k = 0$ component of $f_2$ sourced by a perturbation of wavenumber $\bs k'$ along with its conjugate. 
The time evolution of the Fourier components of $f_2(0,\bs w; \bs k')$ satisfies an equation of the form
\begin{equation}
    \label{eq_f2_evol_y}
    \wt y_2'(t) = \wt A e^{-2\mr{Re} qt} + \wt B \Big[e^{-2 \mr{Re} q t} - e^{-(q^* + \varpi )t}\Big] - \wt C t e^{-(q^* + \varpi)t}  - \nu \avg{y_2(t)} ,
\end{equation}
where, as in \eqref{eq_f1_evol_y}, $\wt A$, $\wt B$, and $\wt C$ are constants appearing in the expansion of the kinetic equation to second order in the perturbation amplitude, ${q = i\omega + \mu}$, and ${\varpi = \nu + i \bs k' \cdot \bs w \vth}$. The terms contributing to $\wt A$ come from interactions between two first-order fluid quantities (for example, $\avg{n_1^2}$). The terms contributing to $\wt B$ come from interactions between a first-order fluid quantity and the first-order kinetic perturbation $f_1$ and therefore include both the $\exp(-i\omega - \mu t$) and $\exp(-\nu t - i\bs k' \cdot \bs w t)$ terms of $f_1$. Contributions to $\wt C$ are from those terms proportional to $\partial f_1/\partial w$ in which the velocity derivative acts on the $\exp(-i\varpi t)$ term of $f_1$, pulling down a factor of $(d\varpi/dw)t$.
Imposing the initial condition $\wt y_2(0) = 0$, the solution to \eqref{eq_f2_evol_y} is
\begin{equation}
    \label{eq_f2_evol_y_soln}
    \begin{split}
    \wt y_2(t) =& \frac{\wt A + \wt B}{\nu - 2\mr{Re}q} \Big[ e^{-2 \mr{Re} q t} - e^{-\nu t} \Big] 
    - \frac{\wt B}{\nu - q^* - \varpi} \Big[ e^{-(q^*  +\varpi) t} - e^{-\nu t} \Big] 
    \\
    & + \frac{\wt C}{(q^* + \varpi - \nu)^2} \Big[ \big[(q^* + \varpi - \nu) t + 1\big]e^{-(q^* + \varpi)t} - e^{-\nu t}\Big].
    \end{split}
\end{equation}

Let $F_2(\bs w; \bs k')$, $J^{(R,I)}_2(\bs w; \bs k')$, and $M^{(R,I)}_2(\bs w; \bs k')$ describe the amplitude of the averaged second-order distribution such that 
\begin{equation}
    \label{eq_f2_F2_J2}
    \begin{split}
    \avg{f_2} =& F_2\Big[e^{-2\mu t} - e^{-\nu t}\Big] 
    \\ & + \Big[J^{(R)}_2\Big(\cos\big([\omega - \bs k \cdot \bs w \vth] t\big)e^{-(\mu + \nu) t}   - e^{-\nu t}\Big)+ J^{(I)}_2 \sin\big([\omega - \bs k \cdot \bs w \vth] t\big)e^{-(\mu + \nu) t}  \Big]
    \\ & + \Big[( \mu t + 1)M^{(R)}_2 - (\omega - \bs k \cdot \bs w \vth)tM^{(I)}_2\Big]\cos\big([\omega - \bs k \cdot \bs w \vth] t\big)e^{-(\mu + \nu) t} - M^{(R)}_2e^{-\nu t}
    \\ &\hspace{0.0cm} + \Big[(\omega - \bs k \cdot \bs w \vth)tM^{(R)}_2 + (\mu t + 1)M^{(I)}_2  \Big]\sin\big([\omega - \bs k \cdot \bs w \vth] t\big)e^{-(\mu + \nu) t}  .
    \end{split}
\end{equation}
In the case where $\nu \ll \mu$, meaning that the transient behavior can be clearly distinguished from the long-time asymptotic decaying behavior, the only constituent of \eqref{eq_f2_F2_J2} to survive at late times is the $F_2 e^{-2\mu t}$ term. The $F_2 e^{-\nu t}$ term describes the evolution of the plasma from the unperturbed initial state to the asymptotic state. The $J^{(R,I)}_2$ and $M^{(R,I)}_2$ terms describe an additional component of the transient response arising from the ballistic streaming of particles across the wave.

The computation of $F_2$, $J^{(R,I)}_2$, and $M^{(R,I)}_2$ is described in Appendix \ref{sec_app_f2}. For concision in writing the results, it is useful to define the negative of the logarithmic velocity derivative of the collision frequency as
\begin{equation}
    \label{eq_h_def}
    h = -\frac{d \ln \nu}{d \ln w} 
\end{equation}
and the denominator of the prefactor in $\wt F_1$ as
\begin{equation}
    \label{eq_D_f2helper}
    D = N + iw\xi - ic .
\end{equation}
Because the second-order distribution function is sourced, in part, by the variation of the collision frequency with temperature in different parts of the perturbation, we define $N_T = \nu_T/(N k\vth)$. In the terms arising from dissipation of the perturbation, we let
\begin{equation}
    \widehat \mu = \frac{\mu}{\nu} \quad \text{and} \quad \widehat \Gamma = \frac{\Gamma}{\mu} 
\end{equation}
be the normalized damping rate and the normalized rate of change of the second-order temperature perturbation, respectively. 
Finally, it is convenient to define a frequently appearing collection of terms as
\begin{equation}
    \label{eq_X_def}
    X = c^2 - Z\varrho - 1 + \frac{3}{2}\vartheta
\end{equation}
and to define the particle velocity relative to the wave phase velocity as
\begin{equation}
    \label{eq_W_def}
     W = w\xi - c .
\end{equation}

The derivation in Appendix \ref{sec_app_f2} defines an auxiliary function $\mc F_2$ as
\begin{equation}
    \label{eq_F2_mc_def}
    \begin{split}
    \mc F_2 = &\quad i \frac{\paren{-ic^* + i W^*\paren{1 - \frac{3}{2}\vartheta^*}  - N - N_T N \vartheta^* - \half i \vartheta^* w\xi}D^*}{|D|^2} 
    \\ & \hspace{1.5cm} \times \bigg[c\paren{1- \frac{3}{2}\vartheta} + Xw\xi - c w^2\xi^2 - \half \vartheta W w^2 \bigg]
    \\ & + i \bigg( -Z\varrho^* - c^*w\xi + (c^*)^2\bigg)\frac{\paren{-iw^2\xi D + w + ihN\xi} (D^*)^2}{w |D|^4}
     \\ & \hspace{1.5cm} \times \bigg[c\paren{1- \frac{3}{2}\vartheta} + Xw\xi - c w^2\xi^2 - \half \vartheta W w^2   \bigg]
    \\ & + i \bigg( -Z\varrho^* - c^*w\xi + (c^*)^2\bigg)\frac{i D^*}{|D|^2}
    \bigg[ X - 2\paren{1 - \half \vartheta}c w\xi - \vartheta \paren{\xi^2 + \half} w^2 \bigg]
    \\ & + i \half \vartheta^* \bigg( -w^2\xi + c^* w\bigg)\frac{\paren{-iw^2 D +w\xi + ihN} (D^*)^2}{w|D|^4}
     \\ & \hspace{1.5cm} \times \bigg[c\paren{1- \frac{3}{2}\vartheta} + Xw\xi - c w^2\xi^2 - \half \vartheta W w^2 \bigg]
    \\ & + i \half \vartheta^* \bigg( -w^2\xi + c^* w\bigg)\frac{i D^*}{|D|^2}\bigg[ X\xi - 2\paren{\xi^2 - \half \vartheta}c w - \frac{3}{2}\vartheta w^2 \xi \bigg] ,
    \end{split}
\end{equation}
leading to the expression for $F_2$ in \eqref{eq_app_F2_final}, replicated here:
\begin{equation}
    \label{eq_F2_final}
    \begin{split}
    F_2(\bs w; \bs k) = \frac{2}{2\widehat \mu - 1} \mr{Re}\Bigg\{  - \frac{1}{N}\mc F_2  + & \frac{1}{N}\Big(-iZ\varrho + \frac{i}{2}\vartheta^* (1 + c^2 - Z \varrho) \Big) w\xi + \frac{1}{N}\frac{i}{2} c^*\vartheta (w^2-3) 
    \\ & + 2\widehat \mu c w \xi - \half \widehat \mu \widehat \Gamma (w^2-3) + \half \widehat \mu |\vartheta|^2 (w^2-3)
    \Bigg\} |\wt n|^2 \fm .
    \end{split}
\end{equation}

By matching the time-dependent components on both sides of \eqref{eq_J_complex_forms}, we have
\begin{align}
    \label{eq_J2_R_final}
    J^{(R)}_2(\bs w; \bs k) = & 2\mr{Re}\Bigg\{ \frac{i\mc F_2}{c^* - w \xi}
    \Bigg\} |\wt n|^2 \fm ,
    \\
    \label{eq_J2_I_final}
    J^{(I)}_2(\bs w; \bs k) = & 2\mr{Im}\Bigg\{ \frac{-i\mc F_2}{c^* - w \xi}
    \Bigg\} |\wt n|^2 \fm ,
\end{align}
where $\mr{Im}$ denotes the imaginary part. Finally, to compute $M^{(R,I)}_2$, we define an auxiliary function $\mc G_2$ in \eqref{eq_app_G2_mc_def}, replicated here:
\begin{align}
    \label{eq_G2_mc_def}
    \mc G_2 = &\Bigg(\Big[Z\varrho^* + c^* w\xi - (c^*)^2\Big]\Big(ih N \xi + w\Big) + \Big[ \half \vartheta^* w^2 \xi - \half c^* \vartheta^* w \Big]\Big(i h N + w\xi \Big)\Bigg) \nonumber
    \\ & \hspace{0.0cm} \times \frac{-D^*}{w|D|^2}\bigg[c\paren{1- \frac{3}{2}\vartheta} + Xw\xi - c w^2\xi^2 - \half \vartheta W w^2 \bigg]
\end{align}
in terms of which $M^{(R)}_2$ and $M^{(I)}_2$ are
\begin{align}
    \label{eq_M2_R_final}
    M^{(R)}_2(\bs w; \bs k) = & 2\mr{Re}\Bigg\{\frac{-\mc G_2}{(c^* - w\xi)^2}\Bigg\} |\wt n|^2 \fm ,
    \\  
    \label{eq_M2_I_final}
    M^{(I)}_2(\bs w; \bs k) = & 2\mr{Im}\Bigg\{\frac{\mc G_2}{(c^* - w\xi)^2}\Bigg\} |\wt n|^2 \fm .
\end{align}

\begin{figure}
    \centering
    \begin{minipage}{0.48\columnwidth}
        \centering
    \includegraphics[width=\textwidth]{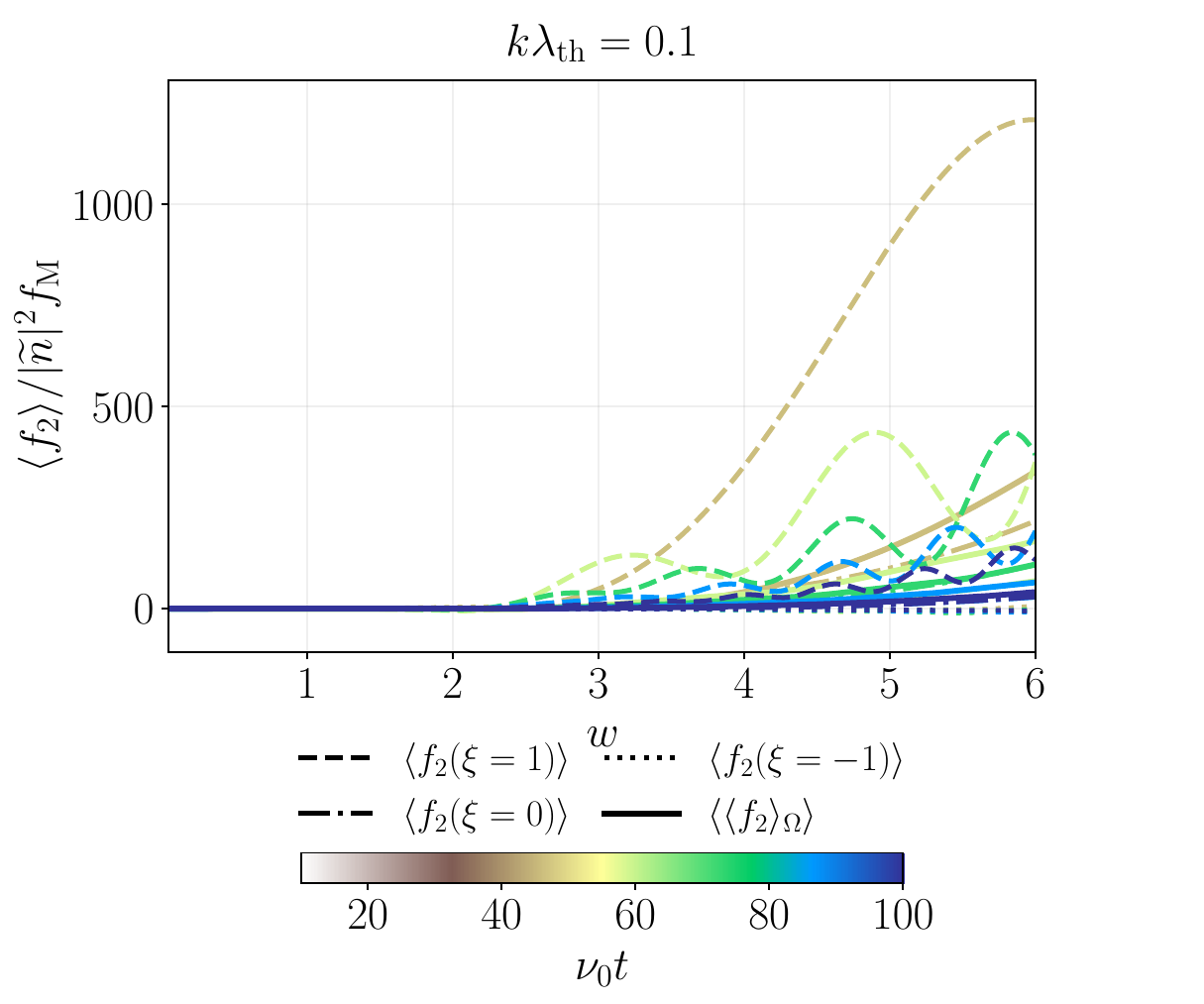}
        % python acoustic_paper_plots.py -p f2 --w-min 0.05 --w-max 6.0 --t-range "10,100,5" -k 0.1 -nw 200 --cmap terrain_r
        %\centerline{(a)}
    \end{minipage}\hfill
    % --- Second Subfigure ---
    \begin{minipage}{0.48\columnwidth}
        \centering
    \includegraphics[width=\textwidth]{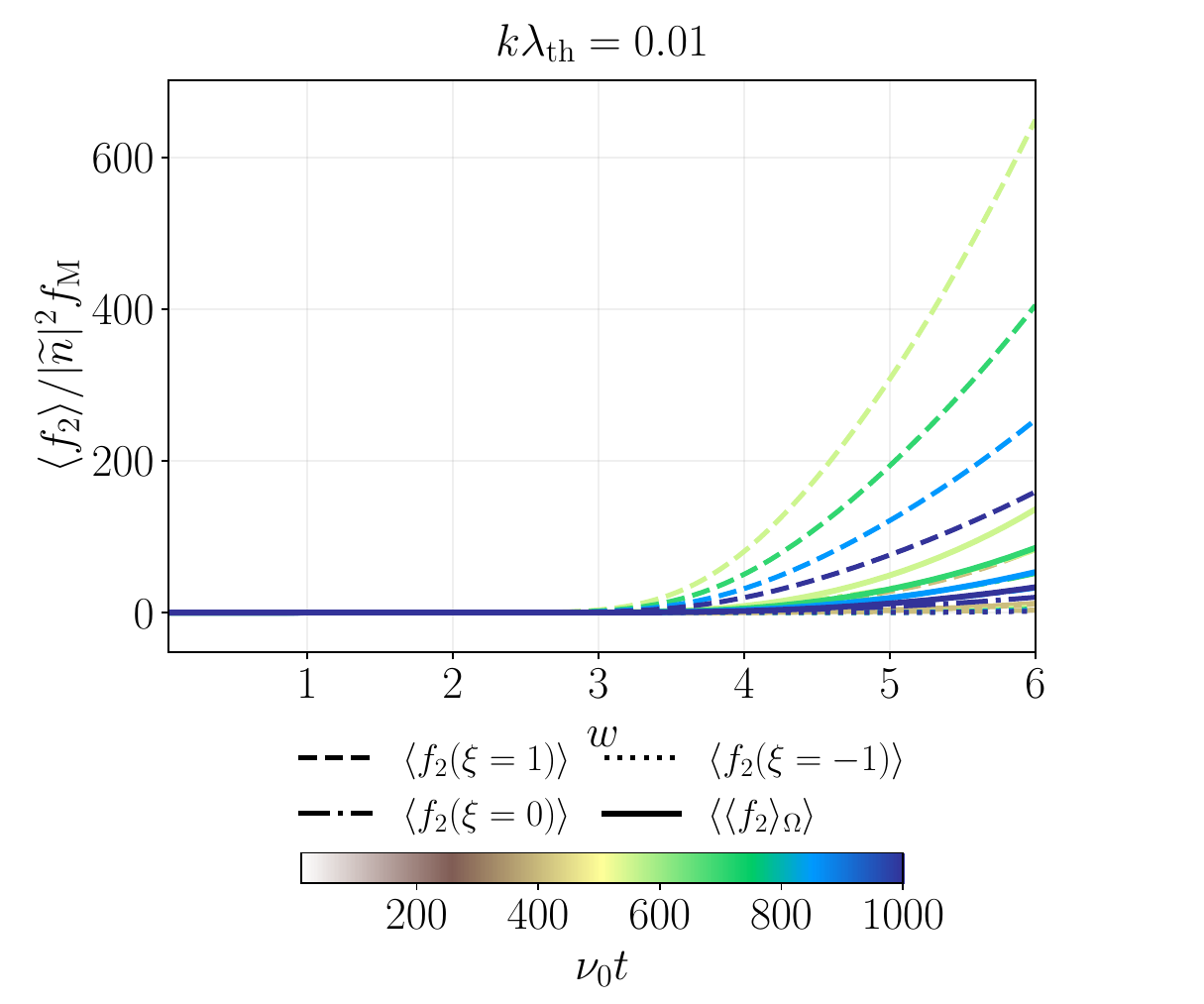}
        % python acoustic_paper_plots.py -p f2 --w-min 0.05 --w-max 6.0 --t-range "10,1000,5" -k 0.01 -nw 200 --cmap terrain_r
        %\centerline{(b)}
    \end{minipage}
    %\vspace{0.5cm}
    \caption{\justifying
    Normalized second-order distribution function $f_2$ in an acoustic wave with $k\lth = 0.1$ (left) and $k\lth = 0.01$ (right), averaged over space and over angles in velocity space and shown at times ranging from $t\nu_0 = 10$ to $t\nu_0 = 100$.
    }
    \label{fig_f2_t_k01_k001}
\end{figure}

\subsection{Exact reactivity integral}
\label{sec_kinetic_reac_int}

Combining \eqref{eq_F2_final}, \eqref{eq_J2_R_final}, \eqref{eq_J2_I_final}, \eqref{eq_M2_R_final}, and \eqref{eq_M2_I_final} via \eqref{eq_f2_F2_J2}, straightforward numerical integration allows computation of the $\reactop[f_0,f_2]$ term of \eqref{eq_K_expansion_f}. 
To compute the remaining reactivity corrections, we need the volume averages of the one-particle and two-particle first-order perturbed distributions. 
The $\reactop[f_0,f_1]$ term is determined by $\avg{n_1 f_1}$, which can be written as
\begin{equation}
    \label{eq_f1_avg_corr}
    \begin{split}
    \avg{ n_1 f_1} =& 2 \mr{Re}\left\{\wt n^* \wt F_1\right\} \Big[e^{-2\mu t} - \cos\big((\omega - \bs k \cdot \bs w \vth)t\big)e^{-(\mu + \nu)t}  \Big] 
    \\ & + 2 \mr{Im}\left\{\wt n^* \wt F_1\right\} \sin\big((\omega - \bs k \cdot \bs w \vth)t\big)e^{-(\mu + \nu)t},
    \end{split}
\end{equation}
for which it is useful to note that, using \eqref{eq_f1_soln_c},
\begin{equation}
    \label{eq_f1_kinetic_V_avg}
    \wt n^* \wt F_1 = \frac{i D^*}{|D|^2} \bigg[ c \paren{1 - \frac{3}{2}\vartheta} + X w\xi - \paren{\xi^2 - \half \vartheta} c w^2  - \half \vartheta w^3 \xi  \bigg] |\wt n|^2  \fm .
\end{equation}
To compute the volume-averaged two-particle distribution, we start by observing that 
\begin{equation}
    \label{eq_f1f1_avg_corr}
    \begin{split}
    \avg{f_1(\bs w')f_1(\bs w)} = & 2\mr{Re} \left\{ \wt F_1^*(\bs w', \bs k) \wt F_1(\bs w, \bs k) \right\}\Big[ 
        e^{-2\mu t} + \cos\big( \bs k \cdot (\bs w - \bs w') \vth t \big)e^{-(\nu + \nu') t} 
    \\ & \hspace{0.5cm} - \cos\big((\omega - \bs k \cdot \bs w \vth)t \big)e^{-(\mu + \nu) t} 
    - \cos\big((\omega - \bs k \cdot \bs w' \vth)t \big)e^{-(\mu + \nu') t}  \Big]
    \\
    & + 2\mr{Im} \left\{ \wt F_1^*(\bs w', \bs k) \wt F_1(\bs w, \bs k) \right\}\Big[ 
        \sin\big( \bs k \cdot (\bs w - \bs w') \vth t \big)e^{-(\nu + \nu') t} 
    \\ & \hspace{0.5cm} + \sin\big((\omega - \bs k \cdot \bs w \vth)t \big)e^{-(\mu + \nu) t} 
    - \sin\big((\omega - \bs k \cdot \bs w' \vth)t \big)e^{-(\mu + \nu') t}  \Big],
    \end{split}
\end{equation}
where $\nu' = \nu(w')$. 
According to \eqref{eq_reac_op_def} and \eqref{eq_K_expansion_f}, and then substituting the results in \eqref{eq_f2_evol_y_soln} and \eqref{eq_f1_avg_corr}, the full reactivity correction is
\begin{equation}
\label{eq_Phi_full_integral}
\begin{split}
    \Phi = 1+&\int d^3 w \int d^3 w' f_M(w) f_M(w') \sigma(|\bs w - \bs w'|) |\bs w - \bs w'| 
    \\
    & \times \bigg[ \bigg( (4+2\alpha\vartheta^*) 2\mr{Re}\Big\{ \wt n^* F_1(w)\Big\} + 2\mr{Re}\Big\{ F_1^*(\bs w') F_1(\bs w)\Big\} + 2F_2(\bs w) \bigg) e^{-2\mu t}
    \\
    & + 2\bigg(
        (2 + \alpha\vartheta^*) 2\mr{Re}\Big\{ -\wt n^* F_1(\bs w)\Big\} - 2\mr{Re}\Big\{ F_1^*(\bs w') F_1(\bs w)\Big\} + J_2^{(R)}(\bs w)
        \\
       & \hspace{0.5cm}  + (\mu t + 1)M_2^{(R)}(\bs w) - [\omega - \bs k \cdot \bs w \vth]t M_2^{(I)}(\bs w)
        \bigg)\cos\big((\omega - \bs k \cdot \bs w \vth)t\big)e^{-(\mu + \nu) t} 
    \\
    & + 2\bigg((2 + \alpha\vartheta^*) 2\mr{Im}\Big\{ \wt n^* F_1(\bs w)\Big\} + J_2^{(I)}(\bs w) + 2\mr{Im}\Big\{ F_1^*(\bs w') F_1(\bs w)\Big\}
        \\
       & \hspace{0.5cm}  + (\mu t + 1)M_2^{(I)}(\bs w) + [\omega - \bs k \cdot \bs w \vth]t M_2^{(R)}(\bs w)
        \bigg)\sin\big((\omega - \bs k \cdot \bs w \vth)t\big)e^{-(\mu + \nu) t} 
    \\
    & + 2\mr{Re}\Big \{ F_1^*(\bs w') F_1(\bs w) \Big \} \cos\big( \bs k \cdot (\bs w - \bs w') \vth t \big)e^{-(\nu + \nu') t}
    \\
    & + 2\mr{Im}\Big \{ F_1^*(\bs w') F_1(\bs w) \Big \} \sin\big( \bs k \cdot (\bs w - \bs w') \vth t \big)e^{-(\nu + \nu') t}
    \\
    & - 2\bigg(F_2 + J_2^{(R)} + M_2^{(R)} \bigg) e^{-\nu t}
    \bigg] \frac{1}{\reactop[f_M,f_M]} ,
\end{split}
\end{equation}
where the symmetry of the integrand under interchange of $\bs w$ and $\bs w'$ has been used to combine some terms, and the $\bs k$ argument has been suppressed for concision.

\begin{figure}
    \centering
    \vspace{0.5cm}
    \includegraphics[width=0.6\columnwidth]{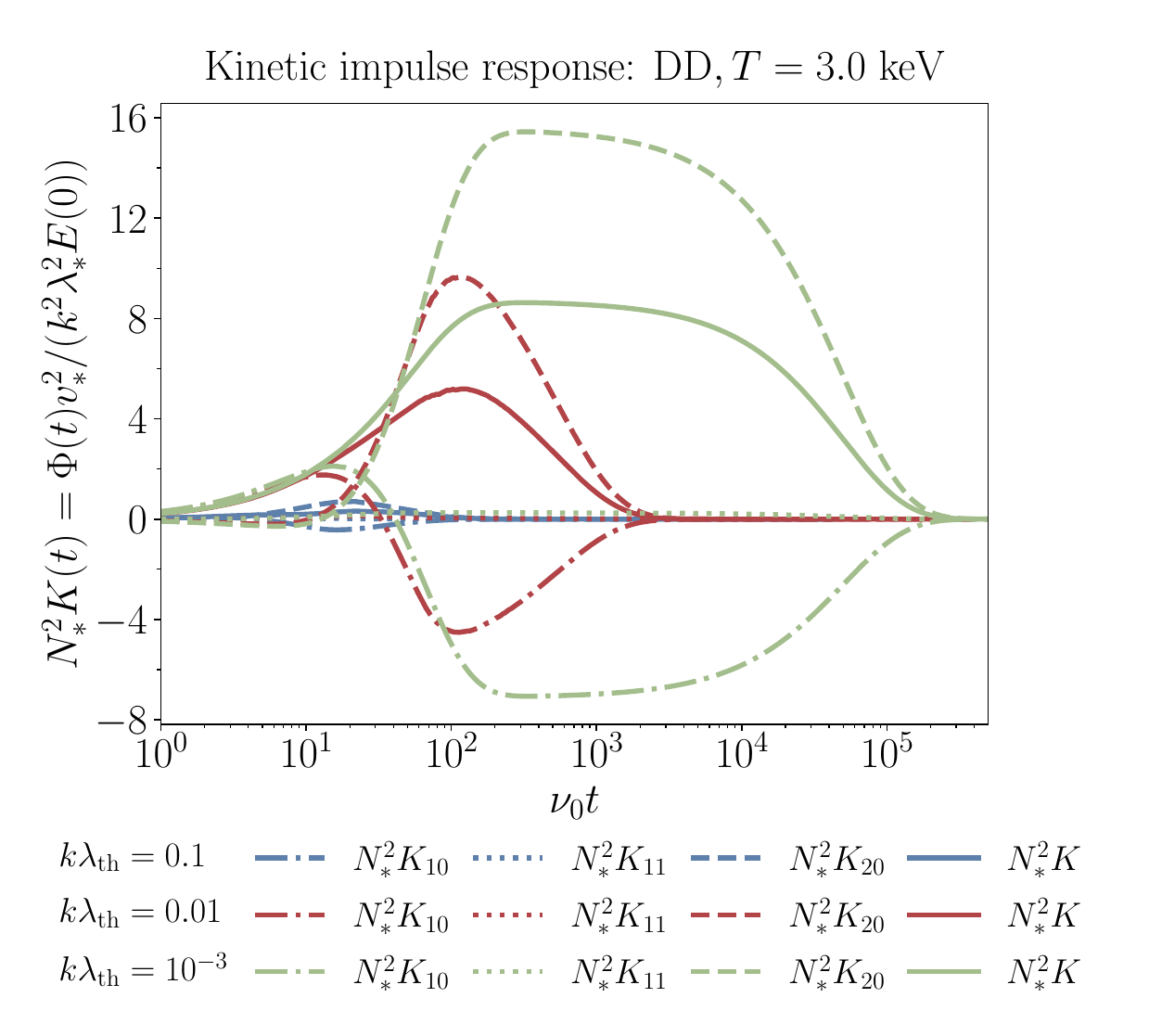}
    \caption{\justifying
    Time evolution of the kinetic fusion-power response obtained by numerical evaluation of \eqref{eq_Phi_full_integral} for acoustic waves in an unmagnetized deuterium plasma at $T = 3~\mr{keV}$.}
    \label{fig_Phi_full_integral}
\end{figure}

The results of direct numerical evaluation of \eqref{eq_Phi_full_integral} are shown in Fig.~\ref{fig_Phi_full_integral}. The reactivity integral is computed by Monte-Carlo sampling of $2^{20}$ points in $(\bs w, \bs w')$ space. The fluctuating quantities in the integrand (the values of $c$, $\vartheta$, and so on that determine $F_1$ and $F_2$) are chosen to correspond to an acoustic wave satisfying the relations derived in Appendix~\ref{sec_app_acoustic} in a pure deuterium plasma at $T = 3~\mr{keV}$. Waves are shown with wavenumbers satisfying $k\lth \in (0.1,0.01, 10^{-3})$. The horizontal axis shows time normalized to the collision time of thermal ions. All curves are normalized to $N_*^2 = (\nu_* / k\vth)^2$ to extract the dominant $1/N_*^2$ scaling of the kinetic corrections at long wavelengths.

\subsection{Approximate reactivity corrections}
\label{sec_kinetic_approx}

To extract the physical content of the results described in \eqref{eq_Phi_full_integral}, we proceed to apply the following three simplifications. 
First, in \S\ref{sec_kinetic_approx_gamow}, the fact that reactivity is determined primarily by a narrow region around the Gamow peak allows us to replace the two-body velocity-space integral in \eqref{eq_reac_op_def} with a simple evaluation of the perturbed distribution at the Gamow velocity. 
%Second, we specialize the results of \eqref{eq_f2_kinetic_simplified_final_re}, \eqref{eq_f1_kinetic_V_avg}, and \eqref{eq_f1f1_kinetic_V_avg} -- derived for arbitrary (small) amplitudes and relative phases of density, temperature, and flow -- to the case of perturbations where dissipation in slow. 
Second, in \S\ref{sec_kinetic_approx_bigN}, we specialize the results in \eqref{eq_Phi_full_integral} to perturbations with long wavelengths, for which $\mr{Kn} \to 0$ and $N \to \infty$; in this limit, dissipation becomes slow and the perturbed distributions simplify considerably, revealing the leading-order kinetic effects that appear as the hydrodynamic approximation breaks down for fast particles. Moreover, at long wavelengths, collisional damping becomes small, allowing a further reduction in complexity of expressions for the reactivity. 
These simplifications lead, in \S\ref{sec_kinetic_asymptotic}, to the asymptotic formula \eqref{eq_Phi} for the kinetic fusion-power response to compressive perturbations, valid in the limit where the Gamow velocity is much greater than the thermal velocity ($b \gg 1$) and where the Gamow-Knudsen number is much less than unity $\mr{Gk} \ll 1$; this formula is the kinetic complement to the general hydrodynamic response formula \eqref{eq_Pf_pert_alpha}.
As a third and final simplification, we consider in \S\ref{sec_comparison} the special cases of isobaric perturbations and adiabatic acoustic waves, deriving the kinetic complements to the hydrodynamic fusion-power responses captured in \eqref{eq_H_par_adiabat} and \eqref{eq_H_par_isotherm}.

\subsubsection{Gamow-peak approximation}
\label{sec_kinetic_approx_gamow}

Recall from \eqref{eq_sigma_model} that the cross sections of non-resonant fusion reactions can be approximated by expressions whose most rapidly varying component an exponential term ${\exp(-b\sqrt{2}\vth/v)}$. When the reactants are nearly Maxwellian, the exponential part of \eqref{eq_reac_op_def} can then be written as $\exp(-b/w_\mr{rel} - \half w_\mr{rel}^2 - \half w_\mr{tot}^2)$, where ${\bs w_\mr{rel} = (\bs w - \bs w')/\sqrt{2}}$ is the normalized relative velocity, ${\bs w_\mr{tot} = (\bs w + \bs w')/\sqrt{2}}$ is the normalized total velocity, and $\bs w$ and $\bs w'$ are the normalized peculiar velocities of the reactants. The Gamow peak, where this exponential term is maximized, is located at $w_\mr{rel} = b^{1/3}$ and $w_\mr{tot} = 0$. Let ${w_* = b^{1/3}/\sqrt{2}}$ be the Gamow velocity in normalized peculiar-velocity coordinates.

Suppose that $f_a$ and $f_b$ are distribution functions that are close to Maxwellian in the sense that their ratio to a Maxwellian with the same moments varies slowly relative to the Maxwellian itself. Let $\varphi$ be the deviation from Maxwellian such that 
\begin{equation}
    \label{eq_varphi_def}
    \varphi_{ab}(\bs w, \bs w') = \frac{f_a(\bs w)f_b(\bs w')}{\fm(\bs w)\fm(\bs w')} .
\end{equation}
The reactivity of these distributions can be approximated by evaluating the integrand of \eqref{eq_reac_op_def} at the Gamow peak; the accuracy of this approximation improves as $b$ becomes larger \citep{Fetsch_Fisch_2025b,Fetsch_Fisch_2026_dpp}. When $b \gg 1$, the reactivity can be approximated by
\begin{equation}
    \label{eq_reac_approx_varphi}
    \reactop[f_a, f_b] \sim \avg{\varphi_{ab}(w_* \bs {\hat r}, -w_* \bs {\hat r})}_\Omega\sqrt{\frac{2}{3}}\frac{A b^{1/3}S\paren{ 2w_*\vth}}{\vth}  e^{-\frac{3}{2}b^{2/3}} ,
\end{equation}
where $\bs {\hat r}$ is an arbitrary unit vector and $\avg{\cdot}_\Omega$ denotes an average over the direction of $\bs{\hat r}$. (The unsubscripted $\avg{\cdot}$ continues to indicate spatial and temporal averages.) From \eqref{eq_K_expansion_f}, it follows that the kinetic reactivity enhancement can be written as
\begin{equation}
    \label{eq_K_expansion_varphi}
    \begin{split}
        \Phi \sim & 1 + 2\avg{\paren{2\frac{n_1}{n_0} + \alpha \frac{T_1}{T_0}} \avg{\varphi_{10}(w_* \bs {\hat r}, -w_* \bs {\hat r})}_\Omega }
        \\ & + \avg{\avg{\varphi_{11}(w_* \bs {\hat r}, -w_* \bs {\hat r})}_\Omega} + 2\avg{\avg{\varphi_{20}(w_* \bs {\hat r}, -w_* \bs {\hat r})}_\Omega} ,
    \end{split}
\end{equation}
where \eqref{eq_K_expansion_varphi} is the approximation of \eqref{eq_Phi_full_integral} obtained by assuming a sufficiently sharp Gamow peak. 
For concision, we label each of the three terms on the right-hand side of \eqref{eq_K_expansion_varphi} as $\Phi_{10}$, $\Phi_{11}$, and $\Phi_{20}$, respectively, such that $\Phi \sim 1 + \Phi_{10} + \Phi_{11} + \Phi_{20}$ and ${K \sim (\Phi_{10} + \Phi_{11} + \Phi_{20})/E}$.

\begin{figure}
    \centering
    \vspace{0.5cm}
    \includegraphics[width=0.6\columnwidth]{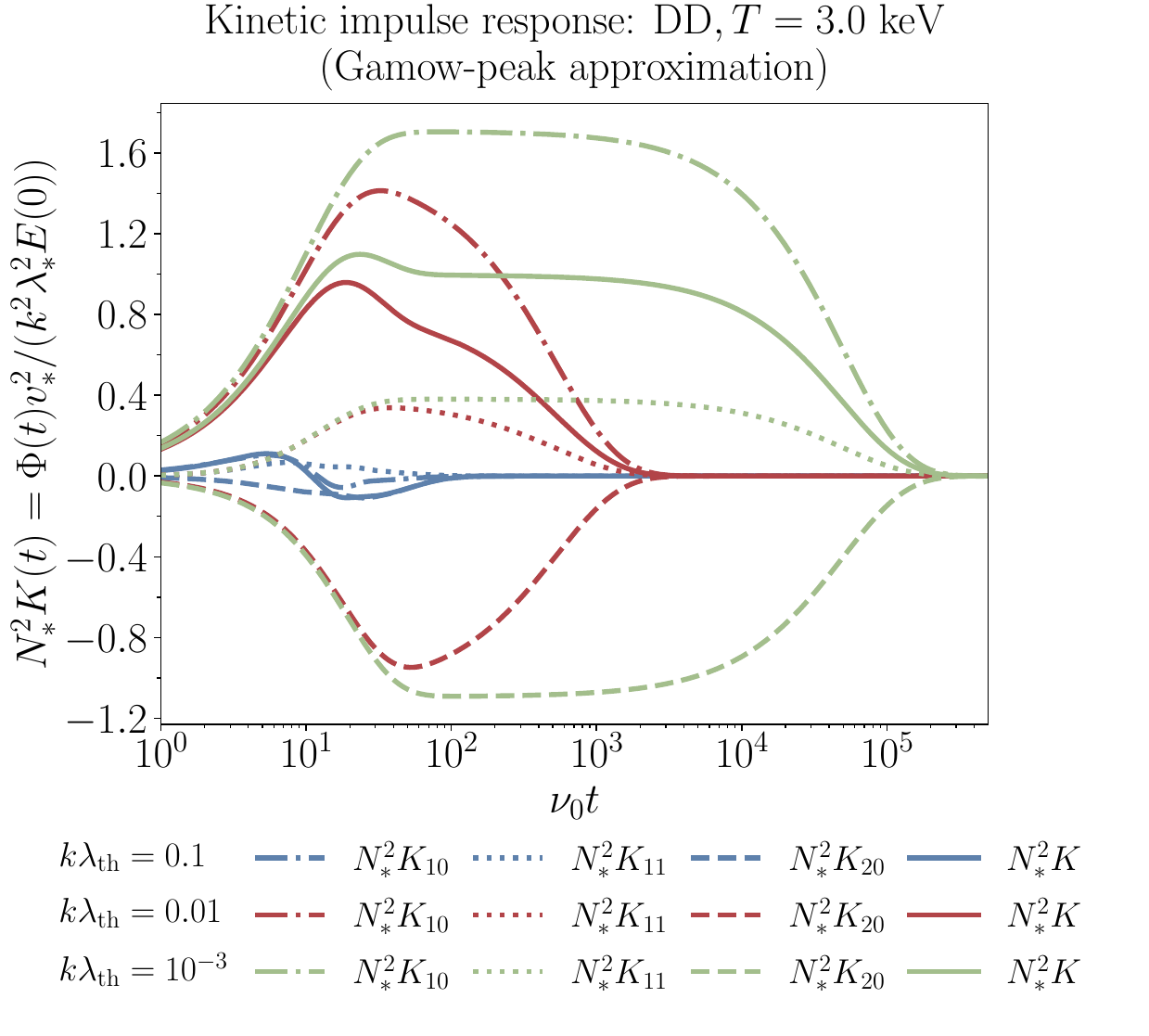}
    \caption{\justifying
    Time evolution of the kinetic fusion-power response obtained by approximating $\Phi$ about the Gamow peak. The curves correspond to acoustic waves with wavenumbers satisfying $k\lth \in (0.1,0.01,10^{-3})$ in an unmagnetized deuterium plasma at $T = 3~\mr{keV}$.}
    \label{fig_Phi_Gamow}
\end{figure}

The simplification afforded by \eqref{eq_K_expansion_varphi} is that, once the perturbed distributions have been averaged over space, time, and angle, their contributions to the perturbed reactivity can approximated by evaluating these distributions at $w_* = b^{1/3}/\sqrt{2}$. 
In \cite{Fetsch_Fisch_2025b}, this approximation was shown to be accurate within a modest order-unity factor when $b \approx 30$, corresponding to DD and DT reactions at temperatures around 3~keV. When $b$ is larger, as is the case for higher-$Z$ reactants and for lower temperatures, the Gamow-peak approximation becomes increasingly accurate. In Fig.~\ref{fig_Phi_Gamow}, the time evolution of the reactivity is shown for the same conditions as in Fig.~\ref{fig_Phi_full_integral}, but with the reactivity computed using the Gamow-peak approximation of \eqref{eq_K_expansion_varphi} instead of the full integral of \eqref{eq_Phi_full_integral}.

\subsubsection{Long-wavelength approximation}
\label{sec_kinetic_approx_bigN}

While the Gamow-peak approximation simplifies the reactivity computation, the expressions for the perturbed distributions are still too complicated for straightforward physical interpretation. To simplify these results, we take the long-wavelength limit, namely 
\begin{equation}
    \label{eq_lim_bigN}
    N_* \to \infty ,
\end{equation}
where ${N_* = \nu(v)/k\vth}$ is related to the inverse Gamow-Knudsen number by $N_*/w_* \sim \mc O(\mr{Gk})$. Naturally, as the wavelength becomes arbitrarily large, kinetic effects stemming from the finite Gamow mean free path vanish. Assuming that the frequency simultaneously becomes small -- as is the case for acoustic waves as the wavelength increases -- the effect of such perturbations on fusion power approaches the purely hydrodynamic behavior of \S\ref{sec_hydro}. 
We concern ourselves here with the case in which $1/N$ is asymptotically small but remains nonzero. Computing the corrections order-by-order in $1/N$ allows us to identify the first kinetic corrections that appear as a perturbation departs from the hydrodynamic limit.

Moreover, damping of the perturbation becomes weak in the long-wavelength limit, allowing a further simplification of the perturbed distributions. To see heuristically why this happens, suppose that the imaginary part of the frequency $\mu$ is produced by diffusive processes such as viscosity or heat conduction, rather than by kinetic processes such as Landau damping; this is the regime corresponding to the small Knudsen-number limit considered in this work. Let $\mc D$ be the effective diffusivity governing the damping of the perturbation, such that $\mu = \mc D k^2$. In unmagnetized plasma, or for field-parallel perturbations in magnetized plasma, assuming that $\mc D$ depends linearly on the ion and electron transport coefficients allows us to write $\mc D = C \nu_0 \lth^2$, where $C$ is a constant. It follows that $\mu \sim C\nu_0\mr{Kn}^2$ and the imaginary part $\mr{Im}$ of the perturbation speed $\mr{Im} \{c\} = \mu/k\vth$ scales as $\mr{Im} \{c\} \sim C \mr{Kn}$. We will assume that the imaginary parts of other fluctuating quantities, such as $\mr{Im} \{\vartheta\}$, are also small in the Knudsen number (see Appendix~\ref{sec_app_acoustic}).

%Noting that $1/N_* \sim \mc O(\mr{Gk})$, and that $\mr{Gk} \gg \mr{Kn}$ when $b \gg 1$, we adopt an ordering where $1/N_* \sim$

While the perturbed distributions were calculated as an expansion in $\mc M \ll 1$, they also constitute an expansion in $1/N$ when $N \gg 1$, corresponding to the standard Chapman-Enskog expansion. Hence, $f_1 \sim \mc O(1/N)$ and $f_2 \sim \mc O(1/N^2)$. We therefore compute $\Phi$ up to $\mc O(1/N^2)$. This calculation is detailed in Appendix~\ref{sec_app_bigN}. 
Combining the $\Phi_{10}$, $\Phi_{11}$, and $\Phi_{20}$ contributions derived in the appendix to compute the total kinetic correction $\Phi$, we find
\begin{equation}
    \label{eq_Phi}
    \begin{split}
    \Phi \sim & 1 + \frac{1}{N_*} c_R\paren{1 - \frac{w_*^2}{3}} \Bigg[ 2\alpha + 6 - 2N_T \Bigg] \frac{\avg{n_1(\bs x) T_1(\bs x - \frac{\bs L}{4})}}{n_0T_0} 
    \\ & +   \widehat \mu \paren{1 - \frac{w_*^2}{3}} \Bigg[(2 - 3\widehat \Gamma)\frac{\avg{n_1^2}}{n_0^2} + (2\alpha - 6 - 2N_T) \frac{\avg{n_1T_1}}{n_0T_0} - (3 \alpha - 3N_T - 3)\frac{\avg{T_1^2}}{T_0^2} \Bigg]
    \\ 
    & -\frac{1}{N_*^2}\Bigg[ \paren{2 + \frac{2}{3}h}c_R^4 + \paren{-3 - \frac{4}{3}h}c_R^2 + \paren{-\frac{1}{3}c_R^4 + \frac{2}{3}c_R^2 + 1 + \frac{2}{5}hc_R^2} w_*^2 - \frac{3}{5}c_R^2w_*^4 \Bigg] \frac{\avg{n_1^2}}{n_0^2}
    \\ 
    & - \frac{1}{N_*^2}\Bigg[\paren{-\frac{27}{4} - \frac{3}{2}h - 3\alpha} c_R^2 + \paren{\frac{11}{2}c_R^2 + \frac{1}{2}hc_R^2 + (c_R^2-1)\alpha - \frac{7}{4} - \frac{1}{2}h}w_*^2 
    \\ & \hspace{1.5cm} + \paren{-\frac{3}{4}c_R^2 + \frac{7}{6} + \frac{1}{6}h+ \frac{1}{3}\alpha}w_*^4 - \frac{1}{12}w_*^6\Bigg] \frac{\avg{T_1^2}}{T_0^2}
    \\ 
    & - \frac{1}{N_*^2}\Bigg[ \paren{9 + 3h + 2\alpha}c_R^2 + \paren{-\frac{29}{3}c_R^2 - \frac{4}{3}hc_R^2 - \frac{2}{3}\alpha(2c_R^2 - 1) - \frac{1}{3} + \frac{1}{3}h}w_*^2 + \paren{\frac{4}{3}c_R^2 + \frac{1}{3}}w_*^4 \Bigg]\frac{\avg{n_1T_1}}{n_0T_0}
    \\ 
    & - \frac{1}{N_*^2}\Bigg[ \paren{-2 - \frac{2}{3}h}(2c_R^2 - 1) + \frac{2}{3}(c_R^2 + 1)w_*^2 \Bigg] \frac{\avg{p_{e,1} n_1}}{n_0^2T_0} 
    \\
    & - \frac{1}{N_*^2}\Bigg[ -\paren{3 + h} + \paren{\frac{10}{3} + \frac{2}{3}h + \frac{2}{3}\alpha}w_*^2 - \frac{1}{3}w_*^4 \Bigg]\frac{\avg{p_{e,1}T_1}}{n_0T_0^2} 
    \\ 
    & - \frac{1}{N_*^2}\Bigg[ \paren{2 + \frac{2}{3}h} - \frac{1}{3}w_*^2 \Bigg] \frac{\avg{p_{e,1}^2}}{n_0^2 T_0^2}
    \end{split}
\end{equation}
Note that the prefactors on the non-dissipative terms of \eqref{eq_Phi} scale as $1/N_*^2 = (k\lambda_*^2)/w_*^2$, illustrating that the parameter controlling the size of the kinetic corrections is the perturbation wavenumber relative to the mean free path of ions near the Gamow peak.

\subsection{Asymptotic kinetic response}
\label{sec_kinetic_asymptotic}

Strictly speaking, the Gamow-peak approximation entails further simplification of $\Phi$ beyond the point reached in \eqref{eq_Phi}. The expansion of the reactivity integrand about $w_*$ is only valid in the asymptotic limit where $b \to \infty$ and drops corrections that are of higher order in $b^{-1/3}$. The leading-order asymptotic expression for $\Phi$ should then, for consistency, retain only the leading-order terms in $w_*$. 
In the following steps, we drop from \eqref{eq_Phi} all sub-leading-order terms within multiplying each fluctuating-quantity correlation term (each line of \eqref{eq_Phi}). To maintain generality for perturbations in which the fluctuations of some quantities may be zero, or small in some independent parameter, we keep the leading-order term multiplying each correlation. 

Assuming that the reaction of interest is non-resonant and that the S factor is slowly varying around the Gamow peak, the logarithmic temperature derivative of the reactivity, to leading order in $b^{-1/3}$, is $\alpha \sim \half b^{2/3}$. 
Furthermore, noting that $\nu \propto T^{-3/2}$, we take $N_T \sim -3/2$. 
Recalling, as noted in \S\ref{sec_kinetic_approx_gamow}, that $w_* = b^{1/3}/\sqrt{2}$, the asymptotic formula for $\Phi$ is
\begin{equation}
    \label{eq_Phi_asymptotic}
    \begin{split}
    \Phi \sim & 1 - k\lambda_* c_R\frac{b}{3\sqrt{2}}\frac{\avg{n_1(\bs x) T_1(\bs x - \frac{\bs L}{4})}}{n_0T_0} 
    \\ & -   \widehat \mu \frac{b^{2/3}}{6}\Bigg[(2 - 3\widehat \Gamma)\frac{\avg{n_1^2}}{n_0^2} + b^{2/3} \frac{\avg{n_1T_1}}{n_0T_0} - \frac{3b^{2/3}}{2}\frac{\avg{T_1^2}}{T_0^2} \Bigg]
    \\ 
    & + k^2\lambda_*^2 \Bigg [c_R^2\frac{3 b^{2/3}}{10}\frac{\avg{n_1^2}}{n_0^2}
     - \frac{b^{4/3}}{16} \frac{\avg{T_1^2}}{T_0^2}
     - \frac{b^{2/3}}{2}\frac{\avg{n_1T_1}}{n_0T_0}
    \\ 
    & - \paren{c_R^2+1}\frac{2}{3}\frac{\avg{p_{e,1} n_1}}{n_0^2T_0} 
    - \frac{b^{2/3}}{6}\frac{\avg{p_{e,1}T_1}}{n_0T_0^2}
     + \frac{1}{3}\frac{\avg{p_{e,1}^2}}{n_0^2 T_0^2} \Bigg] .
    \end{split}
\end{equation}
For concision, we write $\Phi$ instead of $\reacF K$ to report the result \eqref{eq_Phi_asymptotic}, but the kinetic response function is easily retrieved as ${K = (\Phi - 1)/E}$, where E is given by \eqref{eq_app_E_wave_pert}.

\begin{figure}
    \centering
    \vspace{0.5cm}
    \includegraphics[width=0.6\columnwidth]{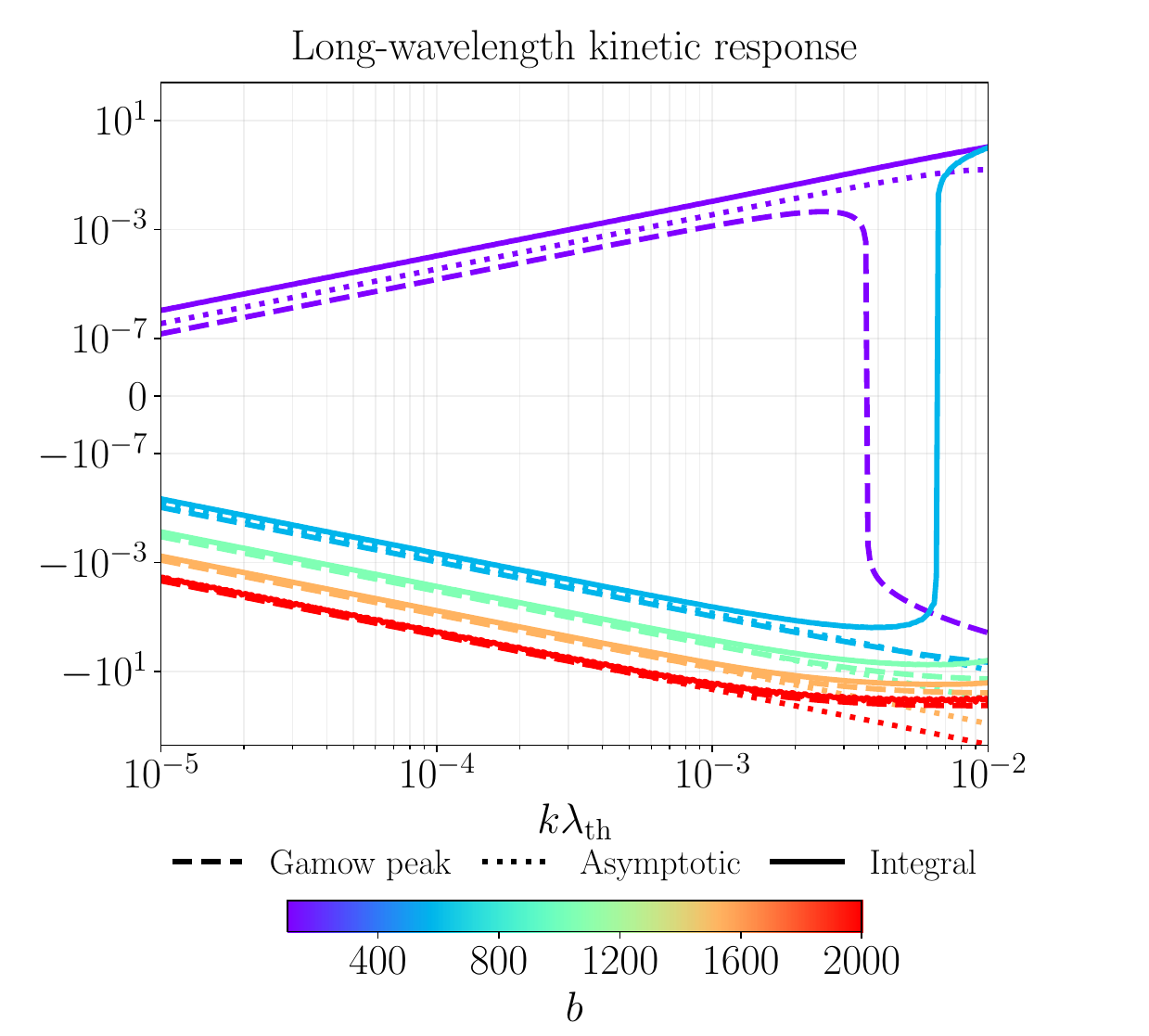}
    \caption{\justifying
    Kinetic fusion-power response function $\reacF K$ for acoustic modes as a function of wavenumber evaluated by direct integration (solid curve), by approximating around the Gamow peak (dashed curve), and by taking the asymptotic limit of $N_* \to \infty$ while applying the Gamow-peak approximation (dotted curve). Curves are shown at values of $b$ ranging from 20 to 800 and are differentiated by color.}
    \label{fig_bigN_comparison}
\end{figure}

\section{Response functions in limiting cases}
\label{sec_comparison}

At this point, we have derived the kinetic fusion-power response function in the low Mach-number limit \eqref{eq_Phi_full_integral} and then further simplified this result in the limit of large Gamow parameter and long wavelength to obtain the asymptotic formula \eqref{eq_Phi_asymptotic}. 
We return now to the basic question posed in \S\ref{sec_fusion}: \textit{If a compressive perturbation is driven in a fusion plasma, what happens to the fusion power?}. This section begins by extracting some of the physical content of \eqref{eq_Phi_asymptotic} through the study of limiting cases relevant to fusion devices and through a comparison between kinetic effects and hydrodynamic effects on fusion power in these cases. 
Generalizing this analysis, a combination of the formulas in \S\ref{sec_hydro}, \S\ref{sec_two_temp}, and \S\ref{sec_kinetic} yields the full fusion-power response to compressive fluctuations in terms of its hydrodynamic, two-temperature, and kinetic components. This result is compared to the fusion-power response to solenoidal perturbations described by the shear flow reactivity enhancement effect.

Despite its restricted regime of validity -- namely the limitations imposed by taking $b$ and $N_*$ to be asymptotically large -- the advantage of \eqref{eq_Phi_asymptotic} is that it permits a direct and informative comparison with the hydrodynamic response functions derived in \S\ref{sec_hydro}. 
Consider, for instance, an isobaric perturbation, whose hydrodynamic response is given by \eqref{eq_H_par_isotherm}. The energy of the perturbation is
\begin{equation}
    \label{eq_E_isobaric}
    E^\mr{isobar} = \frac{5(1+Z)}{4}\frac{\avg{n_1^2}}{n_0^2} ,
\end{equation}
where we are taking the perturbation to be in unmagnetized plasma or parallel to the magnetic field. Assuming that the electron and ion temperatures are equal everywhere, as is reasonable for the case of a slowly evolving isobaric configuration where interspecies equilibration has plenty of time to take effect, and that $c_R \ll 1$, the leading-order kinetic response is
\begin{equation}
    \label{eq_K_isobaric}
    \begin{split}
    K_\parallel^{\mr{(isobar)}} \sim - k^2\lambda_*^2\frac{b^{4/3}}{20(1+Z)}  + \widehat \mu \frac{b^{4/3}}{3(1+Z)} ,
    \end{split}
\end{equation}
where a subscript is added to record that, unlike in \eqref{eq_H_isobaric}, the kinetic treatment does not apply for wavenumbers perpendicular (or oblique) to a magnetic field. 
Both terms in \eqref{eq_K_isobaric} have simple physical explanations. The first term, scaling as $\mr{Gk}^2$, is negative and records the loss of fast ions from the hot regions to their cooler surroundings. Despite the higher density in the cold regions, fast ions streaming through them are less likely to undergo a fusion reaction because, in the Gamow-peak approximation, most fusion reactions are the result of the collision of two ions with equal and opposite velocities. This is the mechanism behind the ``Knudsen layer reactivity reduction effect'' \citep{Molvig_Hoffman_Albright_Nelson_Webster_2012,Albright_Molvig_Huang_Simakov_Dodd_Hoffman_Kagan_Schmit_2013,Yin_Albright_Vold_Nystrom_Bird_Bowers_2019}, which reduces the yield of ICF implosions whose hot spots are are small enough to allow an appreciable fraction of fast ions to escape. Fokker-Planck modeling of the fast-ion distribution near interfaces under conditions characteristic of hot spots has shown qualitatively similar behavior, with the reactivity reduction increasing with Gamow-Knudsen number and decreasing with temperature \citep{McDevitt_Tang_Guo_Berk_2014,Tang_Berk_Guo_McDevitt_2014,Tang_McDevitt_Guo_Berk_2014}; while the geometry in these works was different than that considered here, consisting of a sharp planar interface rather than a small-amplitude sinusoidal profile, the alignment in the qualitative results provides further evidence that the simplified kinetic model adopted here is sufficient to capture the essential physics of systems of interest.

The second term in \eqref{eq_K_isobaric}, scaling as $\widehat \mu b^{4/3}$, is positive and records the fact that, as hot regions of the perturbation cool by thermal conduction, the less-collisional tail of the ion distribution in these regions takes a finite time to ``catch up'' to the cooling bulk. There is therefore a transient excess of fast ions in the hot regions, and a dearth of fast ions in the cold regions, during the decay of the perturbation. Because the hot regions have higher reactivity, the net effect of this transient is, at least to leading order in $b$, to increase the fusion power -- more precisely, since in a decaying isobaric perturbation $\reacF H E$ is a decreasing function of time, the effect of this transient behavior is to moderate this decrease in fusion power through the appearance of a positive component in $\reacF K$. The effect is accentuated by the fact that the hot regions have lower collisionality, by virtue both of their higher temperature and of their lower density, allowing the fast ions in these regions to remain out of equilibrium for a longer time; this behavior, which can be seen in terms containing $N_T$ in \eqref{eq_Phi}, is of lower order in $b$ and so does not appear in \eqref{eq_K_isobaric}.

Considering now an adiabatic acoustic wave and using the energy formula in \eqref{eq_E_acoustic_parallel}, we have
\begin{equation}
    \label{eq_K_acoustic_adiabatic}
    \begin{split}
    K_\parallel^{\mr{(adiabat)}} \sim - k^2\lambda_*^2\frac{b^{4/3}}{60(1+Z)} - \widehat \mu \frac{2b^{2/3}}{3(1+Z)}  ,
    \end{split}
\end{equation}  
where \eqref{eq_app_Gamma_adiabatic} has been used to substitute $\widehat \Gamma = -22/9$. 
Deriving \eqref{eq_K_acoustic_adiabatic} actually requires returning to \eqref{eq_Phi} because the leading-order $\mc O(b^{4/3})$ terms in \eqref{eq_Phi_asymptotic} vanish for an adiabatic acoustic wave, for which ${T_1/T_0 = (2/3)n_1/n_0}$. The dissipative term is in \eqref{eq_K_acoustic_adiabatic} is not at odds with the assumption that the wave is adiabatic -- because the dissipative term is proportional to $\widehat \mu$, it can be large even when $\mu$ is too small relative to $\omega$ to shift significantly the relation between first-order fluctuating fluid quantities. While the dissipative term is usually subdominant relative to the $k^2\lambda_*^2$ term, which scales more strongly with $b$, the former can be relevant in waves with small $\mr{Gk}$ that nevertheless dissipate quickly due, for example, to electron thermal conduction. 

In the asymptotic limit considered in \eqref{eq_K_acoustic_adiabatic}, the kinetic response to an adiabatic acoustic wave is to reduce the fusion power. This can be understood as follows: when $b$ is large, the dominant factor in determining the fusion power is the generation of high-temperature regions, rather than the relatively inefficient clumping of ions. Whenever there is a high-temperature wave peak, some of the fast ions in that region will stream out of it into the cooler troughs. While the fast ions lost from the wave peaks may still produce fusion reactions elsewhere in the wave, their probability of doing so decreases with increasing $b$. Hence, when $b$ is large, the heavy cost (in terms of fusion power) of losing fast ions from the wave peaks dominates the kinetic response, leading $K$ to be negative.

Finally, while the isothermal limit is often not realizable in practice under the conditions of interest in this work (cf. \S\ref{sec_app_acoustic}), it is instructive to consider the kinetic response to isothermal acoustic waves, \textit{viz.}
\begin{equation}
    \label{eq_K_acoustic_isotherm}
    \begin{split}
    K_\parallel^{\mr{(isotherm)}} \sim k^2\lambda_*^2\frac{3b^{2/3}}{10} - \widehat \mu \frac{2b^{2/3}}{3(1+Z)} .
    \end{split}
\end{equation}
In contrast to the adiabatic case, the kinetic response of a sufficiently slowly decaying isothermal wave is to increase the fusion power. In the absence of temperature fluctuations that penalize gradients because fast ions are lost from the hot regions, the leading-order kinetic behavior instead comes from the fluctuating velocities. Note that, while not visible in \eqref{eq_K_acoustic_isotherm} because the isothermal phase velocity $c_R = 1$ has already been substituted, the leading-order (non-dissipative) contribution to $K_\parallel^\mr{(isotherm)}$ comes from a term in \eqref{eq_Phi_asymptotic} proportional to ${c_R^2\avg{n_1^2}}$. The mechanism behind this positive contribution is very similar to that of the SFRE -- when fast ions sampled from one region of a nonuniform velocity field travel to another region, they tend, on average, to broaden the tail of the distribution and increase the reactivity \citep{Fetsch_Fisch_2025a,Fetsch_Fisch_2026_dpp}.

It is interesting to note, per \eqref{eq_H_par_adiabat} and \eqref{eq_H_par_isotherm}, that $\reacF H_\parallel^{\mr{(adiabat)}} > \reacF H_\parallel^{\mr{(isotherm)}}$ when $\alpha$ is large. In other words, when ions are well confined by collisions, the benefit of producing high-temperature regions exceeds the benefit of clumping reactants, but the ability of fast ions to travel between regions of the wave degrades this benefit. 
%These scalings, however, are derived in series of restrictive limits ($b \gg 1$ and $\mr{Gk} \ll 1$). For greater generality, we now compute the full fusion-power response without relying on these limits.

%\subsection{Comparison of asymptotic response functions}

To compare the hydrodynamic and kinetic components of the fusion-power response in greater generality, Table~\ref{tab_coeff_comparison} lists the leading-order coefficients on each of the fluctuating quantities in the formulas for $\reacF H$ and $\reacF K$, taken respectively from \eqref{eq_Pf_pert_alpha} and \eqref{eq_Phi_asymptotic}. The Gamow-peak approximation is adopted, meaning that $\alpha = \half b^{2/3}$. Terms proportional to $c_R^2 \avg{n_1^2}$ are written instead in terms of $\avg{u_1^2}$ to emphasize the importance of the velocity fluctuations. The only positive coefficients in $\reacF K$ are on the $\avg{u_1^2}$ and $\avg{p_{e,1}^2}$ terms, the latter arising from ions accelerated in the electric field present in the slopes of the wave.

\begin{table}
    \centering
    \renewcommand{\arraystretch}{2.2}
    \normalsize
    \begin{tabular}{r||c|l|c}
        \hline\hline
        Correlation& Hydrodynamic $H$ & Compressive kinetic $K$  & Shear-flow kinetic $G$ \\
        \hline
        $\frac{\avg{n_1^2}}{n_0^2 E}$ & $1$ & $ - \widehat{\mu} \frac{1}{6}(2 - 3\widehat{\Gamma})b^{2/3}$ & 0 \\
        $\frac{\avg{u_1^2}}{\vth^2 E}$ & $0$ & $ + k^2\lambda_*^2 \frac{3}{10} b^{2/3}$  & $k^2 \lambda_*^2 \frac{1}{10} b^{2/3}$ \\
        $\frac{\avg{T_1^2}}{T_0^2 E}$ & $\frac{1}{8}b^{4/3}$ & $ - k^2\lambda_*^2 \frac{1}{16}b^{4/3} + \widehat{\mu} \frac{1}{4}b^{4/3}$ & 0 \\
        $\frac{\avg{T_2}}{T_0 E}$ & $\frac{1}{2}b^{2/3}$ & 0 & 0 \\
        $\frac{\avg{n_1 T_1}}{n_0 T_0 E}$ & $b^{2/3}$ & $- k^2\lambda_*^2 \frac{1}{2} b^{2/3} - \widehat{\mu} \frac{1}{6}b^{4/3} $ & 0 \\
        $\frac{\avg{n_1(\bs x) T_1(\bs x - \bs L/4)}}{n_0 T_0 E}$ & $0$ & $-k\lambda_* c_R\frac{1}{3\sqrt{2}}b$ & 0 \\
        $\frac{\avg{p_{e,1} n_1}}{n_0^2 T_0 E}$ & $0$ & $-k^2\lambda_*^2 (c_R^2+1)\frac{2}{3}$ & 0 \\
        $\frac{\avg{p_{e,1} T_1}}{n_0 T_0^2 E}$ & $0$ & $-k^2\lambda_*^2 \frac{1}{6}b^{2/3}$ & 0 \\
        $\frac{\avg{p_{e,1}^2}}{n_0^2 T_0^2 E}$ & $0$ & $+k^2\lambda_*^2 \frac{1}{3}$ & 0 \\
        \hline\hline
    \end{tabular}
    \caption{Comparison of hydrodynamic and kinetic response coefficients for arbitrary longitudinal perturbations in the long-wavelength and large Gamow-parameter limit. Also shown is the response coefficient for the shear flow reactivity enhancement effect.}
    \label{tab_coeff_comparison}
\end{table}

%\subsection{Fast-wave approximation}

%\subsubsection{Pseudosound kinetic response}
%\label{sec_kinetic_pseudosound}

\section{Results and discussion}
\label{sec_discussion}

Moving on from the derivations of the components of the fusion-power response, this section takes a broader view, discussing the relationship among these components and comparing the enhancement to fusion power by compressive and solenoidal fluctuations. We begin, however, by rehearsing several assumptions made in formulating the model adopted here.

\subsection{Limitations of the model}

The model adopted in this work comes with several limitations. The BGK collision operator described in \eqref{eq_krook} neglects the distinction between slowing down, energy-space diffusion, and pitch-angle scattering, instead approximating collisions as a process that reduces the deviation from equilibrium at each point in phase space at a specified rate. Moreover, the fact that the collision frequency is allowed to vary with velocity means that conservation of density, momentum, and energy are not strictly satisfied. In recent work, \cite{Guo_Wu_Zhang_2026} showed that using a Fokker-Planck collision operator with pitch-angle scattering reduces the magnitude of the SFRE relative to that computed with a BGK operator, and an analogous result may be expected to apply to the kinetic fusion-power response to compressive fluctuations. 
Despite these limitations, the advantage of the BGK operator is that it permits relatively straightforward analytical solutions to the kinetic equations. These solutions, developed throughout \S\ref{sec_kinetic}, allow the physical origins of the kinetic fusion-power response to be studied in greater detail than would be feasible in a numerical treatment. Thus, while the use of a more realistic collision operator could improve the quantitative accuracy of the results obtained here, the treatment in this work resolves qualitative features that are likely to persist even as the assumptions made here are relaxed.

This work considered plasmas composed of a single ion species in a single charge state. The reasons for this simplification are twofold: first, considering only single-species fusion reactions bypasses the complexity arising when the distribution functions of different ion species are perturbed in different ways because of the differences in their thermal velocities and mean free paths; cf. \cite{Fetsch_Fisch_2026_dpp}. Second, the dispersion relations for waves in plasmas containing multiple ion species become substantially more complicated \citep{Berger_Valeo_2005}. Many of the results in this work are expressed in terms of correlations between fluctuating quantities and can therefore be applied to fluctuations satisfying arbitrary dispersion relations, but the assumption of a single species allows substantial simplification when specializing to the case of acoustic waves.

The kinetic treatment in \S\ref{sec_kinetic} relied on ordering the Debye length $\lambda_D$ out of the governing equations ($\lambda_D \ll \lth$); combined with the hydrodynamic limit ($k\lth \ll 1$), this removed plasma oscillations from the picture and allowed the electrostatic potential energy to be neglected. While this approach is sensible in weakly coupled plasmas, it may fail in the moderately coupled regime, where the scale separation between $\lambda_D$ and $\lth$ ceases to be very large. Furthermore, in moderately coupled plasmas, the assumption that fusion reactivity is independent of density breaks down as screening corrections become significant \citep{Salpeter_1954,Bahcall_Chen_Kamionkowski_1998}.

In computing the kinetic response to an acoustic wave, some ions will always be resonant with the wave. If the collision frequency around the Gamow peak is too small, the resonant population may become large enough to invalidate the second-order theory advanced in \S\ref{sec_kinetic}, requiring instead a quasilinear treatment. In the long-wavelength limit, where even particles at the Gamow peak are highly collisional, this wave-particle resonance is avoided. At shorter wavelengths, requiring self-consistency of the kinetic treatment in \S\ref{sec_kinetic} could impose stronger constraints on the wave amplitude than the $\mc M \ll 1$ condition used to expand the kinetic equation in the Mach number.

The assumptions of this theory can furthermore be violated by large-amplitude waves if the wave dynamics lead to nonlinear features not described by the dispersion relations in Appendix~\ref{sec_app_acoustic}. Consider, for instance, the momentum equation \eqref{eq_app_fluid_u} to second order in the wave amplitude, and suppose for the sake of simplicity that the terms on the right-hand side are zero at second order. Then \eqref{eq_app_fluid_u} becomes simply 
\begin{equation}
    \label{eq_nonlinear_steepening}
    \partial_t u_2 + \frac{n_1}{n_0} \partial_t u_1 + u_1 \cdot \nabla u_1 = 0.
\end{equation}
If the perturbation consists initially of a pair of modes with wavenumbers $\bs k$ and $- \bs k$, then \eqref{eq_nonlinear_steepening} describes the generation of additional modes with wavenumbers $2\bs k$ and $-2\bs k$, which happens more quickly in flows with larger magnitudes and shorter wavelengths.  This nonlinear steepening eventually generates shocks \citep{Hu_1972,Gupta_Scalo_2018}, which produce kinetic modifications to the fusion reactivity distinct from those considered here \citep{Rinderknecht_et_2017}.

Finally, the initial condition $f(t=0) = \fm$ suppresses the kinetic contribution to the fusion power because the reactivity does not change until ions have had time to travel across the perturbation; in the regime of short-wavelength, rapidly dissipating flows, the fluid background will have relaxed substantially toward equilibrium before this happens. The suitability of alternative initial conditions depends on the particulars of the system under consideration. For instance, the initial conditions imposed on the fluid quantities could be obtained by compressing a system initially containing long-wavelength perturbations and initially producing very little fusion power on account of its lower density and temperature. If this compression were done quickly, the distribution function would not be locally Maxwellian everywhere, and in fact can exhibit significantly enhanced tails \citep{Krook_Wu_1976}, but consideration of these alternative cases is beyond the scope of this work.

\subsection{Total response function}
\label{eq_sec_discussion_total}

\begin{figure}
    \centering
    \vspace{0.5cm}
    \includegraphics[width=0.6\columnwidth]{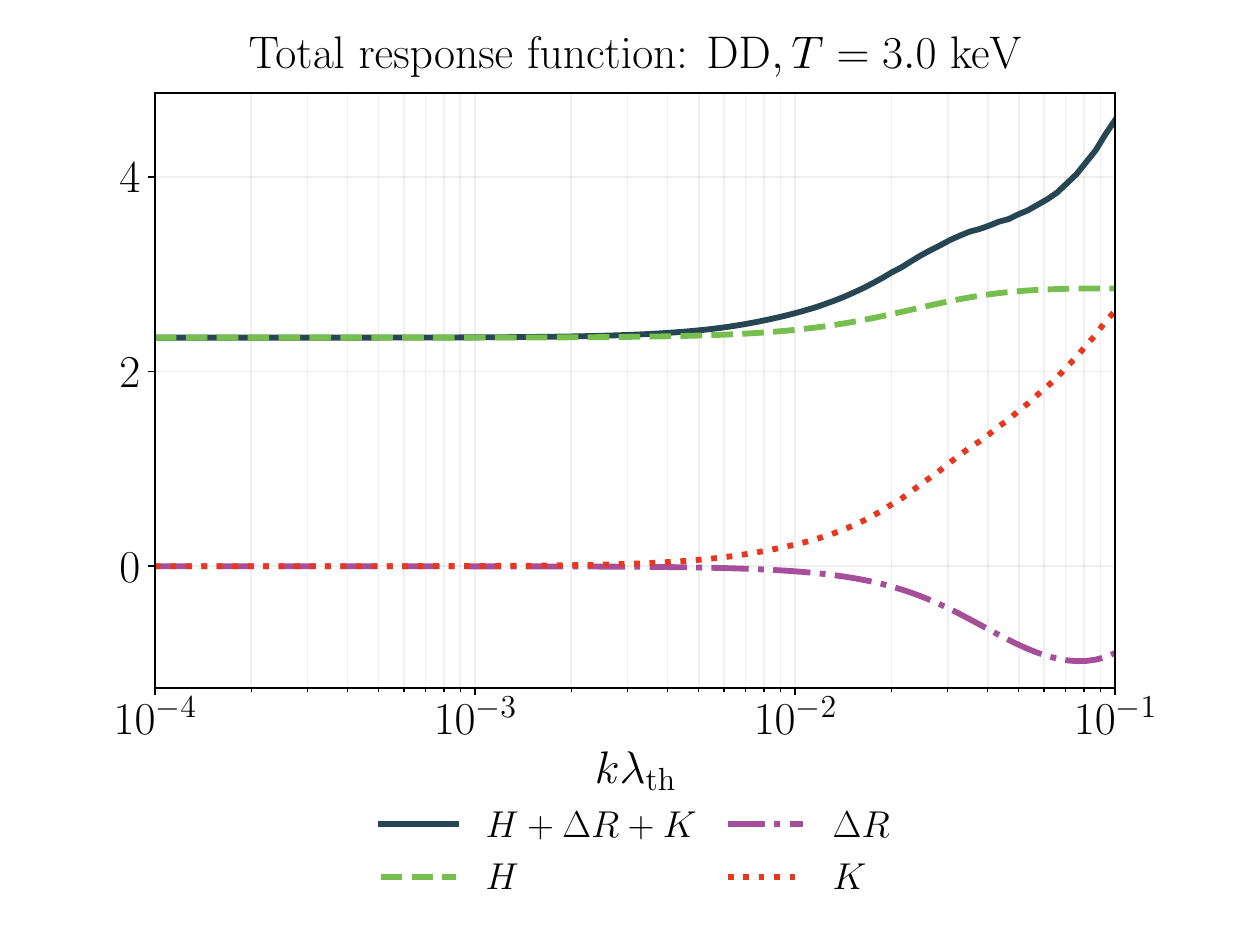}
    \caption{\justifying
    Hydrodynamic and kinetic fusion-power response functions for acoustic waves as a function of wavenumber in a deuterium plasma at $T=3~\mr{keV}$. The kinetic component of the response is computed by numerical integration of \eqref{eq_Phi_full_integral} evaluated at $t = 1/2\mu$.}
    \label{fig_total_response}
\end{figure} 

\begin{figure}
    \centering
    \vspace{0.5cm}
    \includegraphics[width=0.6\columnwidth]{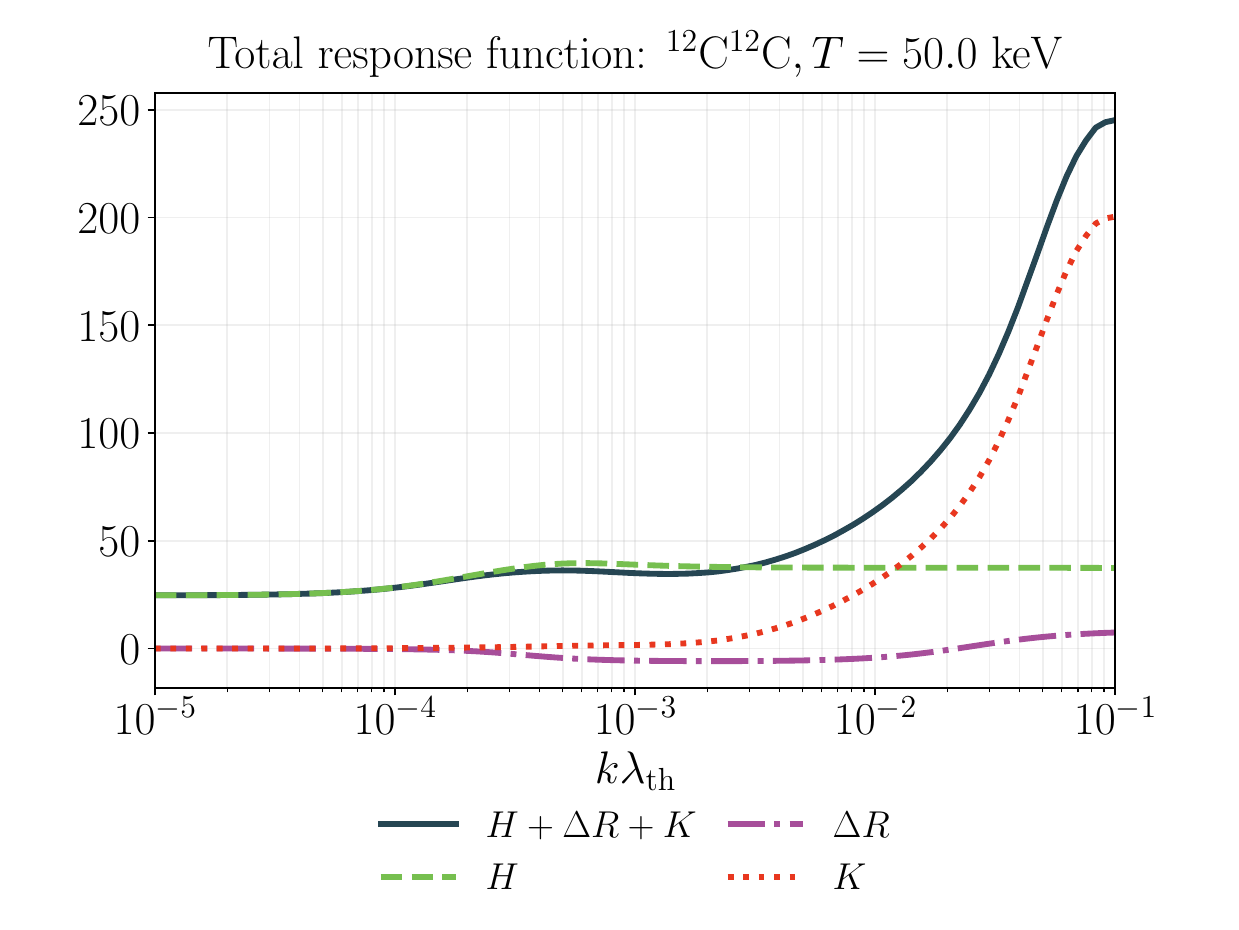}
    \caption{\justifying
    Hydrodynamic and kinetic fusion-power response functions for acoustic waves as a function of wavenumber in a \textsuperscript{12}C plasma at $T=50~\mr{keV}$. The kinetic component of the response is computed by numerical integration of \eqref{eq_Phi_full_integral} evaluated at $t = 1/2\mu$.}
    \label{fig_total_response_C12_T50}
\end{figure}

Recall from \eqref{eq_Pf_pert_setup} that the volume-averaged fusion power can be written in terms of the hydrodynamic response function $\reacF H$, the two-temperature response function $\reacF R$, the kinetic response function $\reacF K$, and the shear-flow response function $\reacF G$. For a purely longitudinal perturbation, $G=0$ and the fusion power is
\begin{equation}
    \label{eq_results_Pf_pert_longitudinal}
    \avg{P_f} \sim P_{f,0}\bigg[1 + \Big(\reacF H + \Delta R + \reacF K \Big)E \bigg] .
\end{equation}
In Fig.~\ref{fig_total_response}, the total response function $\reacF H + \Delta R + \reacF K$ for acoustic waves is plotted as a function of wavenumber in a deuterium plasma at $T=3~\mr{keV}$. As anticipated in \S\ref{sec_fusion}, at long wavelengths the response is dominated by the hydrodynamic component $\reacF H$, with kinetic and two-temperature components providing negligible contributions. While not apparent on the linear scale of Fig.~\ref{fig_total_response}, $\reacF K$ and $\reacF R$ are negative at $k\lth \ll 1$, but the magnitude of these negative contributions is several orders of magnitude smaller than the hydrodynamic component. This is the regime of highly collisional, adiabatic acoustic waves. It is interesting to note that, even in this relatively straightforward regime, the effect of acoustic waves on fusion power can be significant. Suppose that long-wavelength acoustic waves were driven in a deuterium plasma with energy equal to 10\% of the thermal energy. Because the thermal energy per ion is $3T$, the normalized energy of the perturbations is $E = 0.3$. According to the small-$k$ tail of the response function in Fig.~\ref{fig_total_response}, the perturbations increase the fusion power by about 70\% (meaning ${\avg{P_f} / P_{f,0} \approx 1.7}$).

At shorter wavelengths, kinetic effects become dominant. As the wavenumber passes ${k\lth \approx 0.01}$, the hydrodynamic response increases slightly, as noted in \S\ref{sec_hydro}, because in this regime the electrons begin to be able to transmit heat across the wave within a wave period. As the fluctuations of the electron fluid approach the isothermal limit while the ions remain adiabatic, a larger fraction of the wave energy resides in the ions, leading to a larger $\reacF H$. Concomitantly, the electron thermal conduction begins rapidly dissipating the wave energy. Because this dissipation preferentially heats electrons, $\reacF R$ decreases, contributing a negative component to the total response function. Far more significant than either of these effects, however, is the sharp rise in the kinetic response. Notably, the fact that $\reacF K > 0$ in this regime, where the ions remain close to adiabatic, is at odds with the asymptotic picture described in \S\ref{sec_comparison}, illustrating the breakdown of the long-wavelength limit.

As an example of the fusion-power response at larger values of the Gamow parameter, Fig.~\ref{fig_total_response_C12_T50} shows the total response function for \textsuperscript{12}C\textsuperscript{12}C fusion at $T = 50~\mr{keV}$, a regime of interest in some astrophysical systems \citep{Jeet_et_2023}. For simplicity, a constant $S$-factor is assumed. Interestingly, the hydrodynamic response exhibits non-monotonic behavior in Fig.~\ref{fig_total_response_C12_T50}; the dip in $\reacF H$ around $k\lth \approx 10^{-3}$ appears in the regime where the electrons transition from adiabatic to isothermal fluctuations. In this regime, $T_{e,1}$ shifts out of phase from $T_1$ and $n_1$. Local thermal equilibration between the phase-shifted ion and electron temperatures pulls the ion temperature downward in the wave peaks, reducing $\reacF H$. 
All components of the reactivity response are larger in magnitude than in the 3-keV deuterium plasma, but the most striking feature in Fig.~\ref{fig_total_response_C12_T50} is the climb of $\reacF K$ to values in the hundreds at high $k$. This is a consequence of the large Gamow parameter ($b \approx 550$), which leaves the reactivity sensitive to perturbations far out on the tail, where the collision frequency is small. 
The same scaling produces larger values of the SFRE for high-$Z$ reactants, and for low temperatures, where $b$ is large \citep{Fetsch_Fisch_2026_dpp}.

\begin{figure}
    \centering
    \vspace{0.5cm}
    \includegraphics[width=0.6\columnwidth]{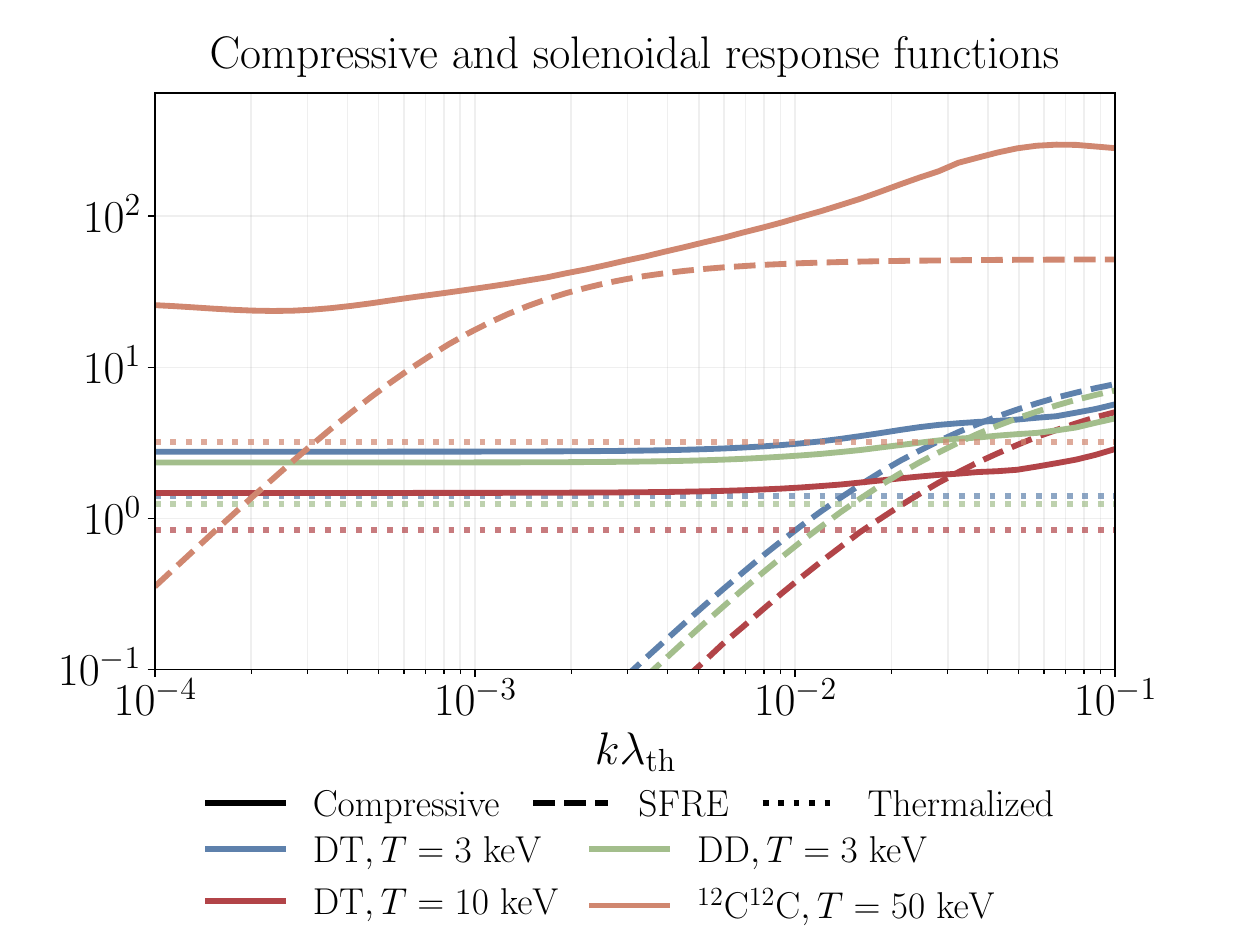}
    \caption{\justifying
    Total response function $H + \Delta R + K$ for acoustic waves as a function of wavenumber, compared to SFRE response function for sinusoidal sheared flows of the same wavenumber, for a range of reactions and temperatures. $K$ is computed by numerical integration of \eqref{eq_Phi_full_integral} evaluated at $t = 1/2\mu$, and $G$ is computed by numerical integration of the reactivity using the distribution functions derived in \cite{Fetsch_Fisch_2026_dpp}. Also shown is the increase in fusion power if the perturbation energy were thermalized.}
    \label{fig_sfre_comp}
\end{figure}

\begin{figure}
    \centering
    \vspace{0.5cm}
    \includegraphics[width=0.6\columnwidth]{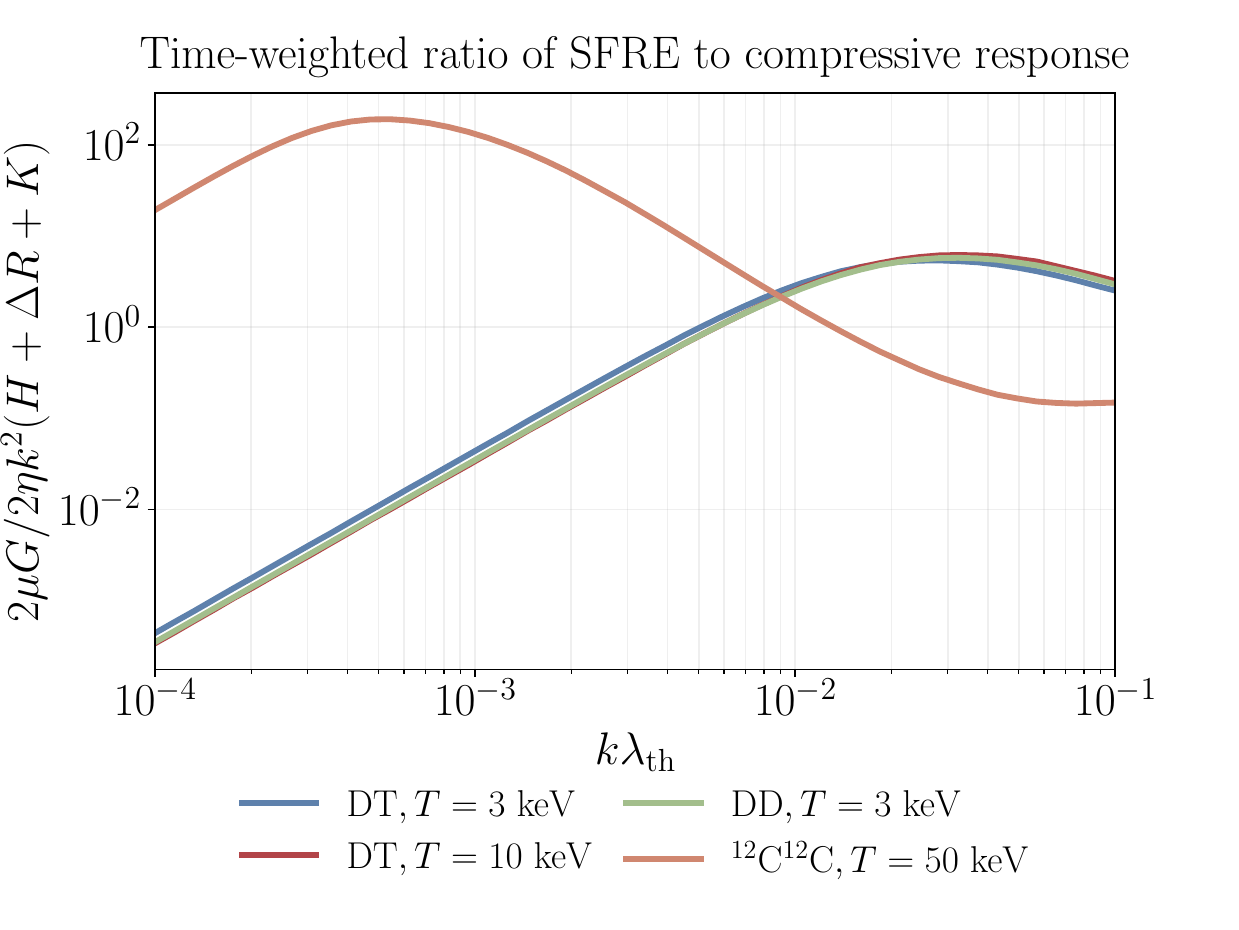}
    \caption{\justifying
    Ratio of the fusion-power response functions for acoustic waves and sinusoidal sheared flows of the same wavenumber, displayed for a range of reactions and temperatures. The ratio is normalized to the characteristic lifetime of each perturbation.}
    \label{fig_sfre_ratio}
\end{figure}

\subsection{Compressible and solenoidal turbulence}

It has been shown by \cite{Fetsch_Fisch_2025a,Fetsch_Fisch_2025b} that the kinetic effect of sheared flows on fine spatial scales can be to produce a large enhancement to the fusion reactivity. In fact, with the proviso that simplifications, such as the BGK operator, made in formulating kinetic models of the SFRE may lead to overestimating the magnitude of the effect \citep{Guo_Wu_Zhang_2026}, it was shown by \cite{Fetsch_Fisch_2026} that replacing a portion of the thermal energy in an ICF hot spot with solenoidal turbulence on optimized scales may permit ignition at lower temperatures and smaller hot-spot areal densities. 

These results for solenoidal flows naturally prompt the question of whether compressive fluctuations can provide similar benefits. The response functions derived in this work allow this question to be answered, at least in an approximate way. In Fig.~\ref{fig_sfre_comp}, the total response function $(H + \Delta R + K)$ for acoustic waves is compared to the SFRE response function $(G)$ for sinusoidal sheared flows of the same wavenumber. Response functions are shown for DD and DT reactions at $T=3~\mr{keV}$, of interest in sub-ignition scale experiments and in hot spots that have yet to ignite; for DT reactions at $T=10~\mr{keV}$, of interest in igniting hot spots; and for \textsuperscript{12}C-\textsuperscript{12}C reactions at $T=50~\mr{keV}$, of interest in some astrophysical systems. In addition to the compressive and solenoidal responses, Fig.~\ref{fig_sfre_comp} shows the fractional increase in fusion power if the energy of the perturbation were instead used to heat the plasma, keeping electrons and ions at the same temperature.

While isolated acoustic wave and planar shear flows represent idealized perturbations unlikely to be realized in most fusion devices, it is worth noting, per \eqref{eq_Pf_pert_setup}, that $H$, $\Delta R$, $K$, and $G$ are linear response functions that can be summed to determine the response to arbitrary small-amplitude perturbations (subject to some additional assumption when multiple modes have the same wavenumber -- see Appendix~\ref{sec_app_averaging}). Thus, if a turbulent plasma is decomposed into its solenoidal and dilatational components \citep{Kovasznay_1960}, the response functions shown in Fig.~\ref{fig_sfre_comp} quantify the modification to fusion power in that plasma, relative to a quiescent plasma with the same temperature and density, as a function of the scale of the turbulent fluctuations and of the partition of its turbulent energy. 
This energy partition is not a function of compressibility or Reynolds number alone, depending also on the nature of the turbulent driving \citep{Donzis_John_2020,Donzis_Sreenivasan_Yeung_2012}. 
Techniques have recently been proposed to diagnose the distribution of eddy sizes in solenoidal turbulence in ICF systems, opening the door to future experiments measuring the effect of turbulence -- at varying degrees of compressibility -- in fusion plasmas \citep{Qu_Fisch_2026}. 
Note that this analysis neglects the pseudosound component generated by solenoidal flows with finite Mach number \citep{Ristorcelli_1997,Wang_Gotoh_Watanabe_2017}, but the general expressions for $H$, $\Delta R$, $K$, derived in this work can readily be applied to pseudosound.

The most important qualitative distinction visible in Fig.~\ref{fig_sfre_comp} is that, while $G$ vanishes at small $k$ (scaling as $k^2$), $H$ remains nonzero at all $k$. Thus, excepting purely sinusoidal modes without density or temperature variations, the hydrodynamic component dominates the fusion-power response to long-wavelength perturbations. As seen in Fig.~\ref{fig_total_response} and Fig.~\ref{fig_total_response_C12_T50}, the response function for acoustic waves increases markedly on short scales, driven primarily by its kinetic component. In the same regime, $G$ increases for a similar reason, with the result that the DD and DT reactions shown in Fig.~\ref{fig_sfre_comp} exhibit similar response functions for acoustic waves and sheared flows at ${k\lth \approx 0.05}$. 
At all wavenumbers, the effect of compressive fluctuations is to increase fusion power by a greater amount than would the same amount of energy if it were instead thermalized. The increase in fusion power by solenoidal fluctuations becomes more efficient than heating at sufficiently high $k$, as anticipated in \cite{Fetsch_Fisch_2026}.

Any design for an ICF experiment seeking to exploit the SFRE must contend with the fact that the short length scales on which sheared flows can produce a large reactivity enhancement are also the scales on which viscosity rapidly dissipates those flows \cite{Fetsch_Fisch_2026_dpp,Fetsch_Fisch_2026}. While this dissipation has certain benefits in that it rapidly heats ions, or in the language of this work, produces $\Delta R > 0$ \citep{Davidovits_Fisch_2016a,Davidovits_Fisch_2016b,Davidovits_Fisch_2019}, it leads to a further question: does the fusion-power amplification by compressive fluctuations persist for a longer or shorter time than that produced by sheared flows? To address this question, Fig.~\ref{fig_sfre_ratio} shows the ratio of the total response function for acoustic waves to the SFRE response function for sheared flows, normalized to the characteristic lifetime of each perturbation. The characteristic lifetime of the acoustic wave is taken to be ${1/2\mu}$, and the characteristic lifetime of the sheared flow is taken to be ${1/2k^2 \eta}$. In effect, each component of the response is thus weighted by the time over which the perturbation persists and acts on the plasma. It is assumed here that the time-independent response functions are an appropriate choice, i.e. that the transients described in \S\ref{sec_fusion} are subdominant. For reactions involving deuterium and tritium under the conditions considered, the ratio of the time-weighted responses to sheared flows and to acoustic waves vanishes at small $k$. Interestingly, however, the ratio becomes greater than unity at $k\lth \approx 5\times 10^{-3}$, a longer length scale than that at which the response functions themselves become comparable (cf. Fig.~\ref{fig_sfre_comp}). This can be explained by the fact that acoustic waves can decay through electron thermal conductivity, whereas sheared flows can dissipate only through viscosity, which is dominated by ions. The smaller mobility of ions means that the decay rate of sheared flows is smaller, and so their time-weighted effect on fusion power is increased. In absolute terms, however, the largest time-weighted effect can be expected to come from long-wavelength acoustic waves, whose lifetime increases with decreasing $k$ and whose response function approaches a constant at small $k$. 
%This analysis neglects, however, the deleterious effects on confinement -- whether in ICF or in MCF devices -- of the perturbations themselves, which may outweigh the increases in fusion power that they produce \cite{Conway_Smolyakov_Ido_2021}.

\section{Conclusion}

In view of recent results indicating an enhancement to fusion reactivity due to the kinetic transport of particles in strongly sheared flows -- the ``SFRE'' described in \cite{Fetsch_Fisch_2025a} -- this work identified the analogous effect in compressive fluctuations. The effect of these fluctuations on fusion power was described in terms of general response functions, which consist of hydrodynamic \eqref{eq_Pf_pert_alpha}, ``two-temperature'' \eqref{eq_Delta_R_t_asymptotic}, and kinetic \eqref{eq_Phi_asymptotic} components. For each of these components, the transient fusion-power response following the sudden driving of a small-amplitude perturbation in a fusion plasma was also quantified. 
Acoustic waves amplify fusion power at long wavelengths through hydrodynamic effects that do not require ions to travel far enough to generate a significant non-Maxwellian population, and the increase in fusion power per unit energy is larger than if the same energy were used for heating the plasma. At shorter wavelengths, kinetic effects analogous to the SFRE provide a further increase in reactivity. 

By taking advantage of the SFRE and driving solenoidal turbulence on carefully chosen scales, some ICF hot spots may be made to ignite at smaller temperatures and areal densities than would otherwise be possible \citep{Fetsch_Fisch_2026}. While the generalization of the ``turbulent ignition criterion'' of \cite{Fetsch_Fisch_2026} to ICF hot spots containing compressive fluctuations is beyond the scope of this work, the large values of the fusion-power response function seen in Fig.~\ref{fig_total_response} suggest that similar benefits may be obtained from compressible turbulence. The response functions derived in this work apply equally to unmagnetized plasmas and to magnetized plasmas where the perturbations are aligned with the magnetic field. Provided that the requirement of the theory that the background plasma is in the hydrodynamic regime is satisfied, the same advantages could, in principle, apply to magnetic-confinement devices, although the deleterious effects of the perturbations on confinement may be so large as to negate these advantages \citep{Conway_Smolyakov_Ido_2021}. 
The ability to produce a large increase in fusion power by driving compressive fluctuations my furthermore be limited by their tendency to steepen through nonlinear effects and produce shocks.  
The SFRE is a purely kinetic effect and therefore vanishes at small $k$; compressive fluctuations, by contrast, produce a fusion-power response with a hydrodynamic component that remains nonzero at arbitrarily small $k$, illustrating a fundamental distinction between the effects of the solenoidal and dilatational components of turbulent flows in fusion plasmas. This scaling suggests that an optimal partition of energy driven in turbulent motion for the purpose of increasing fusion power may consist of compressive fluctuations at large scales and solenoidal fluctuations at small scales. Recently developed techniques for characterizing the effects of material structure on turbulence in ICF devices \citep{Li_Davidovits_2024,Zhang_Davidovits_Fisch_2025}, and on diagnosing the distribution of turbulent energy in dense plasmas \citep{Qu_Fisch_2026,Rocco_et_2022,Rososhek_et_2025}, may facilitate the design of experiments exploring the effects of deliberately driven perturbations in ICF plasmas.

\section*{Acknowledgments}
This work was supported by the Center for Magnetic Acceleration, Compression, and Heating (MACH), part of the U.S. DOE-NNSA Stewardship Science Academic Alliances Program under Cooperative Agreement DE-NA0004148.

\appendix

\section{Averaging procedure}
\label{sec_app_averaging}

When computing the fusion power produced in a plasma containing rapid fluctuations, it is useful, rather than trying to track the fusion power at each moment in time as it responds to the varying plasma conditions, instead to average over the oscillatory dynamics and quote a single spatially and temporally averaged fusion power. In this Appendix, we describe the averaging procedure used in this work to obtain hydrodynamic and kinetic fusion-power response functions.

As described in \eqref{eq_Pf_avg}, the simplest averaging scheme is to integrate the fluctuating quantity over spatial and temporal domains containing an integer number of fluctuation periods and wavelengths and then to divide by the corresponding volume and time interval. This scheme works for exactly periodic fluctuations, and for fluctuations consisting of fast oscillations and slow decay with a wide scale separation between the fast and the slow dynamics. 
%Problems appear, however, when the decay rate of the quantity being averaged ceases to be small relative to its period. 
Most calculations in this work involve averaging functions of the form
\begin{equation}
    \label{eq_app_y_form}
    y(t,x) = \wt y e^{ikx - i\omega t - \mu t} + \wt y^* e^{-ikx + i\omega t + \mu t} .
\end{equation}

Because the quantities under consideration are exactly periodic in space (the background is taken to be homogeneous), the spatial averaging can be performed by a straightforward volume integral as in \eqref{eq_Pf_avg}. When ${\mu \ll \omega}$, the logic of \eqref{eq_Pf_avg} can readily be extended to write
\begin{equation}
    \label{eq_app_avg_window}
    \avg{y}(t) = \frac{1}{L} \int_{-\infty}^{\infty} d\Delta t \int_{0}^{L} dx \, \mc W(\Delta t) y(t+\Delta t,x) ,
\end{equation}
where $\mc W$ is a window function that peaks at ${\Delta t = 0}$ and decays on a scale $\tau$ satisfying $1/\omega \ll \tau \ll 1/\mu$. The precise shape of the window function is unimportant, provided that it acts as a filter that averages over the fast oscillations while leaving the slow decay intact. The averaged quantity then takes the form ${\avg{y} \propto \exp(-\mu t)}$.

When the scale separation between $1/\mu$ and $1/\omega$ ceases to be large, it becomes impossible in general to construct a window function with the required properties. Instead, we perform the temporal average by constructing an ensemble of fluctuations with random phases and then averaging over the ensemble. Consider the transformation
\begin{equation}
    \label{eq_app_avg_transform_phi}
    \wt y \to \wt y_\phi = \wt y e^{i\phi} ,
\end{equation}
where $\phi$ is a real phase. A separate phase is assigned to each Fourier mode of the perturbation such that, if $y$ consists of $N$ modes, \textit{viz.}
\begin{equation}
    \label{eq_app_avg_y_modes}
    y(t,x) = \sum_{j=1}^N \Big[\wt y_j e^{ik_j x - i\omega_j t - \mu_j t} + \wt y_j^* e^{-ik_j x + i\omega_j t + \mu_j t} \Big],
\end{equation}
then each mode characterized by $(k_j, \omega_j, \mu_j)$ is assigned a phase $\phi_j$ such that
\begin{equation}
    \label{eq_app_avg_transform_phi_modes}
    \wt y_{\phi_1,\phi_2,\dots,\phi_N} = \sum_{j=1}^N \Big[\wt y_j e^{i\phi_j} e^{ik_j x - i\omega_j t - \mu_j t} + \wt y_j^* e^{-i\phi_j} e^{-ik_j x + i\omega_j t + \mu_j t} \Big] .
\end{equation}
Several choices are available for imposing a relationship between the phases $\phi_j$. 
One option is to allow the phases to be controlled by a parameter $\psi$ according to the relation 
\begin{equation}
    \label{eq_app_avg_phase_psi}
    \phi_j = \omega_j \psi ,
\end{equation}
and $\psi$ is allowed to range from zero to $\Psi$, where
\begin{equation}
    \label{eq_app_avg_bigPsi}
    \Psi = 2\pi \prod_{j=1}^N \omega_j .
\end{equation}
The average of $y$ can then defined as
\begin{equation}
    \label{eq_app_avg_psi}
    \avg{y}(t) = \frac{1}{L \Psi} \int_0^\Psi d\psi \int_0^L dx \, y_{\phi_1,\phi_2,\dots,\phi_N}(t,x) .
\end{equation}
The effect of \eqref{eq_app_avg_psi} is to allow a nonzero average only for products of modes of the form ${\wt y_{j_1} \dots \wt y_{j_m} \wt y_{\ell_1}^* \dots \wt y_{\ell_n}^*}$ for which ${\omega_{j_1} + \dots + \omega_{j_m} - (\omega_{\ell_1} + \dots + \omega_{\ell_n}) = 0}$. In other words, only resonant products of modes survive the averaging.

As a simpler alternative, the phases can be allowed to vary independently over the interval $[0, 2\pi]$. The average of $y$ is defined as
\begin{equation}
    \label{eq_app_avg_phi}
    \avg{y}(t) = \frac{1}{L (2\pi)^N} \int_0^{2\pi} d\phi_1 d\phi_2 \dots d\phi_N \int_0^L dx \, y_{\phi_1,\phi_2,\dots,\phi_N}(t,x) .
\end{equation}
The effect of the averaging scheme in \eqref{eq_app_avg_phi} is that only terms of the form ${\wt y_j^* \wt y_j = |\wt y_j|^2}$ survive the averaging. (Higher-order products, for example ${\wt y_j^* \wt y_j \wt y_\ell^* \wt y_\ell}$, also survive.) Physically, this corresponds to an ensemble of fluctuations with random phases. If observables -- such as the fusion power -- are computed by averaging over the ensemble, then any terms with a random phase will average to zero; only terms consisting of conjugate pairs, whose phases therefore sum to zero, survive the averaging. 
In this work, we neglect terms smaller than second order in the fluctuation amplitudes. For this purpose, the two averaging schemes outlined here are equivalent. Where it is pertinent, we base discussion of physical interpretations on the conceptually simpler random-phase scheme described in \eqref{eq_app_avg_phi}.

\section{Acoustic waves}
\label{sec_app_acoustic}

Because the fusion-power response functions derived in this work depend on the spectrum of the perturbations, we briefly review in this Appendix some properties of acoustic waves in the regime relevant to this work. 
We derive the acoustic dispersion relation (\ref{sec_app_acoustic_dispersion}), the second-order temperature shift produced by acoustic waves (\ref{sec_app_acoustic_temperature}), general formulas for the wave energy \ref{sec_app_acoustic_energy}, and finally the rate of change of the second-order temperature perturbation in terms of the other fluctuating quantities \ref{sec_app_acoustic_reac}.

\subsection{Acoustic dispersion relation}

\label{sec_app_acoustic_dispersion}

Recall that we consider in this work collisional plasmas ($\mr{Kn} \ll 1$), whose ion and electron populations are close to local thermodynamic equilibrium, excepting, perhaps, the small number of particles in the less collisional tail of the distribution function. As described in \S\ref{sec_kinetic}, we furthermore restrict our attention to perturbations with wavelengths much longer than the Debye length so that quasineutrality can be assumed. We again denote the ion and electron masses by $m$ and $m_e$, the ion and electron number densities by $n$ and $n_e$, the ion and electron temperatures by $T$ and $T_e$, the ion and electron pressures by $p$ and $p_e$, and the $z$ component of the ion flow velocity by $u$. The quantities ${s = \ln(T^{1/(\gamma-1)}/n)}$ and ${s_e = \ln(T_e^{1/(\gamma-1)}/n)}$ are, up to some constants, the entropy per particle of the ion and electron fluids, respectively. 
The relevant governing equations for a one-dimensional longitudinal perturbation then read
\begin{align}
    \label{eq_app_fluid_n}
    &\frac{d n}{d t} + n \partial_z u = 0 ,
    \\
    \label{eq_app_fluid_u}
    &mn\frac{d u}{d t} = - \partial_z p - \partial_z p_e + \frac{4}{3}\partial_z (m n \eta \partial_z u) ,
    \\
    \label{eq_app_fluid_s}
    &nT\frac{d s}{dt} = \partial_z \paren{n \kappa_i \partial_z T} + n\frac{4m\eta }{3}\paren{\partial_z u}^2 + n\nu_\mr{ie}(T_e - T) ,
    \\
    \label{eq_app_fluid_se}
    &ZnT_e\frac{d s_e}{dt} = \partial_z \paren{Z n \kappa_e \partial_z T_e} - n\nu_\mr{ie}(T_e - T) ,
\end{align}
where the ion flow velocity $u$ is taken to be in the $z$ direction, $\gamma = 5/3$ is the adiabatic index, $\eta$ is the ion kinematic viscosity, $\kappa_i$ is the ion thermal diffusivity, $\kappa_e$ is the electron thermal diffusivity, and $\nu_\mr{ie}$ is the characteristic rate of energy exchange between ions and electrons. The electron viscosity is neglected. 
%Note that ${E_z = -\partial_z p_e / Zen}$ for fluctuations slow enough to neglect electron momentum. 
Let the spatially and temporally uniform equilibrium state be denoted by the subscript $0$, and let perturbations be described by \eqref{eq_n_expansion}, \eqref{eq_T_expansion}, and \eqref{eq_u_expansion} as well as $T_e = T_0 + T_{e,1} + T_{e,2} + \dots$. Using $\wt T_e$ to denote the Fourier amplitude of the electron temperature perturbation, and defining the complex frequency $\varOmega = \omega - i\mu$, linearizing the fluid equations about the equilibrium gives
\begin{align}
    \label{eq_app_lin_one}
    & -i\varOmega \wt n + i k \vth \wt u = 0 ,
    \\
    \label{eq_app_lin_two}
    & -i\varOmega \vth \wt u = -ik \vth^2 \paren{\wt n + \wt T + Z\wt p_e} - \frac{4}{3}\eta k^2 \vth \wt u ,
    \\
    \label{eq_app_lin_three}
    & -i\varOmega \frac{\wt T}{\gamma - 1} + i \varOmega \wt n = -\kappa_i k^2 \wt T + \nu_\mr{ie}(\wt T_e - \wt T) ,
    \\
    \label{eq_app_lin_four}
    & -i\varOmega \frac{\wt T_e}{\gamma - 1} + i \varOmega \wt n  = -\kappa_e k^2 \wt T_e - \frac{\nu_\mr{ie}}{Z}(\wt T_e - \wt T) .
\end{align}

From \eqref{eq_app_lin_four} we can solve for $\wt T$ to find
\begin{equation}
    \wt T = \wt T_e - \frac{iZ\varOmega}{\nu_\mr{ie}} \frac{1}{\gamma - 1}\wt T_e + \frac{Z\kappa_e}{\nu_\mr{ie}}k^2 \wt T_e + \frac{iZ\varOmega}{\nu_\mr{ie}}\wt n ,
\end{equation}
and from \eqref{eq_app_lin_three} we can solve for $\wt T_e$ to find
\begin{equation}
    \wt T_e = \wt T - \frac{i\varOmega}{\nu_\mr{ie}} \frac{1}{\gamma - 1}\wt T + \frac{\kappa_i}{\nu_\mr{ie}} k^2 \wt T + \frac{i\varOmega}{\nu_\mr{ie}}\wt n .
\end{equation}
Recalling from \S\ref{sec_kinetic} that $\vartheta = \wt T/\wt n$ and $\varrho = 1 + \wt T_e/\wt n$, we can eliminate $\wt T$ and $\wt T_e$ in turn to find

\begin{equation}
    \vartheta = \frac{i\varOmega \left[ \nu_\mr{ie}(1+Z) - Z \left( \frac{i\varOmega}{\gamma - 1} - \kappa_e k^2 \right) \right]}{\nu_{\mr{ie}} \left[ (1+Z)\frac{i\varOmega}{\gamma-1} - (\kappa_i + Z\kappa_e)k^2 \right] - Z \left( \frac{i\varOmega}{\gamma - 1} - \kappa_e k^2 \right) \left( \frac{i\varOmega}{\gamma - 1} - \kappa_i k^2 \right)}
\end{equation}
and
\begin{equation}
    \varrho = 1 + \frac{i\varOmega \left[ \nu_\mr{ie}(1+Z) - Z \left( \frac{i\varOmega}{\gamma - 1} - \kappa_i k^2 \right) \right]}{\nu_{\mr{ie}} \left[ (1+Z)\frac{i\varOmega}{\gamma-1} - (\kappa_i + Z\kappa_e)k^2 \right] - Z \left( \frac{i\varOmega}{\gamma - 1} - \kappa_e k^2 \right) \left( \frac{i\varOmega}{\gamma - 1} - \kappa_i k^2 \right)} .
\end{equation}
Finally, from \eqref{eq_app_lin_two} we obtain the dispersion relation
\begin{equation}
    \label{eq_app_dispersion}
    \frac{\varOmega^2}{k^2} - Z \varrho - \vartheta + i \frac{4 \eta k^2}{3 \vth^2} \frac{\varOmega}{k} = 0 ,
\end{equation}
which becomes, after substituting all fluctuating quantities except for $\wt n$,
\begin{equation}
    \label{eq_app_dispersion_substituted}
\begin{split}
    \Bigg\{ & \nu_{\mr{ie}} \Bigg[ \left( \varOmega^2 + i \frac{4}{3} \eta k^2 \varOmega - k^2 \vth^2 (1+Z) \right) \left( (1+Z)\frac{i\varOmega}{\gamma-1} - (\kappa_i + Z\kappa_e) k^2 \right) - i\varOmega k^2 \vth^2 (1+Z)^2 \Bigg] \\
    & + Z \Bigg[ - \left( \varOmega^2 + i \frac{4}{3} \eta k^2 \varOmega - k^2 \vth^2 (1+Z) \right) \left( \frac{i\varOmega}{\gamma-1} - \kappa_e k^2 \right) \left( \frac{i\varOmega}{\gamma-1} - \kappa_i k^2 \right) \\ 
    & \qquad + i\varOmega k^2 \vth^2 \left( (1+Z)\frac{i\varOmega}{\gamma-1} - (Z\kappa_i + \kappa_e) k^2 \right) \Bigg] \Bigg\} \wt n = 0 .
\end{split}
\end{equation}

\subsection{Reversible temperature shift}
\label{sec_app_acoustic_temperature}

Computing the fusion-power response functions requires an explicit expression for $\avg{T_2}$. To compute this second-order temperature shift, we begin by expanding by the ion and electron entropy equations \eqref{eq_app_fluid_s} and \eqref{eq_app_fluid_se} to second order in the perturbation amplitude and averaging over space and time. It is useful first to divide by the temperature so that each equation describes the evolution of the ion and electron entropies, with the right-hand side describing the Favre-averaged entropy generation rate. The second-order expansion of $s$ is
\begin{equation}
    \label{eq_app_s2_expansion}
    s_2 = \frac{1}{\gamma - 1} \frac{T_2}{T_0} - \frac{1}{2(\gamma-1)}\frac{T_1^2}{T_0^2} + \half \frac{n_1^2}{n_0^2} ,
\end{equation}
with an analogous expansion for $s_{e,2}$, and the second-order expansion of $ds/dt$ is $\partial_t s_2 + u_1 \partial_z s_1$. Then the entropy equations can be written as 
\begin{align}
    \label{eq_app_2nd_order_s}
    \avg{\partial_t s_2 + u_1 \partial_z s_1 + \frac{n_1}{n_0} \partial_t s_1}
    = & - \avg{\frac{T_1}{T_0} \partial_t s_1} + \frac{4\eta}{3\vth^2} \avg{(\partial_z u_1)^2} 
    \\ \nonumber &+ \nu_\mr{ie} \avg{\paren{2\frac{n_1}{n_0} + \nu_{\mr{ie},T_e}\frac{T_{e,1}}{T_0}} \frac{T_{e,1} - T_1}{T_0}} + \nu_\mr{ie} \avg{\frac{T_{e,2} - T_2}{T_0}} ,
    \\
    \label{eq_app_2nd_order_se}
    \avg{\partial_t s_{e,2} + u_1 \partial_z s_{e,1} + \frac{n_1}{n_0} \partial_t s_{e,1}}
    =& - \avg{\frac{T_{e,1}}{T_0}\partial_t s_{e,1}} 
    \\\nonumber &- \frac{\nu_\mr{ie}}{Z} \avg{\paren{2\frac{n_1}{n_0} + \nu_{\mr{ie},T_e}\frac{T_{e,1}}{T_0}} \frac{T_{e,1} - T_1}{T_0}} - \frac{\nu_\mr{ie}}{Z} \avg{\frac{T_{e,2} - T_2}{T_0}} ,
\end{align}
where ${\nu_{\mr{ie},T_e} = d \ln \nu_{\mr{ie}}/d \ln T_e = -3/2}$. Integrating the ${u_1\partial_z s_1}$ term by parts and using the continuity equation \eqref{eq_app_fluid_n}, the left-hand side of \eqref{eq_app_2nd_order_s} becomes ${\partial_t \avg{s_2 + n_1s_1}}$. Some further manipulation yields
\begin{align}
    \label{eq_app_2nd_order_T}
    & \frac{1}{\gamma - 1}\partial_t \avg{ \frac{T_2}{T_0} - \half\frac{T_1^2}{T_0^2} + \frac{T_1 n_1}{n_0 T_0} - \frac{\gamma-1}{2} \frac{n_1^2}{n_0^2} }
    = \mc L + \nu_\mr{ie} \avg{\frac{T_{e,2} - T_2}{T_0}} ,
    \\
    \label{eq_app_2nd_order_Te}
    & \frac{1}{\gamma - 1}\partial_t \avg{ \frac{T_{e,2}}{T_0} - \half\frac{T_{e,1}^2}{T_0^2} + \frac{ T_{e,1} n_1}{n_0 T_0} - \frac{\gamma-1}{2} \frac{n_1^2}{n_0^2} }
    = \mc L_e - \frac{\nu_\mr{ie}}{Z} \avg{\frac{T_{e,2} - T_2}{T_0}} ,
\end{align}
where the functions $\mc L$ and $\mc L_e$ are the right-hand sides, excepting the final term, of \eqref{eq_app_2nd_order_s} and \eqref{eq_app_2nd_order_se}. In terms of fluctuation amplitudes, $\mc L$ and $\mc L_e$ are given by
\begin{align}
    \label{eq_app_L_def}
    \mc L &= 2 \mr{Re} \Bigg[\frac{\mu}{\gamma - 1} |\vartheta|^2 - i\varOmega \vartheta^* + \frac{4\eta k^2}{3} |c|^2 + \nu_\mr{ie} \Big(2 + \nu_{\mr{ie},T_e} (\varrho^* - 1)\Big) \paren{\varrho - 1 - \vartheta}\Bigg] ,
    \\
    \label{eq_app_Le_def}
    \mc L_e &= 2 \mr{Re} \Bigg[\frac{\mu }{\gamma - 1}|\varrho-1|^2 - i\varOmega (\varrho^* - 1) - \frac{\nu_\mr{ie}}{Z} \Big(2 + \nu_{\mr{ie},T_e} (\varrho^* - 1)\Big) \paren{\varrho - 1 - \vartheta}\Bigg] .
\end{align}

\begin{figure}
    \centering
    \includegraphics[width=0.6\columnwidth]{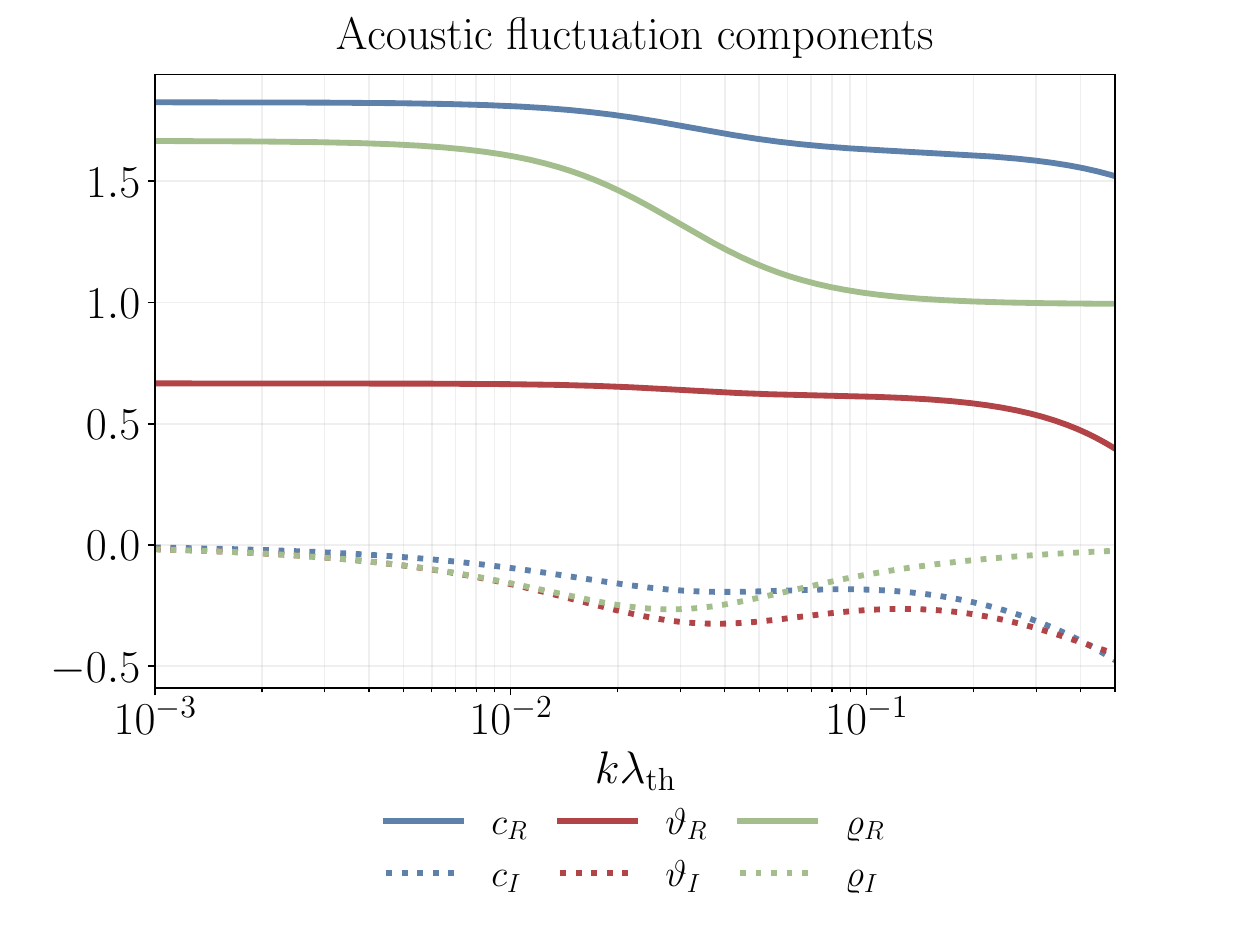}
    \caption{\justifying
    Complex phase velocity $c = (\omega - i\mu)/k\vth$, normalized ion temperature perturbation $\theta = \wt T/\wt n$, and normalized electron temperature perturbation $\varrho = 1 + \wt T_e/\wt n$ for the acoustic wave solution of the dispersion relation \eqref{eq_app_dispersion}, plotted as a function of $k\lambda_\mr{mfp}$ in a deuterium plasma.}
    \label{fig_acoustic_disp}
\end{figure} 
% python acoustic_paper_plots.py -p disp --m-i 2.0 --k-min 1e-4 --k-max 9e-1

Combined with the collisional energy-exchange terms on the right-hand sides of \eqref{eq_app_2nd_order_T} and \eqref{eq_app_2nd_order_Te}, the functions $\mc L$ and $\mc L_e$ record the rate of irreversible heating of the electron and ion fluids. 
The structure of \eqref{eq_app_2nd_order_T} and \eqref{eq_app_2nd_order_Te} suggests a natural decomposition of the second-order temperature shifts into a part associated with the wave itself and a part associated with heating of the background. The former part, which we will denote by $S_2$, is a result of reversible wave dynamics; the latter part, which we will denote by $Q_2$, is a result of irreversible dynamics by which energy leaves the wave and is deposited in the background plasma. Let ${T_2 = S_2 + Q_2}$ and ${T_{e,2} = S_{e,2} + Q_{e,2}}$. The evolution of $S_2$ and $S_{e,2}$ can be determined by setting the right-hand sides of \eqref{eq_app_2nd_order_T} and \eqref{eq_app_2nd_order_Te} to zero, yielding
\begin{align}
    \label{eq_app_S2_soln}
    \avg{S_2} &= \half \frac{\avg{T_1^2}}{T_0} - \frac{\avg{n_1 T_1}}{n_0} + \frac{\gamma-1}{2} \frac{\avg{n_1^2}}{n_0^2}T_0 ,
    \\
    \label{eq_app_Se2_soln}
    \avg{S_{e,2}} &= \half \frac{\avg{T_{e,1}^2}}{T_0} - \frac{\avg{n_1 T_{e,1}}}{n_0} + \frac{\gamma-1}{2} \frac{\avg{n_1^2}}{n_0^2}T_0 ,
\end{align}
or equivalently
\begin{align}
    \label{eq_app_S2_soln_fluctuations}
    \avg{S_2} &= 2\mr{Re}\Bigg[ \half |\vartheta|^2 - \vartheta + \frac{\gamma-1}{2} \Bigg] |\wt n|^2 T_0,
    \\
    \label{eq_app_Se2_soln_fluctuations}
    \avg{S_{e,2}} &= 2\mr{Re}\Bigg[ \half |\varrho-1|^2 - (\varrho-1) + \frac{\gamma-1}{2} \Bigg] |\wt n|^2 T_0.
\end{align}
Finally, in cases where the temperature and density perturbations are in phase, letting ${g = 1+\vartheta \in \mathbb{R}}$ and ${g_e = \varrho \in \mathbb{R}}$ be the effective polytropic indices of the ion and electron fluids, respectively, we have
\begin{align}
    \label{eq_app_S2_soln_g}
    \avg{S_2} &= \Bigg[ (g-1)(g-3) + \gamma-1  \Bigg] |\wt n|^2 T_0,
    \\
    \label{eq_app_Se2_soln_g}
    \avg{S_{e,2}} &= \Bigg[ (g_e-1)(g_e-3) + \gamma-1  \Bigg] |\wt n|^2 T_0.
\end{align}

\begin{figure}
    \centering
    \includegraphics[width=0.6\columnwidth]{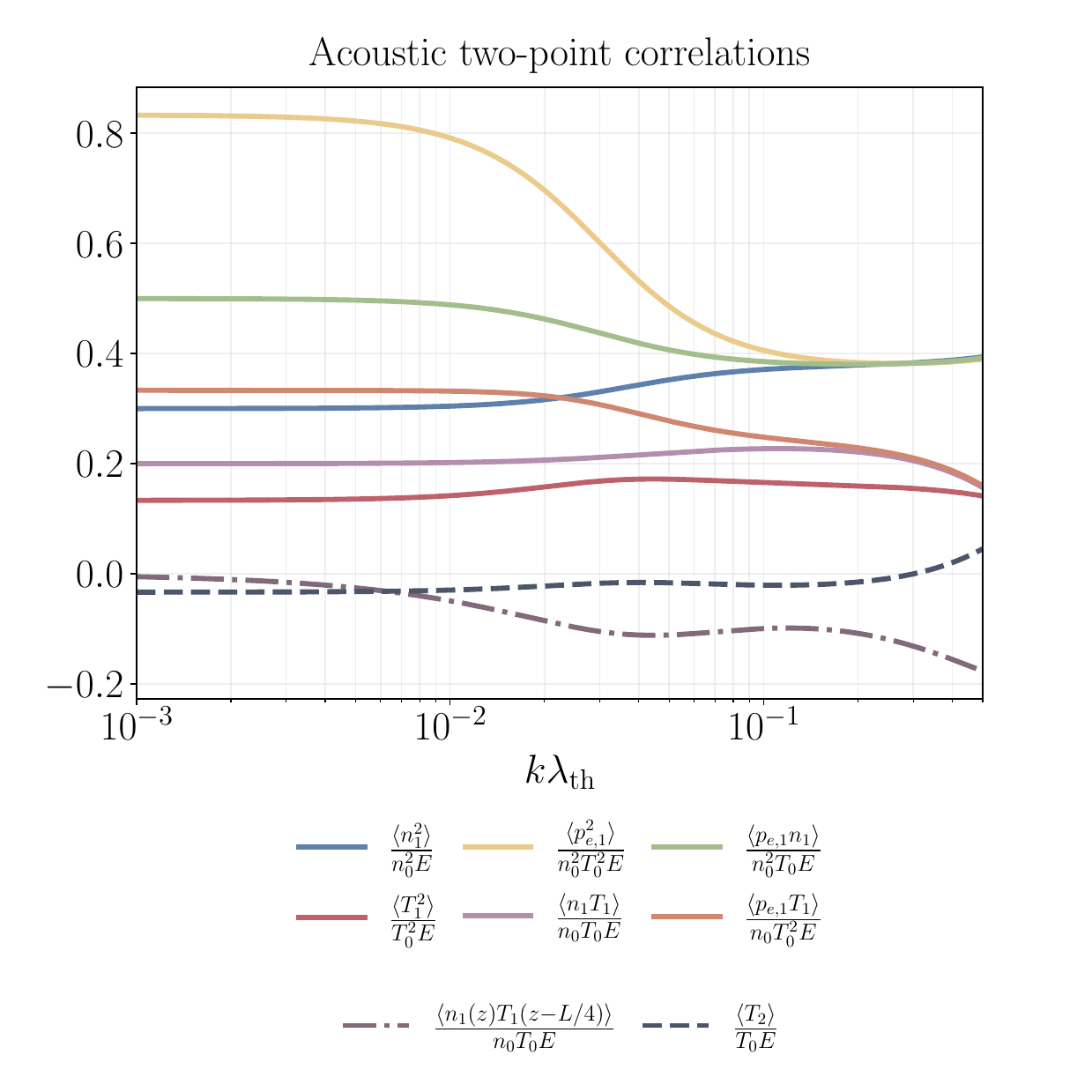}
    \caption{\justifying
    Second-order correlation functions normalized to the wave energy per ion in an acoustic wave satisfiying the dispersion relation \eqref{eq_app_dispersion}, plotted as a function of $k\lambda_\mr{mfp}$ in a deuterium plasma.}
    \label{fig_acoustic_corr}
\end{figure}

In Fig.~\ref{fig_acoustic_corr}, the second-order correlation functions are shown in a perturbation satisfying the acoustic dispersion relation \eqref{eq_app_dispersion}. The six solid curves record the correlations among fluctuations in density, temperature, and electron pressure. The dashed-dotted curve indicates the phase shift between the fluctuations in ion temperature and density (the imaginary part of $\vartheta$), which appear in short-wavelength, dissipative waves. The dashed curve shows the second-order temperature shift $\avg{T_2} = \avg{S_2}$. In the long-wavelength limit, where the waves become adiabatic and so ${\avg{T_2} \to \half(\gamma-1)(\gamma-2)T_0\avg{n_1^2}n_0^2}$ and ${E \to (1+Z)\gamma\avg{n_1^2}/n_0^2}$, the normalized second-order temperature shift asymptotes to ${\avg{T_2}/T_0E \to -1/30}$ for $Z=1$ and $\gamma=5/3$, as reflected in Fig.~\ref{fig_acoustic_corr}.

\subsection{Acoustic energy}
\label{sec_app_acoustic_energy}

From \eqref{eq_app_2nd_order_T} and \eqref{eq_app_2nd_order_Te}, $Q_2$ and $Q_{e,2}$ evolve according to
\begin{align}
    \label{eq_app_Q2_evol}
    \frac{\partial_t \avg{Q_2}}{\gamma - 1} &= \mc L T_0 |\wt n|^2 e^{-2\mu t} + \nu_\mr{ie} \avg{Q_{e,2} - Q_2} , 
    \\
    \label{eq_app_Qe2_evol}
    \frac{\partial_t \avg{Q_{e,2}}}{\gamma - 1} &= \mc L_e T_0 |\wt n|^2 e^{-2\mu t} - \frac{\nu_\mr{ie}}{Z} \avg{Q_{e,2} - Q_2} .
\end{align}
Let $\Delta Q_2 = Q_{e,2} - Q_2$ and $\overline Q_2 = ZQ_{e,2} + Q_2$ be the interspecies temperature separation and the normalized energy deposited into the background plasma, respectively. We set as initial conditions $\Delta Q_2 = 0$ and $\overline Q_2 = 0$ at $t=0$, prescribing that no background heating has occurred at the initial time (or, more accurately, that any background heating prior to $t=0$ is absorbed into the equilibrium temperature $T_0$, and only further heating after $t=0$ determines $Q_2$ and $Q_{e,2}$). Then the solution to \eqref{eq_app_Q2_evol} and \eqref{eq_app_Qe2_evol} is
\begin{align}
    \label{eq_app_Q2_Delta_soln}
    \avg{\Delta Q_2} &= \frac{(\gamma-1)\paren{\mc L_e - \mc L} }{\nu_\mr{eff} - 2\mu}\Big[ e^{-2\mu t} - e^{-\nu_\mr{eff}t}\Big] T_0 |\wt n|^2 ,
    \\
    \label{eq_app_Q2_tot_soln}
    \avg{\overline Q_2} &= \frac{(\gamma-1)(Z\mc L_e + \mc L)}{2\mu} \Big[1 - e^{-2\mu t}\Big] T_0 |\wt n|^2 ,
\end{align}
where ${\nu_\mr{eff} = (\gamma-1)(1 + 1/Z)\nu_\mr{ie}}$ is an effective energy exchange rate, and $\wt n$ is the initial density perturbation. Physically, the apparent singularity in \eqref{eq_app_Q2_Delta_soln} indicates a qualitative shift in the background-temperature evolution as $\mu$ increases. (The solution $\avg{\Delta Q_2}(t)$ remains finite and is replaced by a secular solution at $\mu = \nu_\mr{eff}$.) When the wave decays slowly, collisions are sufficiently fast to keep the ion and electron temperatures clamped together; when $\mu > \half \nu_\mr{eff}$, the wave is heating one species more quickly than collisions can partition the heating between species, so the temperature separation grows, unless $\mc L_e$ and $\mc L$ balance in such a way that the denominator $\nu_\mr{eff} - 2\mu$ is cancelled out, corresponding to a heating process in which heating is already equipartitioned without the need for collisional energy exchange.

It can be shown from \eqref{eq_app_Q2_tot_soln} that this partition between wave energy and background energy is consistent with the energy formula of \cite{Chu_1965}, cf. \eqref{eq_E_pert}. Noting that, for a decaying wave, the initial energy of the wave is equal to the total energy deposited into the background plasma when the wave has entirely decayed, we have 
\begin{equation}
    \label{eq_app_E_wave_lim}
    E = \frac{1}{\gamma-1}\lim_{t \to \infty} \avg{\overline Q_2}
\end{equation}
and therefore
\begin{equation}
    \label{eq_app_E_wave}
    E = \frac{Z\mc L_e + \mc L}{2\mu} |\wt n|^2 .
\end{equation}
In terms of explicit correlations, \eqref{eq_app_E_wave} can be rewritten as
\begin{equation}
    \label{eq_app_E_wave_correl}
    E = \frac{\avg{T_1^2} + Z\avg{T_{e,1}^2}}{2(\gamma-1) T_0^2} 
    - \frac{\avg{n_1 T_1} + Z\avg{n_1 T_{e,1}}}{2 n_0 T_0} 
    - \frac{\omega}{\mu} \frac{\avg{n_1(z) T_1(z - \frac{L}{4})} + Z\avg{n_1(z) T_{e,1}(z - \frac{L}{4})}}{2 n_0 T_0} 
    + \frac{2 \eta k^2}{3 \mu} \frac{\avg{u_1^2}}{\vth^2}  ,
\end{equation}
where $L = 2\pi/k$. While \eqref{eq_app_E_wave_correl} may appear to diverge as $\mu \to 0$, the phase-shifted term and the viscous term approach zero in this limit, keeping the energy finite. It can be verified algebraically or graphically that \eqref{eq_app_E_wave_correl} is equivalent to the two-temperature generalization of \eqref{eq_E_pert}, \textit{viz.}
\begin{equation}
    \label{eq_app_E_wave_pert}
    E = \half \frac{\avg{u_1^2}}{\vth^2} + \frac{1+Z}{2} \frac{\avg{n_1^2}}{n_0^2} + \frac{1}{2(\gamma-1)} \frac{\avg{T_1^2}}{T_0^2} + \frac{Z}{2(\gamma-1)} \frac{\avg{T_{e,1}^2}}{T_0^2} ,
\end{equation} 
when the fluctuating quantities satisfy the acoustic dispersion relation \eqref{eq_app_dispersion}. The wave energy is plotted in Fig.~\ref{fig_acoustic_disp} in a form normalized to the wave amplitude, $\mc E = E/2|\wt n|^2$.

\subsection{Waves and reactivity}
\label{sec_app_acoustic_reac}

In \S\ref{sec_hydro}, we saw that the second-order temperature shift leads to a shift in the fusion reactivity. In the hydrodynamic limit, where collisions are sufficiently fast that the temporal evolution of the perturbation does not shift the distribution function away from a Maxwellian, the reactivity depends only on the instantaneous temperature. To make a clean distinction between the second-order temperature shift and the background heating, it is therefore useful for a given set of plasma conditions and wave amplitude to evaluate $T_2$ at the initial time $t=0$, when no background heating has yet occurred, such that ${T_2 = S_2}$. Any heating that occurred at earlier times is absorbed into the equilibrium temperature $T_0$, and if the reactivity needs to be evaluated at a later time after the wave has undergone some decay, $T_0$ is updated to account for the background heating that occurred in the intervening time, and the time coordinate is shifted so that $t=0$ at the instant where reactivity is being evaluated. 

\begin{figure}
    \centering
    \includegraphics[width=0.6\columnwidth]{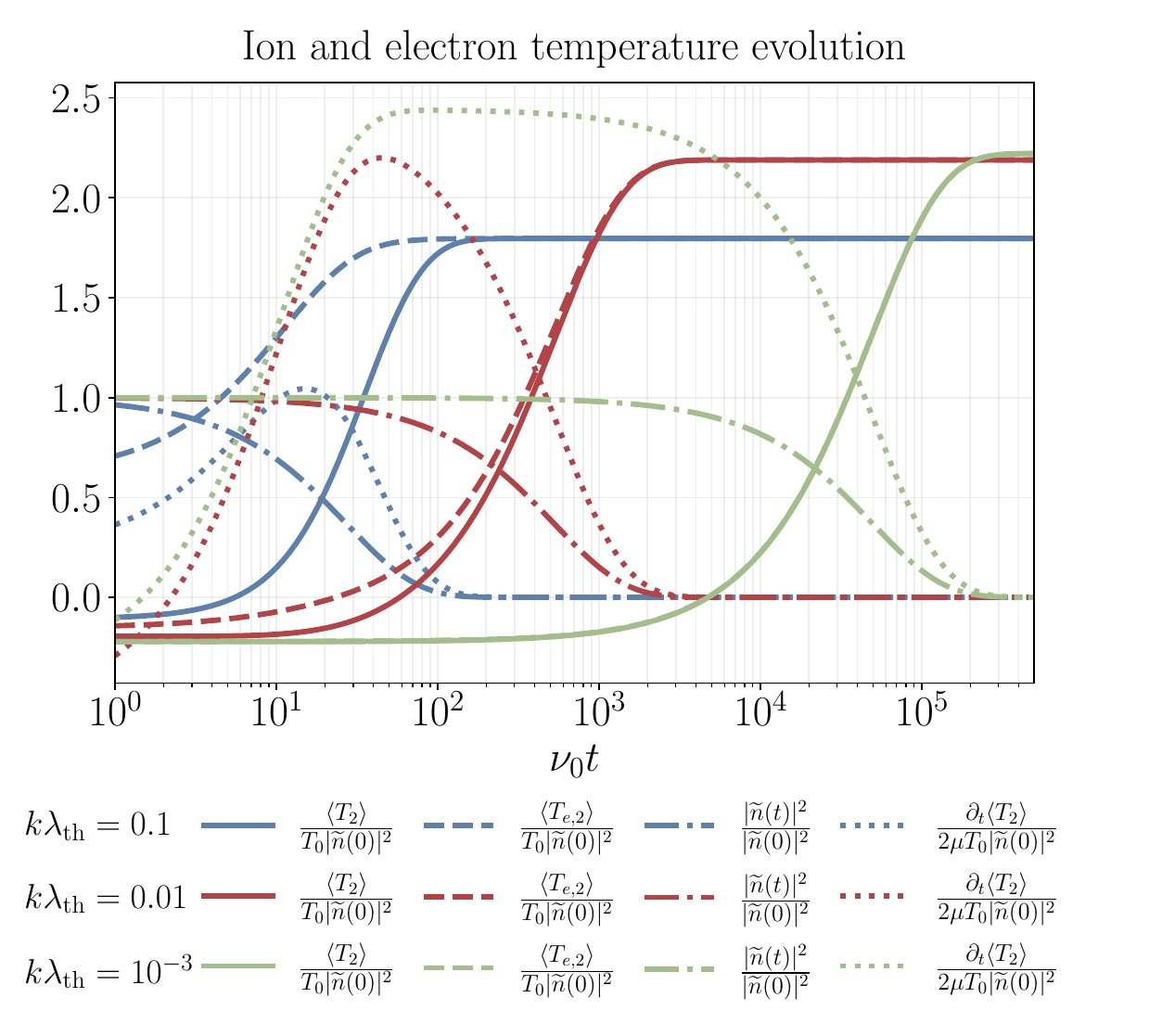}
    \caption{\justifying
        Evolution of the second-order shifts in ion and electron temperatures for decaying acoustic waves of several wavenumbers. The wavenumbers are normalized to the ion mean free path, and time is normalized to the ion-ion collision frequency $\nu$. Temperature shifts are normalized to the background temperature $T_0$ and the initial wave amplitude $|\wt n|^2$. The wave amplitude $|\wt n(t)|^2$ is also plotted for reference, normalized to its initial value. The time derivative of the ion-temperature shift is shown, normalized to the initial wave amplitude and to the decay rate $2\mu$; this derivative is proportional to $\widehat \Gamma$ up to a factor of the decaying wave amplitude.}
    \label{fig_acoustic_temp_evol}
\end{figure}

In \S\ref{sec_kinetic}, we saw that the kinetic corrections to the reactivity depend not only on the current value of the second-order temperature shift but also on its time derivative ${\Gamma = -\partial_t \avg{T_2}/2T_0 |\wt n|^2 }$. Unfortunately, this renders the evaluation of the reactivity substantially more complicated. Using \eqref{eq_app_S2_soln_fluctuations} along with \eqref{eq_app_Q2_Delta_soln} and \eqref{eq_app_Q2_tot_soln}, and noting that ${Q_2 = (\overline Q_2 - Z\Delta Q_2)/(1+Z)}$, we can solve for the time evolution of $T_2$ and find
\begin{equation}
    \label{eq_app_T2_t}
    \begin{split}
    T_2 = (\gamma-1)\Bigg[& \frac{|\vartheta|^2 e^{-2\mu t}}{\gamma-1} - \frac{2\mc \mr{Re}[\vartheta] e^{-2\mu t}}{\gamma-1} + e^{-2\mu t}
     + \paren{\frac{1-e^{-2\mu t}}{2\mu} + Z \frac{e^{-2\mu t} - e^{-\nu_\mr{eff}t}}{\nu_\mr{eff} - 2\mu}}\frac{\mc L}{1+Z} 
    \\& \hspace{4cm} +  \paren{\frac{1-e^{-2\mu t}}{2\mu} - \frac{e^{-2\mu t} - e^{-\nu_\mr{eff}t}}{\nu_\mr{eff} - 2\mu}}\frac{Z \mc L_e}{1+Z}\Bigg] T_0 |\wt n|^2 .
    \end{split}
\end{equation}
%frac{\nu_\mr{eff} - 2\mu + \Big((1+Z)2\mu - \nu_\mr{eff}\Big)e^{-2\mu t} - 2\mu Z e^{-\nu_\mr{eff} t}}{(1+Z)2\mu (\nu_\mr{eff} - 2\mu)} \mc L
The time evolution of the ion and electron temperature shifts is shown in Fig.~\ref{fig_acoustic_temp_evol}, along with the time derivative of the ion temperature, at wavenumbers ${k\lth \in ( 0.1, 0.01, 0.001)}$. At long wavelengths (green curve) the ion and electron temperatures remain closely coupled, and so $\partial_t \avg{T_2}$ decays smoothly with the wave amplitude. On the other hand, at short wavelengths, the electron temperature rises quickly as the wave begins to dissipate. On a somewhat longer time scale, the ions gain energy by collisions with electrons. The result is a delayed rise in $\avg{T_2}$ and a spike in $\partial_t \avg{T_2}$. 
In general, then, $\avg{T_2}$ does not evolve like $\exp(-2\mu t)$, as the other relevant second-order quantities do, and $\widehat \Gamma$ is not a constant. 
This peculiarity of multi-species plasma -- a consequence of the additional degrees of freedom available for dissipated energy to partition itself between species independently of constraints imposed by the first-order dispersion relation -- introduces some ambiguity into the definition of $\avg{T_2}$ and $\widehat \Gamma$ in cases (such as the kinetic formulas of \S\ref{sec_kinetic}) where it is necessary to quote a single value rather than the full time history of these quantities. 

As described in \S\ref{sec_fusion_response}, to obtain a straightforward but representative value for $\widehat \Gamma$, we evaluate $\partial_t \avg{T_2}$ at $t = 1/2\mu$, meaning that temperatures have been allowed to evolve for one wave-energy e-folding time away from the initial conditions imposed at $t=0$. 
In systems where the ions and electrons equilibrate quickly relative to the decay time scale of the wave, this evaluation scheme avoids the transient collisional behavior; the $\exp(-\nu_\mr{eff} t)$ terms vanish when $\mu \ll \nu_\mr{eff}$ because ${\exp(-\nu_\mr{eff}/2\mu) \to 0}$. 
Using this evaluation scheme, $\widehat \Gamma$ is given by
\begin{equation}
    \label{eq_app_Gamma_eval}
    \begin{split}
    \widehat \Gamma \sim & |\vartheta|^2 - 2\mc \mr{Re}[\vartheta] + (\gamma-1) - \frac{\gamma-1}{1+Z}\mc E
    + \frac{-2\mu + \nu_\mr{eff}e^{\nu_\mr{eff}/2\mu+1}}{\nu_\mr{eff} - 2\mu} \frac{(\gamma-1)Z}{(1+Z)2\mu} \paren{\mc L_e - \mc L} ,
    %\widehat \Gamma \sim 1 - 2\mc \mr{Re}[\vartheta] + (\gamma-1) - \paren{1 + Z\frac{-2\mu + \nu_\mr{eff}e^{-\nu_\mr{eff}/2\mu+1}}{\nu_\mr{eff} - 2\mu}} \frac{\gamma - 1}{(1+Z)2\mu}\mc L - \paren{1 - \frac{-2\mu + \nu_\mr{eff}e^{\nu_\mr{eff}/2\mu+1}}{\nu_\mr{eff} - 2\mu}} \frac{(\gamma-1)Z}{(1+Z)2\mu} \mc L_e 
    \end{split}
\end{equation}
where we used \eqref{eq_app_E_wave} to eliminate the $Z\mc L_e + \mc L$ term in favor of $\mc E = E/2|\wt n|^2$. In the adiabatic limit, where ions and electrons are approximately in equilibrium with each other at each point, the dissipation rate is small, and the effective polytropic index is ${g = \gamma = 5/3}$, \eqref{eq_app_Gamma_eval} reduces to
\begin{equation}
    \label{eq_app_Gamma_adiabatic}
    \widehat \Gamma = - \frac{22}{9} .
\end{equation}

\begin{figure}
    \centering
    \vspace{0.5cm}
    \includegraphics[width=0.6\columnwidth]{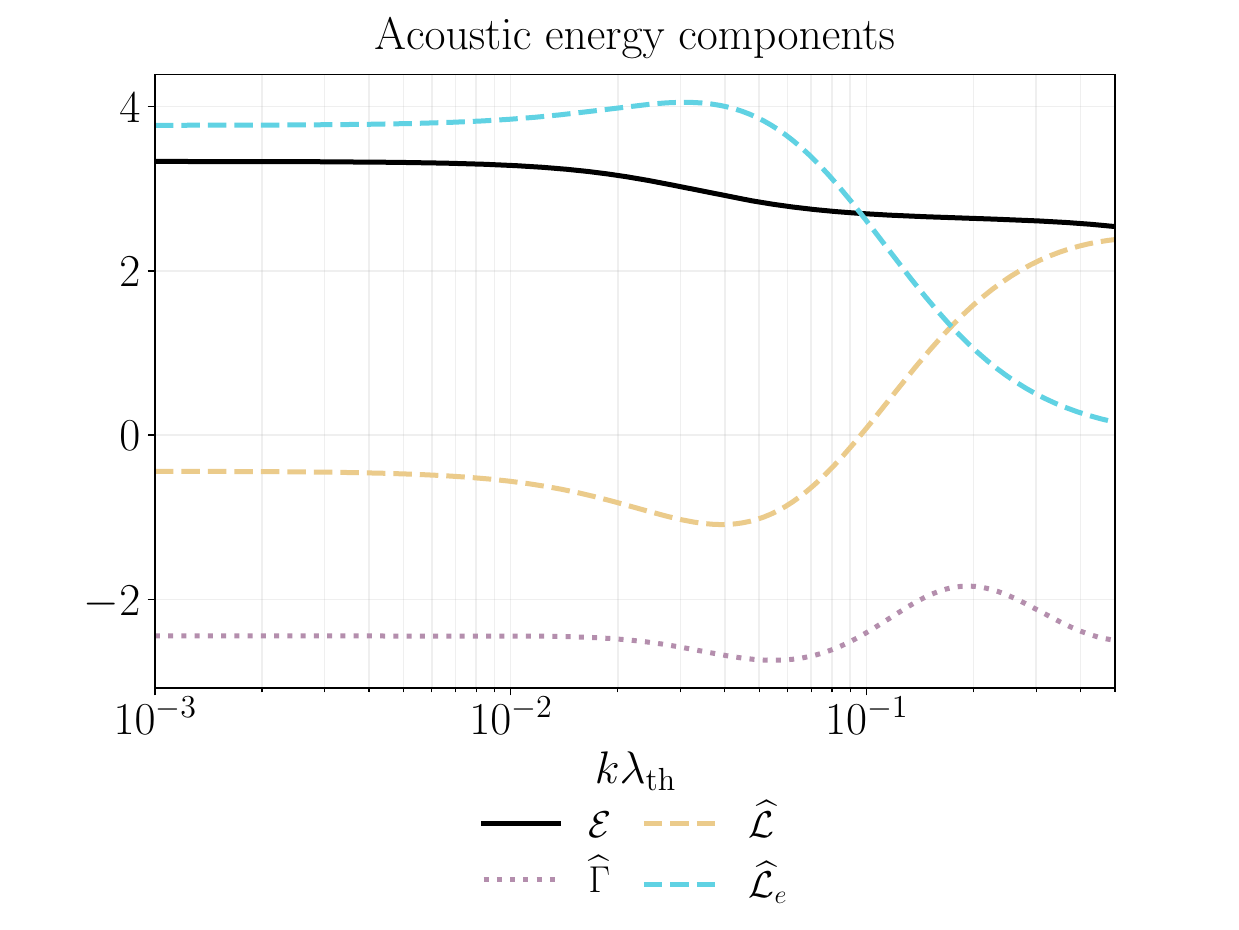}
    \caption{\justifying
    Components of the energy and heating of ions and electrons in an acoustic wave: the normalized energy $\mc E = E/(2|\wt|^2)$ is given in \eqref{eq_app_E_wave} and the normalized heating rate $\widehat \Gamma = \partial_t\avg{T_2} / (2T_0 |\wt n|^2)$ is given in \eqref{eq_app_Gamma_eval}. Both are plotted as a function of $k\lambda_\mr{mfp}$ in a deuterium plasma. Also shown are $\widehat{\mc L} = \mc L/4|\wt n|^2$ and $\widehat{\mc L}_e = \mc L_e/4|\wt n|^2$, representing the entropy generation rates in the ion and electron fluids.}
    \label{fig_acoustic_energy}
\end{figure} 
% python acoustic_paper_plots.py -p disp --m-i 2.0 --k-min 1e-4 --k-max 9e-1

\section{Second-order distribution function}
\label{sec_app_f2}

This Appendix provides the intermediate steps in the calculation of the second-order distribution function $f_2$ in \S\ref{sec_kinetic}. We begin with the second-order kinetic equation \eqref{eq_f2_kinetic_start} along with the form of $\avg{f_2}$ in \eqref{eq_f2_F2_J2} where all time dependence has been made explicit, and the rest of the variation of $\avg{f_2}$ is captured by the functions $F_2$, $J_2^{(R,I)}$, and $M_2^{(R,I)}$. 
Matching the form of \eqref{eq_f2_evol_y} to \eqref{eq_f2_kinetic_start}, let the coefficients $A$, $B$, and $C$ corresponding to each mode satisfy ${A = \wt A + \wt A^*}$, ${B = \wt B + \wt B^*}$, and ${C = \wt C + \wt C^*}$. 
Then the coefficients are given by
\begin{align}
    \label{eq_app_y2_evol_A}
    A & = -\Bigg\langle \Bigg[-\half \partial_t \paren{\frac{n_1^2}{n_0^2}} + \vth \widehat{\bs u}_1 \cdot \nabla \paren{\frac{n_1}{n_0}} + \vth \half \frac{T_1}{T_0} \bs w \cdot \nabla \paren{\frac{n_1}{n_0}} 
    \\ 
    & \nonumber \qquad + \frac{3}{4}\partial_t\paren{\frac{T_1}{T_0}}^2 - \frac{3}{2}\partial_t\paren{\frac{T_2}{T_0}} - \frac{3}{2}\vth \widehat{\bs u}_1 \cdot \nabla \paren{\frac{T_1}{T_0}}  - \vth \frac{3}{4} \frac{T_1}{T_0} \bs w \cdot \nabla \paren{\frac{T_1}{T_0}}\Bigg] f_0 \Bigg\rangle  
    \\ \nonumber & + \Bigg\langle\Bigg[\frac{1}{\vth m n_0}\paren{-\half\frac{T_1}{T_0} - \frac{n_1}{n_0}}\nabla p_{e,1}
    - \half\frac{T_1}{T_0} \vth \partial_t \widehat{\bs u}_1 + \vth \partial_t \widehat{\bs u}_2
    + \half \frac{\partial_t T_2}{T_0} \bs w 
    \\ \nonumber & \qquad - \frac{1}{4}\partial_t \paren{\frac{T_1}{T_0}}^2 \bs w + \half \vth \widehat{\bs u}_1 \cdot \nabla \paren{\frac{T_1}{T_0}} \bs w  \Bigg] \cdot \frac{\partial f_0}{\partial \bs w}\Bigg\rangle
\end{align}
from the terms sourced by interactions between the unperturbed $f_0$ and second-order perturbed fluid quantities,
\begin{align}
    \label{eq_app_y2_evol_B}
    B =& -\nu \avg{\frac{n_1}{n_0} f_1} - \nu_T \nu \avg{\frac{T_1}{T_0} f_1} - \avg{\paren{\half \frac{T_1}{T_0} \vth \bs w + \vth \widehat{\bs u}_1} \cdot \nabla f_1} 
    \\ \nonumber &- \avg{\paren{\partial_t \paren{\frac{n_1}{n_0}} + \vth \bs w \cdot \nabla \paren{\frac{n_1}{n_0}} - \frac{3}{2}\partial_t\paren{\frac{T_1}{T_0}} - \frac{3}{2} \vth \bs w \cdot \nabla \paren{\frac{T_1}{T_0}}      } f_1}
    \\ \nonumber & + \Bigg\langle \Bigg[ \frac{1}{\vth m n_0} \nabla p_{e,1} + \partial_t \widehat{\bs u}_1 + \vth \bs w \cdot \nabla \widehat{\bs u}_1 + \half \partial_t \paren{\frac{T_1}{T_0}} \bs w  + \half \vth \bs w \cdot \nabla \paren{\frac{T_1}{T_0}} \bs w\Bigg] 
    \cdot \paren{\frac{\partial F_1}{\partial \bs w}}\Theta \Bigg\rangle
\end{align}
from the interactions between the time-independent terms of $f_1$ and the first-order fluid perturbations, and
\begin{align}
    \label{eq_app_y2_evol_C}
    C =&  \Bigg\langle\Bigg[ \frac{1}{\vth m n_0} \nabla p_{e,1} + \partial_t \widehat{\bs u}_1 + \vth \bs w \cdot \nabla \widehat{\bs u}_1 + \half \partial_t \paren{\frac{T_1}{T_0}} \bs w  + \half \vth \bs w \cdot \nabla \paren{\frac{T_1}{T_0}} \bs w\Bigg] 
     \cdot \paren{\frac{\partial \Theta }{\partial \bs w}} F_1\Bigg\rangle .
\end{align}
In \eqref{eq_app_y2_evol_B} and \eqref{eq_app_y2_evol_C}, $f_1$ has been decomposed into its time-independent part $F_1$ and its time-dependent part $\Theta$, cf. \eqref{eq_f1_evol_y_soln}.

We begin by computing $F_2$. Noting that the spatial averaging leaves only conjugate pairs of Fourier modes, some simplification of \eqref{eq_f2_kinetic_start} yields
\begin{comment}
\begin{equation}
    \label{eq_f2_kinetic_SIMPLIFIED}
    \begin{split}
    (2\mu - \nu) F_2 = 2 \mr{Re}\Bigg\{ & \bigg[\half ik\vth \vartheta^*  w\xi + i k\vth c^* + (i\omega - \mu) - i k \vth w \xi 
    \\ & - \frac{3}{2}(i\omega - \mu) \vartheta^* + \frac{3}{2} ik \vth \vartheta^* w \xi
    + \nu + \nu_T\nu \vartheta^* \bigg]\wt n^* \wt F_1  
    \\ & + \bigg[\mu + i k \vth c^* + \half ik \vth \vartheta^* w\xi  - \frac{3}{2}\mu |\vartheta|^2 + \frac{3}{2}\Gamma  - \frac{3}{2} ik\vth c^* \vartheta  \bigg] |\wt n|^2 \fm % note dropped purely imaginary term - \frac{3}{4} ik\vth |\vartheta|^2 w\xi 
    \\
    & + \bigg[ -ik  \vth w\xi \paren{1 + \half \vartheta^*} Z \varrho  + \half i k \vth \vartheta^* c^2 w \xi + ik \vth w\xi |c|^2 
    \\
    & \qquad + 2\mu c w \xi -\half \Gamma w^2 + \half \mu |\vartheta|^2 w^2  + \half ik \vth c^* \vartheta w^2 \bigg] |\wt n|^2 \fm
    \\
    & - \bigg[ -ik \vth Z \varrho^* - ik\vth c^* w\xi + (i\omega - \mu) c^*  \bigg] \wt n^* \frac{\partial \wt F_1}{\partial w_z} 
    \\
    & - \bigg[ - \half ik\vth \vartheta^* w^2 \xi +  \half(i\omega - \mu) \vartheta^* w \bigg]  \wt n^* \frac{\partial \wt F_1}{\partial w}\Bigg\} ,
    \end{split}
\end{equation}
\end{comment}
\begin{equation}
    \label{eq_app_f2_kinetic_SIMPLIFIED_TWO}
    \begin{split}
    (\nu - 2\mu) F_2 = -2 \mr{Re}\Bigg\{ & \bigg[i\omega + \half ik \vth \vartheta^* w\xi - \frac{3}{2}\mu |\vartheta|^2 + \frac{3}{2}\Gamma  - \frac{3}{2} ik\vth c^* \vartheta \bigg] |\wt n|^2 \fm
    \\
    & + \bigg[ -ik \vth w \xi Z \paren{1 + \half \vartheta^*} \varrho  + \half i k \vth \vartheta^* c^2 w \xi + ik \vth |c|^2 w\xi 
    \\
    & \qquad + 2\mu c w \xi
     -\half \Gamma w^2 + \half \mu |\vartheta|^2 w^2 + \half ik \vth c^* \vartheta w^2  \bigg] |\wt n|^2 \fm
    \\
    & + \bigg[\half ik \vth \vartheta^* w \xi + 2i k\vth c^* - i k \vth  w \xi + \frac{3}{2} ik\vth (w\xi - c^*)\vartheta^* 
    + \nu + \nu_T\nu \vartheta^* \bigg]\wt n^* \wt F_1  
    \\
    & - \bigg[ -ik\vth Z \varrho^* - ik\vth c^* w\xi + ik\vth (c^*)^2  \bigg] \wt n^* \frac{\partial \wt F_1}{\partial w_z} 
    \\
    & - \bigg[ - \half ik\vth \vartheta^* w^2 \xi +  \half ik\vth c^* \vartheta^* w \bigg] \wt n^*  \frac{\partial \wt F_1}{\partial w}\Bigg\} .
    \end{split}
\end{equation}
% $\vartheta_2 = \avg{T_2}/T_0$ is the normalized second-order temperature perturbation. 
Because gradients in the first-order distribution act as a source for the second-order distribution, it is necessary to compute $\partial \wt F_1/\partial w$ and $\partial \wt F_1/\partial w_z$, which, using \eqref{eq_f1_soln_c}, are
\begin{equation}
\label{eq_app_dF1_dp}
\begin{split}
    \frac{\partial \wt F_1}{\partial w} =& -w \wt F_1 + \frac{1}{w}\frac{-iw\xi + hN}{N + iw\xi - ic}\wt F_1  
    \\ & + \frac{i \wt n \fm}{N + iw\xi - ic} \Bigg[ X\xi - 2\paren{\xi^2 - \half \vartheta} c w  - \frac{3}{2} \vartheta w^2 \xi \Bigg] ,
\end{split}
\end{equation}
and
\begin{equation}
\label{eq_app_dF1_dpz}
\begin{split}
    \frac{\partial \wt F_1}{\partial w_z} =& -w\xi \wt F_1 + \frac{1}{w\xi}\frac{-iw\xi + \xi^2 h N}{N + iw\xi - ic} \wt F_1 
    \\ & + \frac{i \wt n \fm}{N + iw\xi - ic} \Bigg[ X - 2\paren{1 - \half \vartheta}c w \xi   - \vartheta \paren{\xi^2 + \half } w^2 \Bigg] ,
\end{split}
\end{equation}
where the (negative, logarithmic) velocity derivative of the collision frequency is 
\begin{equation}
    \label{eq_app_h_def}
     h = -\frac{d\ln\nu}{d\ln w} .
\end{equation}
Furthermore, let
\begin{equation}
    \label{eq_app_D_f2helper}
    D = N + iw\xi - ic
\end{equation}
be the denominator of $\wt F_1$ and its derivatives and let $N_T = \nu_T/(N k\vth)$. 
Then \eqref{eq_app_f2_kinetic_SIMPLIFIED_TWO} simplifies to
\begin{equation}
    \label{eq_app_F2_final}
    \begin{split}
    F_2 = \frac{-2}{\nu - 2\mu} \mr{Re}\Bigg\{  & ik\vth\paren{|c|^2 - Z \varrho + \half \vartheta^*(1 + c^2 - Z \varrho)} w\xi + \half ik\vth c^*\vartheta (w^2-3) 
    \\ & + 2\mu c w \xi - \half \Gamma (w^2-3) + \half \mu |\vartheta|^2 (w^2-3) %removed purely imaginary    i\omega term
    - k \vth \mc F_2    \Bigg\} ,
    \end{split}
\end{equation} 
where, in the interest of concision, we defined an auxiliary function
\begin{equation}
    \label{eq_app_F2_mc_def}
    \begin{split}
    \mc F_2 = &\quad i \frac{\paren{-ic^* + i W^*\paren{1 - \frac{3}{2}\vartheta^*}  - N - N_T N \vartheta^* - \half i \vartheta^* w\xi}D^*}{|D|^2} 
    \\ & \hspace{1.5cm} \times \bigg[c\paren{1- \frac{3}{2}\vartheta} + Xw\xi - c w^2\xi^2 - \half \vartheta W w^2 \bigg]
    \\ & + i \bigg( -Z\varrho^* - c^*w\xi + (c^*)^2\bigg)\frac{\paren{-iw^2\xi D + w + ihN\xi} (D^*)^2}{w |D|^4}
     \\ & \hspace{1.5cm} \times \bigg[c\paren{1- \frac{3}{2}\vartheta} + Xw\xi - c w^2\xi^2 - \half \vartheta W w^2   \bigg]
    \\ & + i \bigg( -Z\varrho^* - c^*w\xi + (c^*)^2\bigg)\frac{i D^*}{|D|^2}
    \bigg[ X - 2\paren{1 - \half \vartheta}c w\xi - \vartheta \paren{\xi^2 + \half} w^2 \bigg]
    \\ & + i \half \vartheta^* \bigg( -w^2\xi + c^* w\bigg)\frac{\paren{-iw^2 D +w\xi + ihN} (D^*)^2}{w|D|^4}
     \\ & \hspace{1.5cm} \times \bigg[c\paren{1- \frac{3}{2}\vartheta} + Xw\xi - c w^2\xi^2 - \half \vartheta W w^2 \bigg]
    \\ & + i \half \vartheta^* \bigg( -w^2\xi + c^* w\bigg)\frac{i D^*}{|D|^2}\bigg[ X\xi - 2\paren{\xi^2 - \half \vartheta}c w - \frac{3}{2}\vartheta w^2 \xi \bigg] 
    \end{split}
\end{equation}
so that $\wt B =  k\vth \mc F_2 $. 
We next compute $J_2^{(R,I)}$. From \eqref{eq_f2_F2_J2} and \eqref{eq_app_y2_evol_B}, we have
\begin{equation}
    \label{eq_J_complex_forms}
    \begin{split}
        &\Big[J^{(R)}_2\Big(\cos\big([\omega - \bs k \cdot \bs w \vth] t\big)e^{-(\mu + \nu) t}   - e^{-\nu t}\Big)+ J^{(I)}_2 \sin\big([\omega - \bs k \cdot \bs w \vth] t\big)e^{-(\mu + \nu) t}  \Big]
        \\
        & = 2\mr{Re} \Bigg\{ \frac{-k\vth \mc F_2}{i\omega - \mu -i \bs k \cdot \bs w \vth} \bigg[ e^{(i\omega - \mu -i \bs k \cdot \bs w \vth)t} - e^{-\nu t} \bigg]\Bigg\} .
    \end{split}
\end{equation}
Recalling that $\bs k \cdot \bs w = k w \xi$ and $\omega - i\mu = k\vth c$, the right-hand side (RHS) of \eqref{eq_J_complex_forms} can be expanded into
\begin{equation}
    \label{eq_J_complex_RHS_expanded}
    \text{RHS} = 2\mr{Re} \Bigg\{ \frac{i \mc F_2}{c^* - k w \xi} \bigg[ \cos\big([\omega - k w \xi]t\big) + i\sin\big([\omega - k w \xi]t\big) - e^{-\nu t} \bigg]\Bigg\} ,
\end{equation}
then matching the sine, cosine, and pure exponential terms of \eqref{eq_J_complex_forms} yields \eqref{eq_J2_R_final} and \eqref{eq_J2_I_final}. 

Finally, we compute $M_2^{(R,I)}$. Note first that the time derivative acting on the exponential terms in \eqref{eq_f1_evol_y_soln} gives
\begin{equation}
    \frac{\partial e^{-\nu t -i \bs k \cdot \bs w \vth t}}{\partial t}  = \paren{\frac{h \nu \bs w}{w^2} - i \bs k \vth} e^{(i \bs k \cdot \bs w \vth - \nu)} .
\end{equation}
Then, based on \eqref{eq_app_y2_evol_C}, we define an auxiliary function $\mc G_2$ according to
\begin{align}
    \label{eq_app_G2_mc_def}
    \mc G_2 = &\Bigg(\Big[Z\varrho^* + c^* w\xi - (c^*)^2\Big]\Big(ih N \xi + w\Big) + \Big[ \half \vartheta^* w^2 \xi - \half c^* \vartheta^* w \Big]\Big(i h N + w\xi \Big)\Bigg) \nonumber
    \\ & \hspace{0.0cm} \times \frac{-i D^*}{w|D|^2}\bigg[c\paren{1- \frac{3}{2}\vartheta} + Xw\xi - c w^2\xi^2 - \half \vartheta W w^2 \bigg]
\end{align}
such that $\wt C = k^2 \vth^2 \mc G_2$. Then, from \eqref{eq_f2_F2_J2}, we have
\begin{equation}
    \label{eq_M_complex_forms}
    \begin{split}
        & \Big[( \mu t + 1)M^{(R)}_2 - (\omega - \bs k \cdot \bs w \vth)tM^{(I)}_2\Big]\cos\big([\omega - \bs k \cdot \bs w \vth] t\big)e^{-(\mu + \nu) t} - M^{(R)}_2e^{-\nu t}
        \\ &\hspace{0.0cm} + \Big[(\omega - \bs k \cdot \bs w \vth)tM^{(R)}_2 + (\mu t + 1)M^{(I)}_2  \Big]\sin\big([\omega - \bs k \cdot \bs w \vth] t\big)e^{-(\mu + \nu) t}
        \\
        & = 2\mr{Re} \Bigg\{ \frac{k^2 \vth^2 \mc G_2 }{(-i\omega + \mu + i \bs k \cdot \bs w \vth)^2} \bigg[ \big([-i\omega + \mu + i \bs k \cdot \bs w \vth]t + 1\big) e^{(i\omega - \mu -i \bs k \cdot \bs w \vth)t} - e^{-\nu t} \bigg]\Bigg\} .
    \end{split}
\end{equation}
Matching the sine, cosine, and pure exponential terms of \eqref{eq_M_complex_forms} yields \eqref{eq_M2_R_final} and \eqref{eq_M2_I_final}.

\section{Long-wavelength kinetic response}
\label{sec_app_bigN}

This appendix provides the intermediate steps in the calculation of the long-wavelength kinetic response in \S\ref{sec_kinetic}. 
As described in \S\ref{sec_kinetic_approx_bigN}, the expansion in long perturbation wavelengths requires keeping terms to second order in $1/N$. Noting that
\begin{equation}
    \label{eq_varphi10_term_avg}
    \avg{\paren{4n_1/n_0 + 2\alpha T_1/T_0}\varphi_{10}} = 2\mr{Re} \{(4 + 2\alpha \vartheta^*) \wt n^*\wt F_1 \} \frac{1}{\fm} ,
\end{equation} 
we have from \eqref{eq_f1_kinetic_V_avg} that
\begin{equation}
    \label{eq_f1_bigN}
    \begin{split}
    \avg{\paren{4\frac{n_1}{n_0} + 2\alpha\frac{T_1}{T_0}}\varphi_{10}} \sim & 2\mr{Re}\Bigg\{\paren{\frac{i}{N} + \frac{W}{N^2}}  \paren{4 + 2\alpha \vartheta^*} 
    \\ & \times \bigg[ c \paren{1 - \frac{3}{2}\vartheta} + X w\xi - \paren{\xi^2 - \half \vartheta} c w^2  - \half \vartheta w^3 \xi  \bigg]   \Bigg\} |\wt n|^2 .
    \end{split}
\end{equation}
\begin{equation}
    \label{eq_f1_bigN}
    \begin{split}
    \avg{\paren{4\frac{n_1}{n_0} + 2\alpha\frac{T_1}{T_0}}\varphi_{10}} \sim & \frac{1}{N}2 \mr{Re} \Bigg\{i(4 + 2\alpha \vartheta^*)
    \\ & \times \bigg[ c_R \paren{1 - \frac{3}{2}\vartheta} + X w\xi - c_R\paren{\xi^2 - \half \vartheta}  w^2  - \half \vartheta w^3 \xi  \bigg] \Bigg\}   |\wt n|^2
    \\
    & + \widehat \mu 2 \mr{Re} \left\{(4 + 2\alpha \vartheta^*) \bigg[  \paren{1 - \frac{3}{2}\vartheta} + 2 c_R w \xi - \paren{\xi^2 - \half \vartheta}w^2   \bigg] \right\}   |\wt n|^2 
    \\
    & + \frac{1}{N^2 }2 \mr{Re} \Bigg\{(w\xi - c)(4 + 2\alpha \vartheta^*)
    \\ & \times \bigg[ c_R \paren{1 - \frac{3}{2}\vartheta} + X w\xi - c_R\paren{\xi^2 - \half \vartheta}  w^2  - \half \vartheta w^3 \xi  \bigg] \Bigg\}   |\wt n|^2 ,
    \end{split}
\end{equation}
where $c_R = \mr{Re}\{c\}$ is the real phase velocity.

To make more explicit the analogy to the hydrodynamic response described in \S\ref{sec_hydro}, and particularly in \eqref{eq_Pf_pert_alpha}, it is useful to write the fusion-power enhancements in terms of the fluctuation amplitudes $\avg{n_1^2}$, $\avg{T_1^2}$, and so on. 
Recalling that $X = c^2 - Z\varrho - 1 + 3\vartheta/2$ and averaging over angles, we have
\begin{equation}
    \label{eq_Phi10}
    \begin{split}
        \Phi_{10} \sim &\frac{1}{N_*} \Bigg[ (2\alpha + 6) \paren{1 - \frac{1}{3}w_*^2} c_R\frac{\avg{n_1(\bs x) T_1\paren{\bs x - \frac{\bs L}{4}}}}{n_0 T_0}  \Bigg] 
        \\
        & + \widehat \mu \paren{1 - \frac{w_*^2}{3}} \Bigg[ 4 \frac{\avg{n_1^2}}{n_0^2} - 3\alpha \frac{\avg{T_1^2}}{T_0^2} + (2\alpha - 6) \frac{\avg{n_1T_1}}{n_0 T_0}   \Bigg]
        \\ & - \frac{c_R^2}{N_*^2} \paren{1 - \frac{w_*^2}{3}} \Bigg[ 4 \frac{\avg{n_1^2}}{n_0^2} - 3\alpha \frac{\avg{T_1^2}}{T_0^2} + (2\alpha - 6) \frac{\avg{n_1T_1}}{n_0T_0} \Bigg]
        \\ & + \frac{w_*^2}{N_*^2} \Bigg[ \frac{4}{3}\paren{c_R^2 - 1} \frac{\avg{n_1^2}}{n_0^2} + \alpha\paren{1 - \frac{w_*^2}{3}} \frac{\avg{T_1^2}}{T_0^2} 
        \\ & + \paren{\frac{2}{3}\alpha\paren{c_R^2 - 1} + 2\paren{1 - \frac{w_*^2}{3}}} \frac{\avg{n_1T_1}}{n_0T_0} 
         - \frac{4}{3} \frac{\avg{n_1 p_{e,1}}}{n_0^2 T_0} - \frac{2}{3}\alpha \frac{\avg{T_1 p_{e,1}}}{n_0T_0^2}\Bigg],
    \end{split}
\end{equation}
where $\bs L = 2\pi \bs{e_z}/k$ is a vector in the direction of the perturbation with length equal to the perturbation wavelength, and $\bs x$ is the quantity over which the volume average is taken. Note that, by construction, ${T_1(t - \Delta t, \bs x) = T_1(t, \bs x - \Delta x)}$ when ${\omega \Delta t = k \Delta x}$, so the spatial phase shift in the first line of \eqref{eq_Phi10} can be replaced with an equivalent temporal phase shift.

While it would appear that the leading-order kinetic contribution to the fusion power is of $\mc{O}(1/N)$, obviating the need to calculate the $\mc O(1/N^2)$ contributions from $\avg{\varphi_{11}}$ and $\avg{\varphi_{20}}$, this is only true when some components of the perturbation are out of phase, i.e. when $c$, $\vartheta$, $\varrho$, or $X$ has an imaginary part. Otherwise the leading-order term in the first line of \eqref{eq_Phi10} vanishes. In an acoustic wave, the phase shift between the wave components scales with ${\mu/k\vth}$. The contribution to $\Phi_{10}$ from nonzero $\avg{n_1(\bs x)T_1(\bs x - \bs L/4)}$ (or equivalently from nonzero $\vartheta_I$) is then of $\mc O(\widehat \mu)$.

%When dissipation is driven by diffusive processes, such as viscosity or heat conduction, the dissipation rate scales with the inverse square of the wavelength ($\mu \sim \mc O(1/N^2)$). In these cases, the $i/N$ term is actually of $\mc O(1/N^3)$ and therefore does not constitute the leading-order kinetic contribution to the fusion power. In the non-dissipative limit considered below, the leading-order contribution from \eqref{eq_f1_bigN} comes from the real part of the $W/N^2$ term multiplied by the real part of the rest of the bracketed expression. 
%This means that, in many cases of interest, the $\varphi_{10}$ term of \eqref{eq_K_expansion_varphi} contributes to $K$ at the same order as the $\varphi_{11}$ and $\varphi_{20}$ terms, so all three must be computed to obtain the leading-order kinetic correction to the fusion power.

The leading-order $\Phi_{11}$ contribution, appearing at $\mc O(1/N^2)$, can be found from \eqref{eq_f1f1_avg_corr} simply by replacing $|D|^2$ with $N^2$. When evaluating \eqref{eq_f1f1_avg_corr} at the Gamow peak, $\bs w = -\bs w'$ and $\xi = -\xi'$. After the angular average, we obtain
\begin{equation}
    \label{eq_Phi11_NEW}
    \begin{split}
    \Phi_{11} \sim & \frac{1}{N_*^2} \Bigg[ \paren{c_R^2\paren{1 -\frac{2}{3}w_*^2 + \frac{1}{5}w_*^4} - (c_R^2-1)^2\frac{1}{3}w_*^2}\frac{\avg{n_1^2}}{n_0^2} 
    \\ & + \paren{c_R^2\paren{\frac{9}{4} - \frac{3}{2}w_*^2 + \frac{1}{4}w_*^4} - \frac{3}{4}w_*^2 + \frac{1}{2}w_*^4 - \frac{1}{12}w_*^6} \frac{\avg{T_1^2}}{T_0^2}
    \\ & + \paren{-3c_R^2 + (c_R^2+1)w_*^2 - c_R^2\frac{1}{3}w_*^4 +(c_R^2 - 1)\frac{1}{3}w_*^4} \frac{\avg{n_1T_1}}{n_0T_0}
    \\ & + \frac{2}{3}\paren{c_R^2 - 1}w_*^2 \frac{\avg{n_1 p_{e,1}}}{n_0^2 T_0} + w_*^2\paren{1 - \frac{w_*^2}{3}} \frac{\avg{T_1 p_{e,1}}}{n_0 T_0^2} - \frac{w_*^2}{3} \frac{\avg{p_{e,1}^2}}{n_0^2 T_0^2} \Bigg] .
    \end{split}
\end{equation}

%\clearpage

The $\Phi_{20}$ contribution follows from \eqref{eq_F2_final}. We assume an ordering ${\widehat \Gamma \sim \mc{O}(1)}$. Then we have

\begin{equation}
    \label{eq_Phi20}
    \begin{split}
    \Phi_{20} \sim & -2\frac{N_T}{N_*} c_R\paren{1 - \frac{w_*^2}{3}} \frac{\avg{n_1(\bs x) T_1(\bs x - \frac{\bs L}{4})}}{n_0T_0} 
    \\ & +   \widehat \mu \paren{1 - \frac{w_*^2}{3}} \Bigg[-2N_T \frac{\avg{n_1T_1}}{n_0T_0} + 3(N_T + 1)\frac{\avg{T_1^2}}{T_0^2} - (2 + 3\widehat \Gamma) \frac{\avg{n_1^2}}{n_0^2}  \Bigg]
    \\ 
    & - \frac{2}{N_*^2}\Bigg[ \paren{1 + \frac{1}{3}h}c_R^4 +  \paren{-3 - \frac{2}{3}h} c_R^2 + \paren{-\frac{1}{3}c_R^4 + \frac{5}{3}c_R^2 - \frac{1}{3} + \frac{1}{5}hc_R^2} w_*^2 - \frac{1}{5}c_R^2w_*^4\Bigg] \frac{\avg{n_1^2}}{n_0^2}
    \\ 
    & - \frac{2}{N_*^2}\Bigg[\paren{-\frac{9}{4} - \frac{3}{4}h} c_R^2 + \paren{2c_R^2 - \frac{5}{4} + \frac{1}{4}hc_R^2 - \frac{1}{4}h}w_*^2 + \paren{\frac{5}{6} - \frac{1}{4}c_R^2 + \frac{1}{12}h}w_*^4 - \frac{1}{12}w_*^6\Bigg] \frac{\avg{T_1^2}}{T_0^2}
    \\ 
    & - \frac{2}{N_*^2}\Bigg[\paren{6 + \frac{3}{2}h }c_R^2 
         -\paren{\frac{16}{3}c_R^2 + \frac{2}{3}hc_R^2 - \frac{4}{3} - \frac{1}{6}h}w_*^2 + \frac{1}{3}(2c_R^2 - 1)w_*^4
    \Bigg]\frac{\avg{n_1T_1}}{n_0T_0}
    \\ 
    & - \frac{2}{N_*^2}\Bigg[ -\paren{1 + \frac{1}{3}h}(2c_R^2 - 1) + \frac{2}{3}(c_R^2-1)w_*^2  \Bigg] \frac{\avg{p_{e,1} n_1}}{n_0^2T_0}
    \\
    &- \frac{2}{N_*^2}\Bigg[-\frac{3}{2}\paren{1 + \frac{1}{3}h} + \paren{\frac{13}{6} + \frac{1}{3}h}w_*^2 - \frac{1}{3}w_*^4 \Bigg]\frac{\avg{p_{e,1}T_1}}{n_0T_0^2}
    \\ 
    & - \frac{2}{N_*^2}\Bigg[ \paren{1 + \frac{1}{3}h} - \frac{1}{3}w_*^2\Bigg] \frac{\avg{p_{e,1}^2}}{n_0^2 T_0^2}    
    \end{split} .
\end{equation}
With these pieces in place, \S\ref{sec_kinetic_approx_bigN} derives a formula for the full kinetic fusion-power response in the long-wavelength limit.

\bibliography{jpp_acoustic}

\end{document}